\documentclass[twocolumn]{aastex61}

\usepackage{graphicx}	
\usepackage{amsmath}	
\usepackage{amssymb}	
\usepackage{enumitem}

\newbox\rotatingbox
\def\rotatetablecbw{\setbox\rotatingbox=\vbox\bgroup}
\def\endrotatetablecbw{\egroup\rotatebox{90}{\hbox to \textheight{\vbox{\unvbox\rotatingbox}}}}

\newcommand{\ssz}{224}
\newcommand{\bsz}{39}
\newcommand{\tsz}{263}
\newcommand{\gbmgii}{113}

\def\nodata{ ~$\cdots$~ }

\newcommand{\zabs}{$z_{\rm abs}$}

\newcommand{\cmm}{cm$^{-2}$}

\newcommand{\mnhi}{N_{\rm H\,I}}
\newcommand{\mnhii}{N_{\rm H\,II}}
\newcommand{\lnhi}{$\log N_{\rm H\,I}$}
\newcommand{\mlnovi}{\log N_{\rm O\,VI}}

\newcommand{\mlnmgii}{\log N_{\rm Mg\,II}}

\newcommand{\mlnhi}{\log N_{\rm H\,I}}

\newcommand{\xh}{\ensuremath{{\rm [X/H]}}}
\newcommand{\ah}{\ensuremath{{\rm [\alpha/H]}}}
\newcommand{\ca}{\ensuremath{{\rm [C/\alpha]}}}
\newcommand{\km}{${\rm km\,s}^{-1}$}

\newcommand{\hi}{\ion{H}{1}}

\newcommand{\cii}{\ion{C}{2}}
\newcommand{\ciii}{\ion{C}{3}}

\newcommand{\nii}{\ion{N}{2}}
\newcommand{\niii}{\ion{N}{3}}
\newcommand{\niv}{\ion{N}{4}}

\newcommand{\oi}{\ion{O}{1}}
\newcommand{\oii}{\ion{O}{2}}
\newcommand{\oiii}{\ion{O}{3}}
\newcommand{\oiv}{\ion{O}{4}}

\newcommand{\ovi}{\ion{O}{6}}

\newcommand{\Siii}{\ion{S}{3}}

\newcommand{\siii}{\ion{Si}{2}}
\newcommand{\siiii}{\ion{Si}{3}}
\newcommand{\siiv}{\ion{Si}{4}}

\newcommand{\mgii}{\ion{Mg}{2}}

\newcommand{\feii}{\ion{Fe}{2}}

\newcommand{\znii}{\ion{Zn}{2}}

\newcommand{\nhi}{$N_{\rm H\,I}$}

\newcommand{\mathNHI}{N_{\rm H\,I}}
\newcommand{\logNHI}{\log N_{\rm H\,I}}

\newcommand{\mathnmgii}{N_{\rm Mg\,II}}

\newcommand{\lya}{Ly\,$\alpha$\relax}

\newcommand{\hst}{{\it HST}}
\newcommand{\fuse}{{\it FUSE}}

\newcommand{\logU}{\ensuremath{\log U}}

\defcitealias{lehner13}{{\rm L13}}
\defcitealias{lehner18}{{\rm Paper I}}
\defcitealias{lehner18b}{{\rm Paper III}}
\defcitealias{wotta16}{{\rm W16}}
\defcitealias{haardt96}{{\rm HM05}}
\defcitealias{haardt12}{{\rm HM12}}

\newcommand{\strongLLSfLowmetPLLSLLS}{56\%}

\newcommand{\strongLLSlogUgaussianTest}{0.79} 
\newcommand{\PLLSMCMCmedian}{-1.35}
\newcommand{\PLLSMCMCmedianep}{1.36} 

\shortauthors{Wotta et al.}
\shorttitle{CCC II: Metallicities of the pLLSs and LLSs at $\lowercase{z}\lesssim 1$}

\begin{document}

\title{The COS CGM Compendium. II: Metallicities of the Partial and Lyman Limit Systems at $\lowercase{z}\lesssim 1$}

\author{Christopher B. Wotta}
\affiliation{Department of Physics, University of Notre Dame, Notre Dame, IN 46556, USA}
\author{Nicolas Lehner}
\affiliation{Department of Physics, University of Notre Dame, Notre Dame, IN 46556, USA}
\author{J. Christopher Howk}
\affiliation{Department of Physics, University of Notre Dame, Notre Dame, IN 46556, USA}
\author{John M. O'Meara}
\affiliation{Department of Chemistry and Physics, Saint Michael's College, Colchester, VT 05439, USA}
\affiliation{W.M. Keck Observatory 65-1120 Mamalahoa Highway Kamuela, HI 96743, USA}
\author{Benjamin D. Oppenheimer}
\affiliation{CASA, Department of Astrophysical and Planetary Sciences, University of Colorado, Boulder, CO 80309, USA}
\author{Kathy L.~Cooksey}
\affiliation{Department of Physics and Astronomy, University of Hawai`i at Hilo, HI 96720, USA}

\begin{abstract}
We present the results from our COS circumgalactic medium (CGM) compendium (CCC), a survey of the CGM at $z\la 1$ using \hi-selected absorbers with $15<\mlnhi <19$. We focus here on 82 partial Lyman limit systems (pLLSs, $16.2 \le \mlnhi < 17.2$) and 29 LLSs ($17.2 \le \mlnhi < 19$). Using Bayesian techniques and Markov-chain Monte Carlo sampling of a grid of photoionization models, we derive the posterior probability distribution functions (PDFs) for the metallicity of each absorber in CCC. We show that the combined pLLS metallicity PDF at $z\la 1$ has two main peaks at  $\xh \simeq -1.7 $ and $-0.4$, with a strong dip at $\xh \simeq -1$. The metallicity PDF of the LLSs might be more complicated than an unimodal or bimodal distribution. The pLLSs and LLSs probe a similar range of metallicities $-3 \la \xh \la +0.4$, but the fraction of very  metal-poor absorbers with $\xh \la -1.4$ is much larger for the pLLSs than the LLSs. In contrast, absorbers with $\mlnhi \ge 19$ have mostly $-1 \la \xh \la 0$ at $z\la 1$. The metal-enriched gas probed by pLLSs and LLSs confirms that galaxies that have been enriching their CGM over billions of years. Surprisingly, despite this enrichment, there is also  abundant metal-poor CGM gas (41--59\% of the pLLSs have $\xh \la -1.4$), representing a reservoir of near-pristine gas around $z\la 1$ galaxies. We  compare our empirical results to recent cosmological zoom simulations, finding some discrepancies, including an overabundance of metal-enriched CGM gas in simulations.
\end{abstract}
\keywords{cosmology: observations --- galaxies: abundances --- galaxies: evolution --- galaxies: halos --- intergalactic medium --- quasars: absorption lines}

\section{Introduction}
\label{s-strongLLS_intro}

It is generally accepted that accretion and galactic-scale outflows strongly influence the fate of galaxies, and they are often invoked to explain observable properties of galaxies \citep[e.g.,][]{keres05,dekel06,faucher-giguere11,sommerville15,tumlinson17}. The direct signatures of these inflows and outflows can be observed in the circumgalactic medium (CGM) of galaxies. However, the actual means by which a galaxy acquires additional material or ejects matter and energy into its CGM and beyond are still very much under debate. Both their limited resolution and the complexity of the physical processes at play forces current cosmological simulations to use sub-grid prescriptions when modeling the processes that drive the large-scale flows in the CGM \citep[e.g.,][]{sommerville15,tumlinson17}. Cosmological and zoom simulations have been quite successful in  reproducing  galaxy properties in reasonable agreement with the observations, such as, e.g., the  galaxy stellar-mass function, galaxy clustering, and in some cases the \ovi\ distribution in the CGM of galaxies  \citep[e.g.,][]{hopkins14,schaye15,sommerville15,oppenheimer16,springel18,nelson18}. The feedback prescriptions are, however, different in all these simulations and comparison with galaxy properties alone appear not to be enough to differentiate between them \citep[e.g.,][]{hummels13}. The next frontier for simulations is to improve the feedback modeling and resolution to attempt to produce simulated galaxies that match both real galaxies' stellar properties and those of the CGM \citep[e.g.,][]{scannapieco15,tumlinson17,hopkins18,mccourt18,vandevoort18,peeples18,vandevoort18,suresh18}. To test and confront these ever-improving simulations and to constrain the phenomena that shape galaxies and their evolution, a robust empirical characterization of CGM gas properties is critical.

Over the last few years, as a result of the installation of the Cosmic Origins Spectrograph (COS) on the {\em Hubble Space Telescope} and ground-based observing campaigns of galaxies, our knowledge  of the CGM and its relationship to galaxies at $z\la 1$ has been redefined \citep[e.g.,][]{tumlinson11a,werk13,borthakur13,stocke13,bordoloi14,liang14,kacprzak15,lehner15,johnson15,burchett16,chen17,keeney17,prochaska17}. COS has transformed our understanding of the CGM and IGM field thanks to its high sensitivity and resolution in the ultraviolet wavelengths where a rich suite of diagnostics is available. The COS QSO archive has grown to a level where unprecendented large samples of high resolution ($R\sim 17,000$) QSO spectra can be assembled to study the gaseous content of the universe  from $z=0$ to $z\la 1.5$ (see, e.g., the recent survey of the Milky Way halo by \citealt{richter17} where 270 QSO sightlines were used). 

Our group has embarked on a survey, the COS CGM Compendium (CCC), where our main aim is to characterize the properties of the cool gas in the $z\la1$ CGM using a sample of \hi-selected absorbers with \hi\ column densities $15 <\mlnhi <19$ \citep[see ][hereafter \citetalias{lehner18}]{lehner18}. Absorbers with these \hi\ column densities are known to probe largely the CGM gas at low redshift (e.g., \citealt{lanzetta95,penton02,bowen02,chen05,prochaska11a,prochaska11c,lehner13} and see also \citetalias{lehner18}). With CCC and in this work, we focus on deriving the metallicity of these CGM absorbers since the enrichment levels of the gas help differentiate between the plausible origins of the gas, including metal-enriched outflows or accretion of metal-poor gas. We approach this problem using an \hi\ selection and requiring good coverage of strong metal-line transitions to minimize observational biases in the metallicities of these absorbers. We are therefore able to use our results to directly determine the metallicity distributions of the cool CGM.

CCC follows two previous surveys led by our group at $z\la 1$ (\citealt{lehner13,wotta16}, hereafter \citetalias{lehner13} and \citetalias{wotta16}, respectively). In these works we investigated the metallicity distribution of cold $z\la 1$ CGM absorbers probed by partial Lyman limit systems (pLLSs) and LLSs\footnote{We adopt the  \citetalias{lehner18} definition for the absorbers in CCC: pLLSs and LLSs have \hi\ column densities  $16.2 \la \mlnhi <17.2 $ and $17.2 \le \mlnhi <19$, respectively. Damped \lya\ absorbers (DLAs) have $\mlnhi \ge 20.3$ and  super LLSs (SLLSs, a.k.a. sub-DLAs) have $19.0 \le \mlnhi <20.3$. Finally, absorbers with $15 \la \mlnhi <16.2$, which are the focus of \citet{lehner18b} (hereafter \citetalias{lehner18b}), are defined as the strong \lya\ forest systems (SLFSs). See \citetalias{lehner18} for the motivation of these various \nhi\ intervals.} and found that about half of the CGM gas has a  low metallicity with $\xh \la -1$.\footnote{We use the standard notation $\xh \equiv \log N_{\rm X}/N_{\rm H} - \log {\rm X/H}_\sun$. In our case X is typically an $\alpha$-element, unless otherwise stated.} This suggests that the $z\la 1$ CGM is host to a large reservoir of metal-poor gas, which was previously unrecognized. Furthermore, we also found that the metallicity distribution of the pLLSs was strikingly different from those of the higher-\nhi\ absorbers, the SLLSs and DLAs ($\mlnhi > 19$). Our early results strongly imply that there are two populations of gas probed by the pLLSs: gas at $\sim$2\% solar metallicity and below and gas at $\sim$40\% solar metallicity and above, with a paucity of gas around $\sim$10\% solar metallicity. 

Our initial survey \citepalias{lehner13} consisted of 28 absorbers (23 pLLSs and 5 LLSs). Our second survey doubled the size of our initial sample (44 pLLSs and 11 LLSs), confirming our initial results and tentatively showing a change in the shape of the metallicity distribution between the pLLSs and LLSs as well as a lower frequency of metal-poor LLSs compared to pLLSs \citepalias{wotta16}. While our surveys have increased by an order of magnitude the number of pLLSs and LLSs where the metallicities have been estimated at $z\la 1$ compared to the status prior to the installation of the COS onboard \hst\ (see \citetalias{lehner18}), the sample of absorbers was still relatively small, especially when the pLLSs and LLSs are separately considered.

In our first paper describing CCC \citepalias{lehner18}, we built the largest sample to date of \hi-selected absorbers with $15<\mlnhi < 19$ at $z\la 1$. The sample presented in \citetalias{lehner18} was largely from COS G130M and/or G160M spectra. These spectra were drawn from the first data release of the HST Spectroscopic Legacy Archive (HSLA, \citealt{peeples17} with some complementary in-house data reduction, see \citetalias{lehner18}) available at the \textit{Barbara A. Mikulski Archive for Space Telescopes} (MAST).  Additional pLLSs and LLSs were included from the literature, including absorbers observed with \fuse, \hst /STIS and \hst /COS G140L, including our own COS G140L survey \citepalias{wotta16}. In this second paper we focus on the metallicity distribution of the pLLSs and LLSs at $z\la 1$. We increase the sample of pLLSs by a factor $\sim$2 (from 44 in \citetalias{wotta16} to 82) and of LLSs by a factor $\sim$3 (from 11 in \citetalias{wotta16} to 29). Overall we have increased the pLLS+LLS sample by a factor 4 from the initial sample of \citetalias{lehner13}. 

As shown in \citetalias{lehner18}, the gas in the $15 < \mlnhi < 19$ range is largely ionized, i.e., $\mnhii \gg \mnhi$. Hence, large ionization corrections are required when comparing the column density of \hi\ with those of singly- and doubly-ionized species to derive the metallicities.\footnote{Neutral metal lines such as \oi\ that would require much smaller corrections are typically not detected, and the upper limits on their column densities do not typically offer interesting constraints.} In our previous surveys---and for literature metallicities derived prior to \citetalias{lehner13}---the ionization corrections were assessed by comparing ionization model outputs with measured column densities in a rough way that led to ill-defined (or even unreported) uncertainties. Most often the range of valid models was assumed to be those that violated no observational constraints at the $1\sigma$ level, and the uncertainties were mapped to that range. Metallicity limits were often more crudely defined, and the inadequacies of this approach led to complications in characterizing the metallicity distributions or comparing with the results from cosmological simulations. Furthermore, not all absorbing systems compiled for assessing metallicity distributions were modeled with the same ionizing EUV background (EUVB), introducing an additional systematic uncertainty \citepalias{wotta16}. 

To derive the metallicity in CCC, we adopt and extend the methodology developed by \citet[see also \citealt{crighton15}]{fumagalli16} that combines a Bayesian formalism and Markov Chain Monte Carlo (MCMC) techniques with grids of photoionization models to determine the ionization corrections necessary to derive gas-phase metal abundance. The Bayesian MCMC approach allows us to robustly assess the posterior probability distribution functions (PDFs) of the metallicities of the absorbers (and hence lower or upper limits are now robustly characterized by a posterior PDF). It also allows us to reliably estimate the confidence intervals on the metallicities of our absorbers, and thus on the resulting distribution of $z\lesssim1$ pLLS and LLS metallicities. Here we apply the Bayesian MCMC methods to the entire CCC survey and use the same EUVB so that our analysis is consistent across datasets; we also test different EUVB radiation fields to explore the effects from an uncertain ionizing background on the metallicity estimates.  

Our paper is organized as follows. In \S\ref{s-sample}, we briefly describe the sample of pLLSs and LLSs as well our newly assembled sample of DLAs and \hi-selected SLLSs from the literature, which we use to study the evolution of the metallicity as a function of \nhi.  In \S\ref{s-strongLLS_ioncorrect_and_MCMC_metallicity}, we present our methodology and assumptions for estimating the metallicity; we compare the newly derived metallicities with our previous estimates and discuss their uncertainties and limitations. Our main results are presented in \S\ref{s-met-pdf-cgm} and discussed in \S\ref{s-disc}, in particular in the context of recent results from cosmological and zoom simulations \citep{hafen17,rahmati18}. In \S\ref{s-strongLLS_summary}, we summarize our main conclusions.


\section{Samples}\label{s-sample}
\subsection{Sample of pLLSs and LLSs}
\label{s-sample-lls}
In \citetalias{lehner18}, we assembled a sample of \ssz\ \hi-selected absorbers  with $15<\mlnhi < 19$ at $z\la 1$. The data were retrieved from the \hst/COS G130M and/or G160M archive with supplementary ground-based observations of \mgii\ and \feii\ with high resolution Keck/HIRES and VLT/UVES spectra. All these were uniformly analyzed to derive the column densities of \hi\ and metal atoms and ions. We refer the reader to \citetalias{lehner18} for more details. As noted in \citetalias{lehner18}, an additional \bsz\ absorbers are available from our two first surveys \citepalias{lehner13,wotta16}. Most of these extra absorbers were observed with \hst/STIS, \fuse, and \hst/COS G140L, and their column densities were not re-assessed in \citetalias{lehner18} (but see \S \ref{s-strongLLS_mgii_saturation_mods_vs_hires}). While we have not revisited the column densities for these  \bsz\ absorbers (except for the few discussed in \S\ref{s-strongLLS_mgii_saturation_mods_vs_hires}), we nevertheless wanted to use the same methodology and EUVB to estimate the metallicity from the column densities in order to avoid introducing spurious effects that could affect the metallicity distribution. We therefore estimate in this work the metallicity of the entire sample of  \tsz\ absorbers. In this paper, we focus solely on the pLLSs and LLSs, while in \citetalias{lehner18b} we will present the metallicities of the strong Ly$\alpha$ forest systems (SLFSs; $15<\mlnhi < 16.2$). 


\subsection{Low Resolution Observations of \mgii }
\label{s-strongLLS_mgii_saturation_mods_vs_hires}
As part of our survey we acquired new high-resolution observations of \mgii\ (and \feii; see \citetalias{lehner18}) for \gbmgii\ absorbers. Data for three of these absorbers were also initially obtained with low-resolution spectra ($R\sim 4,000$), which allow us to assess potential systematic errors from using low-resolution spectra to estimate the column densities. For these 3 absorbers, the  low resolution spectra suggested that the saturation of \mgii\ is relatively mild when comparing the weak and strong transitions of the \mgii\ doublet. However, in each of these cases, the high-resolution \mgii\ spectra imply that the \mgii\ is more saturated than initially realized with the low-resolution data. For these absorbers, the equivalent widths are similar in the low- and high-resolution spectra, but the apparent column densities differ significantly. The absorption in these absorbers is quite narrow (rather than being spread in several components), certainly helping hide unresolved saturation effects. 

\begin{figure}[tbp]
\epsscale{1.15}
\plotone{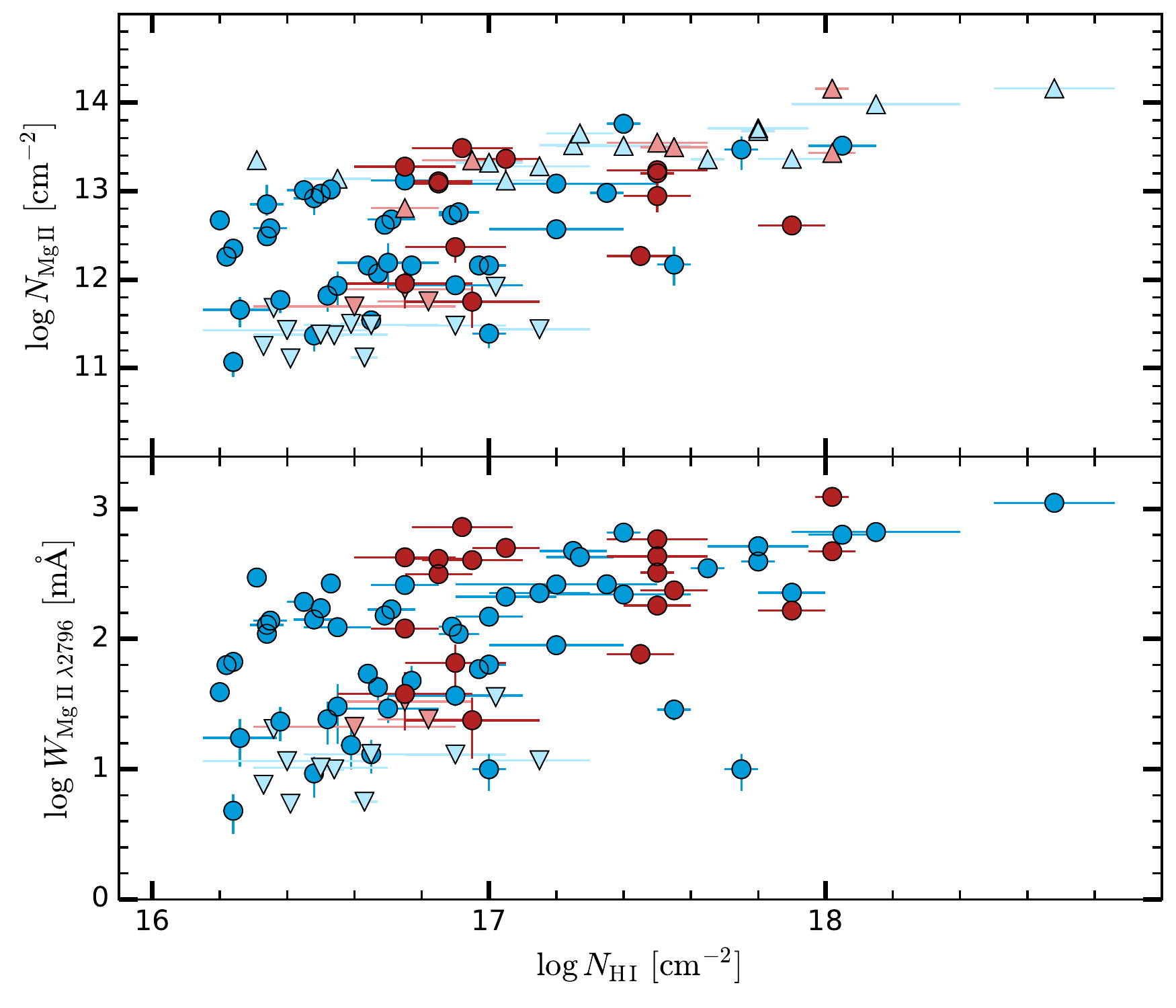}
    \caption{\mgii\ column densities and equivalent widths as a function of \nhi. Red points denote observations with a low-resolution spectrograph ($R\sim 4,000$), while blue points indicate \mgii\ observations made with high resolution ($R\ga 30,000$). Lower limits on \mgii\ are indicated by triangles, while downward triangles show $2\sigma$ upper limits.
    \label{f-strongLLS_mgii_vs_hi_2panels}}
\end{figure}

Owing to this finding, we feel the need to revisit the low-resolution results from \citetalias{wotta16}, where a large fraction of the \mgii\ observations were acquired using relatively low-resolution spectra ($R\sim 4,000$). This is even more critical since in that survey \mgii\ is the sole ion that was used to determine the metallicity using our ``low resolution method'' (see \citetalias{wotta16}). To assess the possible effects from unresolved saturation in low-resolution \mgii\ spectra, in Fig.~\ref{f-strongLLS_mgii_vs_hi_2panels} we adapt Fig.~10 from  \citetalias{lehner18} where we added the data from \citetalias{wotta16} (blue data being from high-resolution observations; red from low-resolution observations). For $\mlnmgii \ga 13.1$ ($W_{\rm Mg\,II \lambda 2796}\ga 250$ m\AA), 14/18 ($78\%$) of the high-resolution \mgii\ absorption lines were identified as saturated, with only a lower limit derived on the column density. There is also no correlation between the saturation of \mgii\ and the \nhi\ within the observed \nhi\ range. In contrast, 5/12 ($42 \%$) of the low-resolution \mgii\ absorption lines are saturated, with a lower limit on $N$ (not including the 3 original absorbers from the \citetalias{lehner13} sample). Based on this result, we treat hereafter any low-resolution \mgii\ absorption with  $\log\mathnmgii \ga 13.1$ as a lower limit to be conservative. We note this affects only 7 absorbers from the \citetalias{wotta16} sample. We also observe that there is otherwise no systematic effect in the low- and high-resolution data, and the upper limits are typically at about the same levels.

\subsection{Comparison Samples}
\label{s-sample-comp}
One of the goals of our survey is to determine the overall trend of the metallicity with \nhi. While our survey focuses on the absorbers with $15< \mlnhi <19 $, we also want to determine how the metallicity changes with \nhi, including in the high column density regime probed by  the SLLS ($19\le \mlnhi < 20.3$) and DLA ($\mlnhi \ge 20.3$) regimes. For this purpose, we construct a sample of DLAs and SLLSs at $z\la 1$ for which the metallicities have been estimated in a way that avoids obvious biases.  Below we discuss our compilation of the literature results and provide some caveats to keep in mind when comparing the literature metallicity estimates to those derived here for lower column density systems ($\mlnhi < 19$).

\subsubsection{Comparison Sample: DLAs}
\label{s-sample-dla}

For the DLAs at $z\la 1$, several compilations have been made over the years, and here we compare the compilations summarized by \citetalias{lehner13}, \citet{rafelski12}, and \citet{quiret16}. We found and corrected some small discrepancies in each of these compilations in order to compare DLA abundances with those derived in our work. In Table~\ref{t-dla}, we summarize the DLA sample and list in the last column the references to the original measurements. We note all the abundances have been re-scaled to the solar abundances from \citet{asplund09}. In total we found 35 measurements of DLAs at $z\la 1$ (the maximum redshift was chosen to match the sample of pLLSs and LLSs).  

For the DLAs (contrary to lower-\nhi\ absorbers), no ionization correction is needed as long as neutral or singly-ionized species are used. However, the reported metallicities of several DLAs were not based on an $\alpha$-element (e.g., O, Si, Mg), which are the most-commonly measured elements in lower column density absorbers at $z\la 1$. Instead, they were derived using 6 $\alpha$ measurements, 9 \feii\ measurements, and 19 \znii\ measurements (see Table~\ref{t-dla}). As noted in \citetalias{lehner13}, $[\alpha/{\rm Zn}] \approx 0 \pm 0.2$ based on the relative abundance analysis in stars and DLAs \citep{nissen07, rafelski12, skuladottir17, dasilveira18}; i.e., we can directly compare the metallicity distributions derived from Zn and $\alpha$-elements. 

The situation is more complicated for Fe, however. The gas-phase abundances of Fe in DLAs can be significantly affected by dust depletion \citep{jenkins09, de-cia18}. Additionally, there can be non-solar $\alpha$/Fe ratios \citep[typically enhancements; see, e.g.,][]{de-cia16} owing to the different nucleosynthetic origins of these elements, which can affect the comparison of DLA metallicities with those in lower column density systems ($\mlnhi \le 19$) for which metallicities are based on $\alpha$-elements. We apply corrections for the non-solar element ratios, adjusting measurements of [Fe/H] in DLAs into $\ah$ (although we use a more generic \xh\ throughout). The DLA compilations of \citet{rafelski12} and \citet{quiret16} both measure directly a ratio [Fe/$\alpha] = -0.32 \pm 0.18$ (where the uncertainty is the standard deviation in values). Those works applied this correction to DLA metallicities derived solely from \feii, and this serves both as an empirical correction for $\alpha$-enhancement and dust depletion; thus their metallicities were corrected such that $ [{\rm X/H}] = [{\rm Fe/H}]+0.3$. \citet{de-cia16} derive similar corrections based upon detailed modeling of the relative abundances and dust depletion patterns observed in DLAs finding  a typical $[{\rm Fe}/\alpha] \approx -0.3$ at low metallicities. However, this correction is valid only for DLAs with $\ah \le -1.5$. For higher metallicities---particularly important for DLAs at $z\la 1$ where typically $\ah \ga -1.5$---the ${\rm {Fe/\alpha}}$ ratio decreases further. Using the \citet{rafelski12} sample, we derive $\langle [{\rm Fe}/\alpha]\rangle = -0.5 \pm 0.3$; the larger correction here compared with lower metallicities is likely due to additional dust depletion on the top of the general $\alpha$-enhancement due to nucleosynthesis. This correction is quite similar to \citet{rao05}, who derive a $0.4$ dex correction based on a smaller sample. Therefore in this work we applied the following correction to the metallicity derived using \feii\ in DLAs (unless otherwise stated in Table~\ref{t-dla}): $ [{\rm X/H}] = [{\rm Fe/H}]+0.5$.

The above correction is a statistical one, and, indeed, in some cases where both $\alpha$ and Fe measurements exist, the depletion of Fe can be more severe. However, with this adopted +0.5 dex correction, the statistics of the DLA metallicities are now quite similar whether one includes Fe-based metallicities or not. When including the corrected Fe-based values, we find the mean (plus or minus the standard deviation) and the median of the DLA metallicities are $-0.69\pm 0.35$ and $-0.62$ (sample size 35); excluding the Fe-based metallicities yields $-0.61\pm 0.33$ and $-0.59$ (sample size 26).

Finally, as discussed in \citetalias{lehner13}, although the majority of $z\la1$ DLAs are metal-line selected, rather than \hi-selected, this should not bias the metallicity distribution of the DLAs (see also the discussion in \citealt{rao06} and \citealt{rao17}). The metal-line selection of low-redshift DLAs typically uses the NUV transitions of \mgii\ and/or \feii . The \mgii\ equivalent width cut-off for identifying DLAs (typically $W_\lambda \ge 0.6$ \AA) is low enough to select even low-metallicity DLAs at least at $z\la 1$. Indeed, $W_\lambda \ge 0.6$ \AA\ for \mgii\ $\lambda$2796 yields $\mlnmgii \ga 13.6$--14.0. The metallicity implied by $\mlnmgii \simeq 14$ would be $\xh \simeq -1.8$ for a DLA with $\mlnhi = 20.3$; if the \nhi\ is larger, that threshold would decrease to even lower values. Thus the selection via \mgii\ absorption should not bias the metallicity distribution of the DLAs at $z\la 1$.

\startlongtable
\begin{deluxetable*}{llccccc}
\tablecaption{Sample of DLAs \label{t-dla}}
\tablehead{
\colhead{Name}  & \colhead{Original}  & \colhead{$z_{\rm abs}$} & \colhead{$\mlnhi$} & \colhead{\xh} & \colhead{Ion} & \colhead{Ref.} \\
\colhead{} & \colhead{Name} & \colhead{} & \colhead{[\cmm]} & \colhead{} & \colhead{}& \colhead{}
}
\startdata
\hline
J154420.29+591227.1 &          HS1543+5921  &	0.0100  & $  20.42  \pm  0.04  $  & $ -0.35  \pm   0.06 $  & \ion{S}{2} 	& 1	  \\ 
J074110.70+311200.2 &   	 Q0738+313  &	0.0912  & $  21.18  \pm  0.05  $  & $ -1.14  \pm   0.24 $  & \ion{Fe}{2}	& 2, 3	  \\ 
J161916.54+334238.3 &   	J1619+3342  &	0.0963  & $  20.55  \pm  0.10  $  & $ -0.61  \pm   0.13 $  & \ion{S}{2} 	& 4	  \\ 
J100902.06+071343.8 &   	J1009+0713  &	0.1140  & $  20.68  \pm  0.10  $  & $ -0.57  \pm   0.16 $  & \ion{S}{2} 	& 4	  \\ 
J012236.76-284321.5 &  		  B0120-28  &   0.1856  & $ 20.50   \pm  0.10  $  & $  -1.19 \pm   0.20 $  & \ion{S}{2} 	& 5	  \\ 
J074110.70+311200.2 &   	 Q0738+313  &	0.2212  & $  20.90  \pm  0.07  $  & $ -0.70  \pm   0.18 $  & \ion{Zn}{2}     	& 2 	  \\ 
J095456.83+174331.2 &   	 Q0952+179  &	0.2378  & $  21.32  \pm  0.05  $  & $ -1.34  \pm   0.24 $  & \ion{Fe}{2}	& 2	  \\ 
J113007.04-144927.4 &   	 Q1127-145  &	0.3127  & $  21.70  \pm  0.08  $  & $ -0.73  \pm   0.15 $  & \ion{Zn}{2}     	& 6	  \\ 
J161649.43+415416.4 &   	J1616+4154  &	0.3213  & $ 20.60   \pm  0.20  $  & $ -0.37  \pm   0.23 $  & \ion{S}{2}		& 4	  \\ 
J123200.01-022404.8 &   	 Q1229-021  &	0.3950  & $ 20.75   \pm  0.07  $  & $ -0.41  \pm   0.12 $  & \ion{Zn}{2}	& 7	  \\ 
J081336.05+481302.6 &   	     3C196  &	0.4370	& $ 20.36   \pm  0.05  $  & $ -0.58  \pm   0.12 $  & \ion{Zn}{2}	& 4	  \\ 
J083052.08+241059.8 &   	 Q0827+243  &	0.5249  & $ 20.30   \pm  0.04  $  & $ -0.89  \pm   0.19 $  & \ion{Zn}{2}	& 3	  \\ 
J023838.93+163659.2 &   	 Q0235+164  &	0.5264  & $ 21.65   \pm  0.10  $  & $ -0.60  \pm   0.40 $  & \ion{Fe}{2}	& 8	  \\ 
J163145.25+115602.9 &   	   4C12.59  &	0.5310	& $ 20.70   \pm  0.09  $  & $ -1.42  \pm   0.22 $  & \ion{Fe}{2}	& 9	  \\ 
J143120.53+395241.5 &   	J1431+3952  &	0.6019  & $ 21.20   \pm  0.10  $  & $ -0.77  \pm   0.20 $  & \ion{Zn}{2}	& 10	  \\ 
J171532.48+460640.1 &   	Q1715+4606  &	0.6511  & $ 20.44   \pm  0.10  $  & $ -0.48  \pm   0.21 $  & \ion{Fe}{2}	& 3	  \\ 
J235321.62-002840.6 &   	J2353-0028  &	0.6043  & $ 21.54   \pm  0.15  $  & $ -0.89  \pm   0.32 $  & \ion{Zn}{2}	& 11	  \\ 
J013209.76-082349.8 &       J013209-082349  &	0.6470	& $ 20.60   \pm  0.12  $  & $ -0.62  \pm   0.23 $  & \ion{Fe}{2}	& 8	  \\ 
J232820.37+002238.2 &   	J2328+0022  &	0.6520	& $ 20.32   \pm  0.07  $  & $ -0.45  \pm   0.17 $  & \ion{Zn}{2}	& 12	  \\ 
J162439.08+234512.1 &   	 Q1622+238  &	0.6561  & $ 20.36   \pm  0.10  $  & $ -0.75  \pm   0.23 $  & \ion{Fe}{2}	& 13	  \\ 
J112442.87-170517.5 &   	 Q1122-168  &	0.6820	& $ 20.45   \pm  0.05  $  & $ -0.60  \pm   0.13 $  & \ion{Si}{2}	& 14	  \\ 
J133108.29+303032.9 &   	 Q1328+307  &	0.6920	& $ 21.25   \pm  0.06  $  & $ -1.09  \pm   0.14 $  & \ion{Zn}{2}	& 6	  \\ 
J132323.79-002155.2 &   	J1323-0021  &	0.7156  & $ 20.54   \pm  0.15  $  & $ +0.29  \pm   0.19 $  & \ion{Zn}{2}	& 11, 12  \\ 
J113709.49+390723.4 &   	J1137+3907  &	0.7200	& $ 21.10   \pm  0.10  $  & $ -0.27  \pm   0.11 $  & \ion{Zn}{2}	& 3	  \\ 
J025607.25+011038.6 &   	J0256+0110  &	0.7252  & $ 20.70   \pm  0.15  $  & $ -0.07  \pm   0.17 $  & \ion{Zn}{2}	& 12	  \\ 
J110729.03+004811.2 &   	J1107+0048  &	0.7405  & $ 21.00   \pm  0.05  $  & $ -0.50  \pm   0.16 $  & \ion{Zn}{2}	& 12	  \\ 
J122556.61+003535.1 &   	J1225+0035  &	0.7731  & $ 21.38   \pm  0.12  $  & $ -0.67  \pm   0.21 $  & \ion{Fe}{2}	& 3	  \\ 
J045647.17+040052.9 &   	 Q0454+039  &	0.8597  & $ 20.69   \pm  0.06  $  & $ -0.75  \pm   0.13 $  & \ion{Zn}{2}	& 15	  \\ 
J051410.91-332622.4 &         HE0512-3329A  &	0.9313  & $ 20.49   \pm  0.08  $  & $ -1.00  \pm   0.21 $  & \ion{Fe}{2}	& 16	  \\ 
J172739.02+530229.2 &   	J1727+5302  &	0.9465  & $ 21.16   \pm  0.05  $  & $ -0.53  \pm   0.06 $  & \ion{Zn}{2}	& 11, 17  \\ 
J173322.99+553300.8 &   	J1733+5533  &	0.9981  & $ 20.70   \pm  0.04  $  & $ -0.73  \pm   0.08 $  & \ion{Si}{2}	& 3	  \\ 
J045214.27-164016.3 &   	Q0449-1645   &	1.0072  & $ 20.98   \pm  0.06  $  & $ -0.93  \pm   0.10 $  & \ion{Zn}{2}	& 18	  \\ 
J030449.86-221151.9 &   	Q0302-223   &	1.0095  & $ 20.36   \pm  0.11  $  & $ -0.47  \pm   0.13 $  & \ion{Zn}{2}	& 13	  \\ 
J172739.02+530229.2 &   	J1727+5302  &	1.0306  & $ 21.41   \pm  0.03  $  & $ -1.33  \pm   0.12 $  & \ion{Zn}{2}	& 11, 17  \\ 
J100715.53+004258.3 &   	Q1007+0042  &	1.0373  & $ 21.15   \pm  0.24  $  & $ -0.48  \pm   0.24 $  & \ion{Zn}{2}	& 11	  \\ 
\enddata
\tablecomments{The ion column gives the ion that was used to estimate the metallicity. In the case of \feii, a correction of $+0.5$ dex was applied to correct for dust depletion and/or $\alpha$-enhancement (see text for more details), except for the DLA toward Q0235+164 where a correction of $+1$ dex was applied \citep{chen05}. Note that all the abundances were re-scaled to the solar abundance from \citet{asplund09}.}
\tablerefs{1: \citet{bowen05}; 2: \citet{kulkarni05}; 3: \citet{meiring06}; 4: \citet{battisti12}; 5: \citet{oliveira14}; 6: \citet{kanekar14}; 7: \citet{boisse98}; 8: \citet{chen05}; 9: \citet{quiret16}; 
10: \citet{ellison12}; 11: \citet{nestor08}; 12: \citet{peroux06}; 13: \citet{churchill00}; 14: \citet{delavarga00}; 15 ; \citet{pettini00}; 16: \citet{lopez05}; 17: \citet{turnshek04}; 
18: \citep{peroux08}.    }
\end{deluxetable*}

\subsubsection{Comparison Sample: SLLSs\label{s-sample-slls}}

For the SLLSs, the situation is more complicated. A large fraction of the SLLS sample at $z\la 1$ was initially selected based on the strength of \mgii. While that selection does not bias the DLA metallicity distribution, we argue it may affect the SLLSs. Using the same argument as for the DLAs, a \mgii -based selection with $W_\lambda (2796) \ge 0.6$ \AA\ leads to a lack of sensitivity to gas below $\xh \la -1.8$ to $-1.0$ for SLLSs in the \hi\ column density range $19 \la \mlnhi \la 19.7$. This sensitivity depends on the ionization correction, which can be variable in the SLLS regime. Some systems require significant corrections \citep[see, e.g.,][]{milutinovic10,fumagalli16}---which actually improves the metallicity sensitivity for a given \mgii\ column; others require very little or no corrections (for example, comparing the metallicities derived by \citealt{fumagalli16} and \citealt{quiret16}, see below). Thus, the metallicity sensitivity depends on the nature of the absorbers being studied, even in the same column density regime. And, indeed, using our previous compilations that included \mgii-selected SLLSs (\citetalias{lehner13,wotta16} and see also the updates by \citealt{fumagalli16,quiret16}), we find that the metallicities for the SLLSs are distributed toward higher values than the metallicities of the DLAs, with a substantial sample of even super-solar SLLSs. This could be a real effect as some argued \citep[][]{som13,som15}. However, if there was no bias in the dominantly \mgii-selected sample, a \hi-selected sample of SLLSs should have a similar metallicity distribution.

Using the results from \citet{fumagalli16} (where the metallicities were estimated with an ionization correction) and from \citet{quiret16} (where the metallicities were estimated without any correction for ionization), the mean (and median) metallicities of the SLLSs at $z \la 1$ are $\xh = -0.39$ ($-0.31$, sample size 27) and $-0.29$ ($-0.11$, sample size 36), respectively. Relying only on an \hi-selected sample of SLLSs---those serendipitously observed in UV spectra---decreases the sample to 8 SLLSs; this sample is summarized in Table~\ref{t-slls}. The mean (and standard deviation) and median of the metallicity of the \hi-selected sample of SLLSs are $\xh = -0.75 \pm 0.50$ and $-0.55$, both significantly lower than the results for the dominantly \mgii-selected SLLS sample. This suggests a bias is at work in the metal-selected results and thus in our previous compilations. In this work we therefore choose to use only \hi-selected SLLSs, even though this decreases the comparison SLLS sample substantially. We note that \citet{dessauges-zavadsky09} came to a similar conclusion on the metallicity bias  in \mgii-selected SLLSs in view of the apparent steep evolution in metallicity of the SLLSs from high (mostly \hi-selected) to low (most \mgii-selected) redshift. We finally note that the \hi-selected sample of SLLSs have metallicity statistics that are equivalent to the DLAs, implying no differences between these two populations of \hi\ absorbers.  

\begin{deluxetable*}{llccccc}
\tablecaption{Sample of \hi-selected SLLSs \label{t-slls}}
\tablehead{
\colhead{Name}  &\colhead{Original}  & \colhead{$z_{\rm abs}$} & \colhead{$\mlnhi$} & \colhead{\xh} & \colhead{Ion} & \colhead{Ref.} \\
\colhead{} &\colhead{Name} & \colhead{} & \colhead{[\cmm]} & \colhead{} & \colhead{}& \colhead{}
}
\startdata
\hline
J121920.93+063838.4 &   PG1216+069     &	0.01 & $  19.32  \pm   0.03 $  & $ -1.69  \pm	0.10 $  & \ion{O}{1}	& 1      \\ 
J155304.92+354828.5 &   J1553+3548     &	0.08 & $  19.55  \pm   0.15 $  & $ -1.12  \pm	0.16 $  & Multiple Ions	& 2       \\ 
J092837.97+602521.0 &   J0928+6025     &	0.15 & $  19.35  \pm   0.15 $  & $ +0.01  \pm	0.17 $  & Multiple Ions	& 2       \\ 
J143511.53+360437.2 &   J1435+3604     &	0.20 & $  19.80  \pm   0.10 $  & $ -0.41  \pm	0.16 $  & Multiple Ions	& 2       \\ 
J092554.70+400414.1 &   J0925+4004     &	0.25 & $  19.55  \pm   0.15 $  & $ -0.29  \pm	0.17 $  & \ion{O}{1}	& 2       \\ 
J100102.55+594414.3 &   J1001+5944     &	0.30 & $  19.32  \pm   0.10 $  & $ -0.37  \pm	0.10 $  & \ion{O}{1}	& 2       \\ 
J011014.45-021657.6 &   Q0107-0232     &	0.56 & $  19.50  \pm   0.20 $  & $ -0.72  \pm	0.32 $  & \ion{O}{1}	& 3       \\ 
J001855.22-091351.1 &   J0018-0913     &	0.58 & $  20.11  \pm   0.10 $  & $ -1.19  \pm	0.13 $  & \ion{Fe}{2}	& 4       \\ 
\enddata
\tablecomments{The ion column gives the ion that was used to estimate the metallicity. For \oi, no ionization correction was applied. When ``Multiple Ions" is listed, an ionization correction was applied to estimate the metallicity based on several ions (including, \ion{S}{2}, \ion{Si}{2}, \ion{Si}{3}, \ion{Fe}{2}, \ion{Fe}{3}, see text for more details). In the case of \ion{Fe}{2}, we did {\it not}\ apply an ionization correction (owing to the large \nhi\ value for this absorber), but a correction of $+0.5$ dex was applied to correct for dust depletion and $\alpha$-enhancement (see text for more details). Note that all the abundances were re-scaled to the solar abundance from \citet{asplund09}.}
\tablerefs{1: \citet{tripp05}; 2: \citet{battisti12}; 3: \citet{crighton13}; 4: \citet{quiret16}. }
\end{deluxetable*}

Another issue with the SLLSs compared to the DLAs is that an ionization correction is sometimes needed to determine the metallicity when singly-ionized species are compared to \hi, especially if $\mlnhi < 20$ \citep{battisti12,fumagalli16,lehner16}. For 4 of the SLLSs, the metallicities were derived using \oi, providing a reliable estimate of the metallicity as long as the ionization is not extreme, which is unlikely to be the case.  In this case, we only adjusted the original measurements to match the solar oxygen abundance from \citet{asplund09}.\footnote{Note that \citet{prochaska17} ran ionization models on these absorbers that are also in the COS-Halos survey, but we elected to use to the [\oi/\hi] estimate because the \nhi\ in the COS-Halos models departed largely from the original measurements and because COS-Halos used a different EUVB.} For the 3 absorbers marked ``Multiple Ions'' in Table~\ref{t-slls}, ionization models were run following the procedure described in \S\ref{s-strongLLS_ioncorrect_and_MCMC_metallicity}, and we find very similar results compared to the original ionization correction by \citet{battisti12}. Finally, for the remaining SLLSs, only \feii\ is available \citep{quiret16}. With  $\mlnhi= 20.11 \pm 0.10$, the \hi\ column density of this SLLS is sufficiently high that the ionization correction must be small. We corrected the abundance by $+0.5$ dex following the DLA method (see \S\ref{s-sample-dla}).


\section{Determining the Metallicities of the Absorbers}
\label{s-strongLLS_ioncorrect_and_MCMC_metallicity}

\subsection{Overview and General Assumptions}
As shown empirically in \citetalias{lehner18} and using photionization modeling (\citetalias{lehner13,wotta16}, and see also, e.g., \citealt{crighton13,fumagalli16,lehner16,prochaska17}), pLLSs and LLSs are largely ionized. Since we cannot estimate the column densities for all the ionization states of each species (esp.\ H), we must apply an ionization correction to estimate the metallicities of these absorbers. In \citetalias{lehner18} (and see also \citetalias{lehner13}), we show that the absorption profiles of the low ions (singly-ionized species, e.g., \cii, \mgii, \siii) and {\it often}\ the intermediate ions (singly ionized species with higher ionization energies than \hi, e.g., \oii, and doubly-ionized species, e.g., \ciii, \siiii) have similar profiles to the \hi\ absorption profiles. The good correspondence between the velocities at which peak optical depths of the \hi\ and of the low ions occur suggests these species being co-spatial and can be modeled in a single ionization phase. For each absorber, we are careful to confirm that the overall velocity profiles of the metal ions followed that of the \hi, especially those of the intermediate ions (e.g., \ciii, \oiii) where most of the column density may not always be associated with the low \nhi\ component (see also \citetalias{lehner13}).  

We do not use higher ions (e.g., \oiv, \ovi) in our ionization modeling because often those often trace a different gas-phase as discussed in \citetalias{lehner13}.\footnote{In the early stage of the photoionization modeling, we first included ions as \oiv, but as shown and discussed in \citetalias{lehner13}, this could not be modeled by single-phase photoionization model.} This differs from some previous surveys where a more agnostic approach of modeling ions in all ionization stages has been used \citep[e.g.,][]{prochaska17}. We note that the amount of \hi\ associated with these higher metal ionization stages are most likely factors of 10 to 100 lower than those associated with the lower ionization states in either collisional or photoionization conditions (see, e.g., \citealt{stern18} for the photoionization case and \citealt{gnat07} for the collisional ionization). Observationally  \hi\ absorption lines associated with \ovi\ tend to have \hi\ column densities much lower than those studied here \citep[e.g.,][]{lehner06,tripp08,savage14,pachat16}; specifically \citet{savage14} show for \ovi\ components aligned with \hi\ absorption that $13\la \mlnhi \la 14.6$ and $13 \la \mlnovi \la 14.5$. Thus, this is unlikely to bias our metallicity estimates. 

To determine the metallicity for each absorber we  follow closely the overall method of our previous surveys and in particular make similar assumptions \citepalias{lehner13,wotta16}. In  \citetalias{lehner18}, we empirically show that the properties of \hi-selected absorbers in our sample were consistent with being predominantly photoionized, and {\it a posteriori} we find that the photoionization modeling matches well the observational constraints. We model the photoionization using Cloudy \citep[version C13.02, see][]{ferland13}, assuming a uniform slab geometry in thermal and ionization equilibrium. In all cases the slab is illuminated with a Haardt--Madau EUVB radiation field from quasars and galaxies (HM05 and HM12, see \citealt{haardt96,haardt01,haardt12}). We adopt HM05 \citep[][as implemented in Cloudy]{haardt01} as the fiducial radiation field for CCC, but we make use of the HM12 EUVB to explore systematics associated with uncertainties in the radiation field. We note that absorbers in the column density range of the pLLSs and LLSs tend to lie at projected distances $\rho >30$ kpc from their host galaxies \citepalias{lehner13}, and thus local sources of (galactic) ionizing radiation should not much impact the shape of the radiation field \citep{fox05}. At $z\sim 3$, \citet{fumagalli16} also show that local ionizing sources did not affect much the metallicity estimates even if it does affect the density distribution (and hence the path length of the absorbers). They show that the observations are able to quite precisely constrain the ionization parameter, and thus the metallicity, even though the density is more uncertain because it is more  sensitive to the amplitude of the ionization radiation field owing to a degeneracy between density and radiation. 

We compare each absorber with a grid of photoionization models, effectively searching for models within the grid that produced the best agreement with the observed column densities from \citetalias{lehner18}. The principal parameters of interest are the ionization parameter ---$U \equiv n_\gamma/n_{\rm H}={\rm H}$, the ionizing photon density/total hydrogen number density (neutral + ionized)---and the metallicity, $\xh$. We assume solar relative heavy element abundances from \citet{asplund09}, but allow for possible variation between some species, especially between C and $\alpha$-elements (where $\alpha$ can be O, Si, Mg; see \S\ref{s-mcmc-method}). 

In \citetalias{lehner13}, we modeled the metallicities using a more subjective procedure that attempted to constrain the ionization parameter by excluding models that were inconsistent with observations at the $\sim$1$\sigma$ level. As we show in \S\ref{s-comp-l13}, those results largely stand. However, that approach is subjective and cumbersome to execute on a large scale. Furthermore, it does not explore robustly the various parameters to produce reliable confidence intervals / uncertainties. Here we adopt a Markov-chain Monte Carlo (MCMC) technique that employs Bayesian statistics  \citep[see][]{crighton15,cooper15,fumagalli16} to robustly determine a posterior probability density function (PDF) for the metallicity of each absorber, marginalized over the other parameters. Not only does this method provide more robust evaluations of the full PDFs (summarized with confidence intervals), but it also treats more rigorously the lower and upper limits, as those are now themselves described by PDFs. 

In \S\ref{s-strongLLS_MCMC_metallicity}, we give a brief overview of the Bayesian MCMC technique and explain how we apply it to our absorbers to derive metallicities. In \S\ref{s-strongLLS_MCMC_example}, we  give a detailed example for a single absorber. Given the improvements provided by this technique, we apply it to our entire sample (i.e., we reassess the metallicities in our previous samples) to provide a homogeneous analysis across datasets.

We explain our overall strategy for implementing the MCMC techniques in \S\ref{s-mcmc-method}. As we detail, some absorbers remain not well-constrained by the metal ion measurements. In these cases, we use the ``low-resolution'' approach that we developed in \citetalias{wotta16}, now incorporated into the Bayesian formalism used here. We show that with prior knowledge of the $U$ distribution, we can robustly estimate the metallicities of the pLLSs and LLSs at $z\la 1$.


\startlongtable
\begin{deluxetable}{lccc}
\tabcolsep=3pt
\tablecolumns{4}
\tablewidth{0pc}
\tablecaption{Cloudy Ionization Parameters\label{t-mcmc_parameters}}
\tabletypesize{\footnotesize}
\tablehead{\colhead{Parameter} & \colhead{Minimum} & \colhead{Maximum} & \colhead{Step size$^a$} }
\startdata
$\logNHI$ [cm$^{-2}$]          &  12.0    &  20.0  &  0.25  \\
$z$                            &  0.0     &  5.0   &  0.25  \\
$[{\rm X/H}]$                  &  $-$5.0  &  +2.5  &  0.25  \\
$\log\;n_{\rm H}$ [cm$^{-3}$]  &  $-$4.5  &  0.0   &  0.25  \\
$[{\rm C/}\alpha]$             &  $-$1.0  &  1.0   &  0.20  
\enddata
\tablecomments{$^a$We note that while the Cloudy grid was initially computed using the step sizes listed here, the grid was interpolated within the MCMC code}
\end{deluxetable}

\subsection{Bayesian MCMC Technique and Cloudy Grid}
\label{s-strongLLS_MCMC_metallicity}

To derive the metallicities of the ionized gas probed by the pLLSs and LLSs in our sample, we compare the observed column densities for each absorber with a grid of photoionization models in a Bayesian context (applying a range of priors to the parameters, although in some cases just the maximum extent of the parameters listed in Table~\ref{t-mcmc_parameters}). We assess the normalized probabilities of the models using an MCMC sampling of the grid parameter space. The distribution values taken on by the output ``walkers'' in this approach provide a robust PDF for each parameter. A recent review of MCMC methods and their application to astronomical topics can be found in \citet{sharma17}. 

For CCC, we make use of the publicly available codes described in \citet[see also \citealt{prochaska17}]{fumagalli16}.\footnote{The code presented in \citet{fumagalli16} has been incorporated into the PyIGM Python package and is available at \url{https://github.com/pyigm/pyigm} \citep{prochaska17a}.} These codes utilize the {\tt emcee}\footnote{For more information on the {\tt emcee} package, see \url{http://dfm.io/emcee}.} Python module from \citet{foreman-mackey13}, which is based on the ``affine invariant ensemble sampler'' developed by \citet{goodman10}. We computed a new grid of photoionization models using the range of conditions summarized in Table \ref{t-mcmc_parameters} for \nhi, redshift, metallicity, hydrogen number density ($n_{\rm H}$; and therefore the ionization parameter $U$), and $\ca$ ratio. The photoionization models used the Cloudy software with the assumptions detailed above. Compared to the grid used in \citet{fumagalli16}, we extend \nhi\ to lower values (critical for the pLLSs and SLFSs) and the metallicities to lower and higher values. We also include a new grid variable, $\ca$. At both low and high redshifts, we have found that $\ca$ ranges from about $-0.6$ and $+0.4$ dex, with no dependence on the metallicity (\citetalias{lehner13}; \citealt{lehner16}). For that reason, we consider values over $-1\le \ca \le 1$ when possible. 

For the MCMC analysis, we use the pre-computed grid described above, comparing the models' predicted metal ion column densities with the observations. The output walkers provide a posterior PDF for each of the parameters (marginalized over the others). We characterize the PDFs with a central (median) value and confidence intervals (where we adopt 80\% for lower/upper limits on the metallicity, and 68\% otherwise).

We perform the MCMC analysis on an absorber-by-absorber basis (see \S\ref{s-strongLLS_MCMC_example} for a detailed example). For each absorber we apply Gaussian priors on \nhi\ and redshift (these are measurables estimated directly from the data). We nominally adopt flat priors on metallicity, hydrogen number density ($n_{\rm H}$), and the $\ca$ ratio. 

The code inserts 400 ``walkers'' distributed across the (interpolated) photoionization grid.\footnote{We calculated output parameters for each absorber twice, using different initial walker positions. For the first, the walkers' starting positions randomly sample the entire parameter space. This occasionally leads to walkers that get trapped in areas of zero probability, unable to step toward higher-likelihood parameter space. To work around this, we perform a second calculation in which the walkers are distributed around the best results from the first run, from which each walker propagates outward. This prevents walkers from becoming stranded in zero-probability regions. For more discussion, see \url{http://dan.iel.fm/emcee/current/user/faq/\#how-should-i-initialize-the-walkers}. The same approach was used in \citet{fumagalli16} and \citet{prochaska17}.} The code calculates the likelihood of each point by comparing the model's predicted column densities with those of the observed metal ions. The likelihoods are based on the assumptions that the detected metal ions are described by Gaussian distributions centered on the observed column densities, with standard deviations based on their measurement errors; for lower (upper) limits, the likelihoods are based on a rescaled Q-function (cumulative distribution function) \citep[see][]{fumagalli16}. The likelihood is recalculated after each walker step. We allowed each walker to take 400 steps through the parameter space to allow it to converge to the highest-likelihood models.\footnote{Our choice of the number of walkers and steps is based on testing various combinations, with considerations for total runtime and convergence. In our testing, 400 walkers and 400 steps was sufficient for convergence. We calculated results for one absorber (the absorber at $z=0.463336$ toward the sightline J111132.16+554726.2) with 2000 walkers and 2000 steps and saw no noticeable difference in the outcome.} The ending positions of the walkers define the posterior PDF for each of the input parameters.


\subsection{Applying Prior Knowledge to Photoionization Models}\label{s-mcmc-method}

For many absorbers, the observed ionic and atomic column densities are not always sufficient to constrain ionization models--- and hence the metallicity ---well enough for our purposes if we assume no prior knowledge of the ionization conditions in the absorbers. This does not mean their metallicities are unknowable. As we demonstrated in \citetalias{wotta16}, assuming prior knowledge on the ionization parameter, $U$, can be sufficient to derive good constraints on metallicities even if only one metal ion is measured with \hi. This section describes our approach to adopting prior knowledge in a Bayesian context when modeling the absorbers in our survey. The general strategy is to adopt flat priors on all model parameters when we have sufficient information; when we lack sufficient constraints, we adopt Gaussian priors on \logU\ based on the distributions observed in the well-constrained systems. An additional concern is the potential for non-solar \ca\ ratios. Typically our reported metallicities, \xh, refer to metallicities derived from $\alpha$ elements. For systems where metallicity and ionization constraints come only from C, we adopt a prior on \ca\ to bring these systems onto the same metallicity scale \xh\ used in the other systems.  Below we describe each of these approaches and the data that go into constraining the priors for \tsz\ absorbers in our sample (i.e., including the SLFSs that will be presented in \citetalias{lehner18b}). 

\subsubsection{Flat Prior on $U$}\label{s-mcmc-flat-u}

For all absorbers, we calculate initial posterior PDFs assuming a flat prior on \logU\ with the constraint $\ca =0$. We examine the diagnostic outputs (convergence of walkers, residuals of predicted column densities compared with observed, and ``corner plots'' showing the posterior PDFs) to assess which absorbers could be reliably modeled with this flat prior on \logU\ and which may require a Gaussian prior on \logU\ (following \citetalias{wotta16}). We also determine which absorbers could be modeled by a flat prior on $\ca$ (i.e., for which we could estimate $\ca$), which require $\ca = 0$ (no or non-constraining measurements of \cii\ and \ciii\ column densities), and which require a Gaussian prior on $\ca$ (those for which \cii\ and/or \ciii\ appeared to drive the solution). For those absorbers that could be modeled with flat priors on $\log U$, we resample the parameter space with initial conditions based on the highest likelihood outputs of the first sampling, as described in \S \ref{s-strongLLS_MCMC_metallicity}.

\subsubsection{Gaussian Prior on \logU}\label{s-mcmc-gauss-u}

\begin{figure}[tbp]
\epsscale{1.2}
\plotone{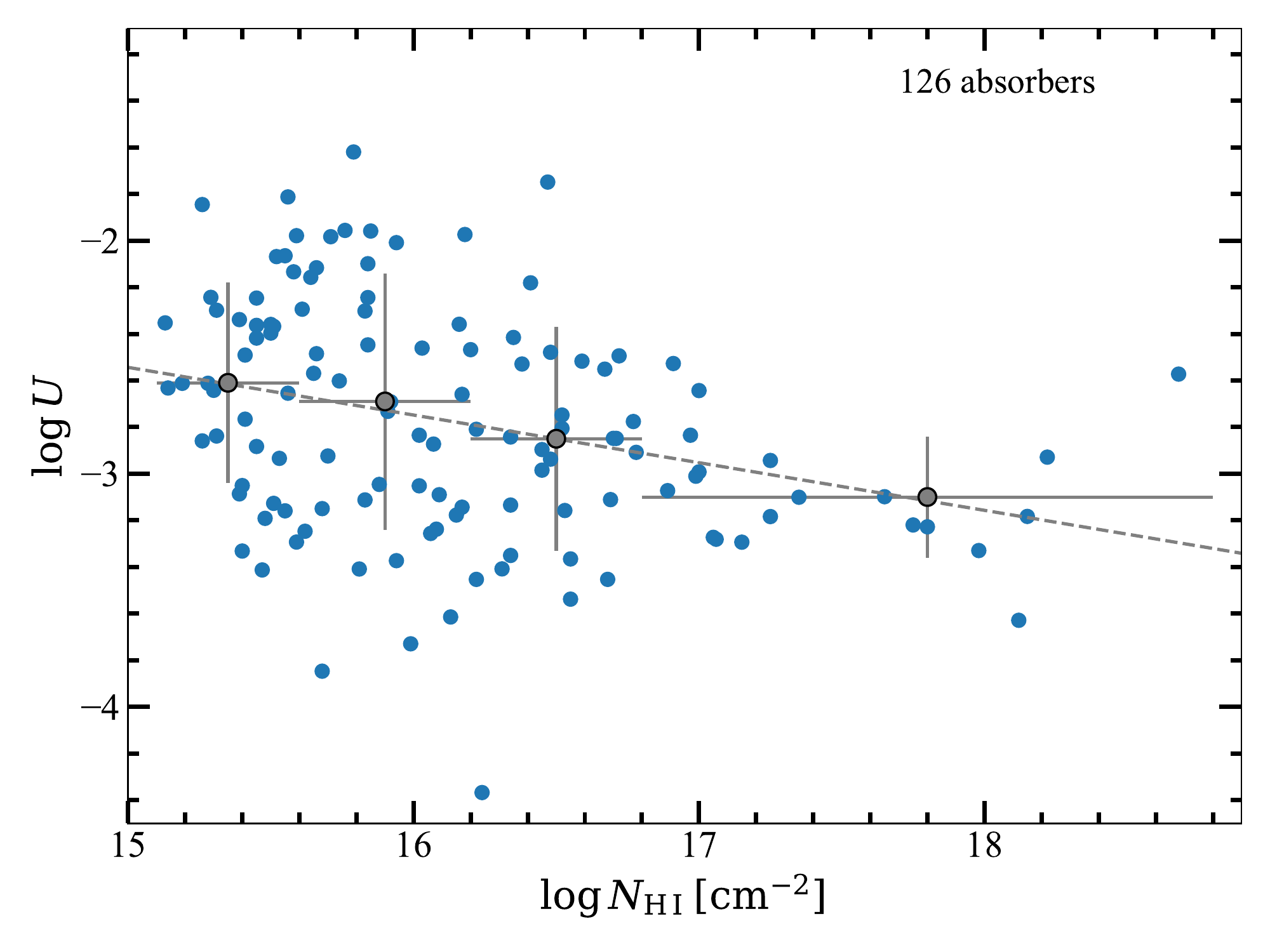}
\caption{The \logU\ median values as a function of \nhi\ for 126 $z\la 1$ absorbers in the CCC sample. The solid shows a linear fit to the data (with a slope $-0.20$ and intercept $+0.52$). The crosses shows the mean values of the median values in each interval \nhi\ interval indicated by the horizontal bar (see Table~\ref{t-mcmc_parameters} for actual values). 
\label{f-strongLLS_logU_vs_h1}}
\end{figure}

As demonstrated in \citetalias{wotta16} (see also \citetalias{lehner13}), the ionization parameters found in low-redshift pLLSs and LLSs is not random, but can be well-characterized by a Gaussian distribution. \citetalias{wotta16} showed that this could be used to constrain the ionization conditions and thus metallicities in such absorbers. Here we apply this prior knowledge to systems without sufficient constraints on the ionization models to allow reliable metallicity estimates. We show in Fig.~\ref{f-strongLLS_logU_vs_h1} the best values of \logU\ versus \nhi\ for the well-constrained absorbers in our sample (i.e., absorbers with enough constraints that we can adopt a flat prior on $U$). These values are derived from the posterior PDFs of $n_{\rm H}$ for each absorber; for an adopted radiation field (HM05), the density of an absorber is directly proportional to the value of $U$ appropriate for that system. Fig.~\ref{f-strongLLS_logU_vs_h1} shows a strong anti-correlation between the ionization parameter and \nhi, which is confirmed by a Spearman ranking order with a correlation coefficient $r_{\rm S} =-1.0$ ($p$-${\rm value}= 0.005$).\footnote{We note that the gradient with \nhi\ is much smaller than that of the phenomenological model of \citet[][assuming a \citetalias{haardt96} EUVB; see their Equation~(7)]{keeney17}.} While \logU\ decreases with increasing \nhi, the change is relatively small, with a decrease of $\sim$0.4 dex in \logU\ over $\sim$3 dex in \lnhi. This implies the ionization conditions for absorbers across this regime are comparable, in agreement with the empirical evidence presented in \citetalias{lehner18}. It is especially significant that there is little difference across the $\log\mathNHI=17.2$ transition between optically-thin and optically-thick gas. There is no evidence that differences in the derived metallicity distributions of SLFS, pLLS, and LLS (see \S\ref{s-met-pdf-cgm} and \citetalias{lehner18b}) are due a disparity in their ionization conditions.

To characterize the distribution of \logU\ appropriate for use as a Bayesian prior, we split the \nhi\ range into four bins (based on the scatter in \logU\ as a function of \nhi, as seen in Fig.~\ref{f-strongLLS_logU_vs_h1}). Within each bin, we combine the \logU\ PDFs from all absorbers into a single, joint distribution. As in \citetalias{lehner13} \citepalias[and][]{wotta16}, we find each PDF is well-fit with a normal distribution; we summarize the properties of these fits in Table~\ref{t-logUconstraint_mean_sigma}.\footnote{We performed Kolmogorov--Smirnov tests and found that the \logU\ distributions are not statistically distinguishable from Gaussian distributions ($p=\strongLLSlogUgaussianTest$). While this test does not prove the distributions' Gaussianity, it suggests that the differences from a normal distribution are small. We therefore approximate them as normal distributions for simplicity, adopting a Gaussian with a mean and standard deviation as defined in Table~\ref{t-logUconstraint_mean_sigma}.} These Gaussian approximations are consistent with fits to the distribution of \citetalias{lehner13}, which has a mean of $\left\langle\log U\right\rangle = -3.1\pm0.6$ \citepalias[and see also][]{wotta16}.  

For poorly-constrained absorbers, we adopt \nhi-dependent Gaussian priors on \logU\ based on the results given in Table \ref{t-logUconstraint_mean_sigma} in our analysis. We otherwise calculate the results in the manner described in \S \ref{s-mcmc-flat-u}. This follows the ``low-resolution'' method described and tested in \citetalias{wotta16} to determine metallicities in pLLSs and LLSs with insufficient information to break the degeneracy between $n_{\rm H}$ and metallicity \citep[see Figure 5 of ][]{fumagalli16}.

\begin{deluxetable}{lccc}
\tabcolsep=3pt
\tablecolumns{3}
\tablewidth{0pc}
\tablecaption{Gaussian Prior Placed on $\log U$ for Low-Resolution MCMC Analysis\label{t-logUconstraint_mean_sigma}}
\tabletypesize{\footnotesize}
\tablehead{\colhead{$\mlnhi$} & \colhead{$m$} & \colhead{$\left\langle\log U\right\rangle$} & \colhead{$\sigma_{\log U}$} }
\startdata
$15.0 < \mlnhi < 15.6$       & 35  &   $-2.61$  &  $0.43$  \\
$15.6 \le \mlnhi < 16.2$     & 39   &  $-2.69$  &  $0.55$  \\
$16.2 \le \mlnhi < 16.8$     & 31   &  $-2.85$  &  $0.48$  \\
$16.8 \le \mlnhi \le 19.0$   & 21   &  $-3.05$  &  $0.26$  \\
\enddata
\tablecomments{$m$ is the number of absorbers in each \nhi\ interval for which we derived the mean and standard deviation of $\log U$.} 
\end{deluxetable}

\begin{figure}[tbp]
\epsscale{1.2}
\plotone{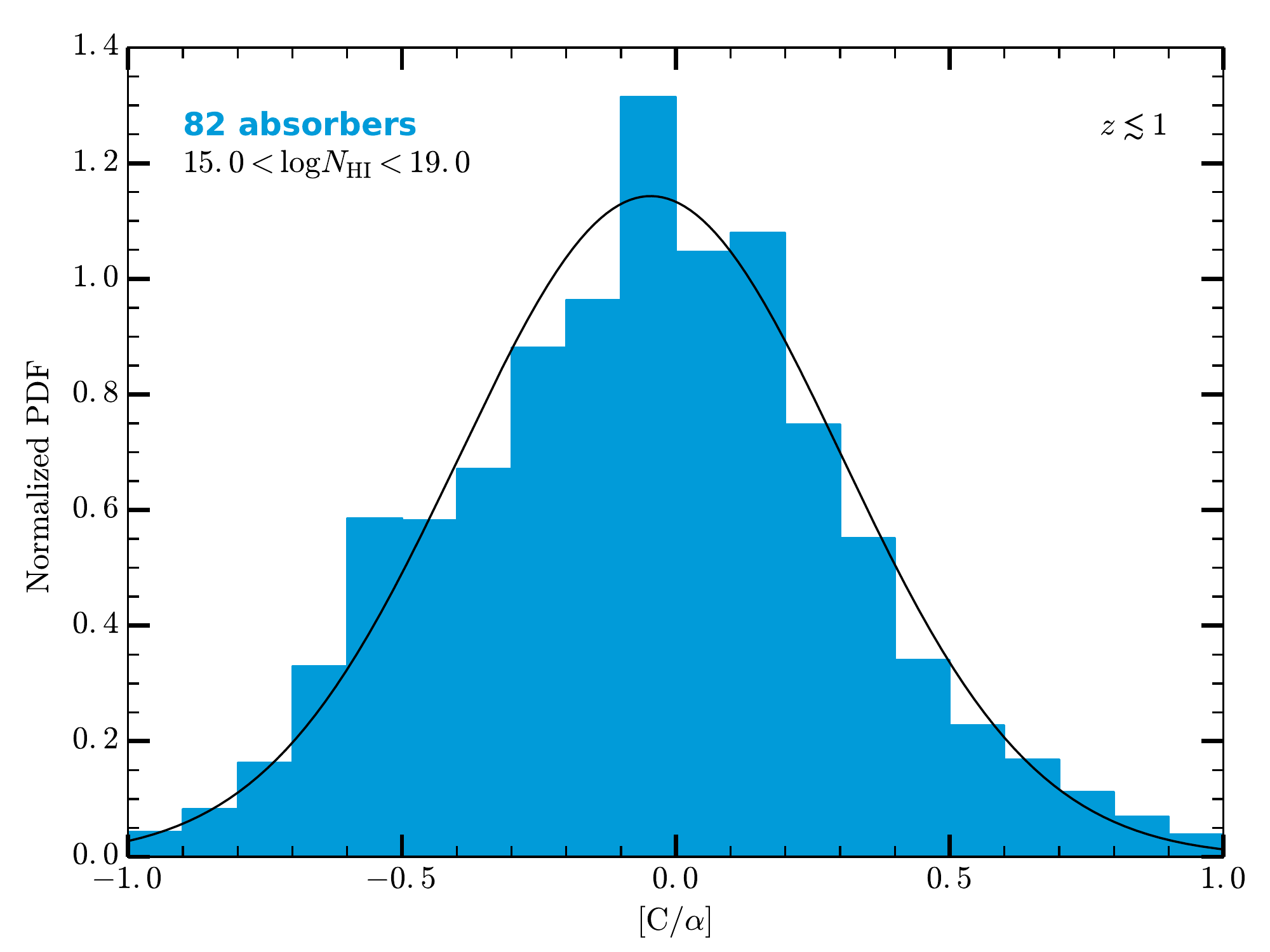}
\caption{Normalized posterior PDF of $\ca$\ for the absorbers with well-constrained $\ca$\ ratios for SLFSs, pLLSs and LLSs at $z\lesssim1$. The solid line shows the Gaussian fit to the PDF.  
\label{f-carbalpha_pdf}}
\end{figure}

\subsubsection{Gaussian Prior on $\ca$ }\label{s-mcmc-gauss-ca}

For a majority of the absorbers at $z \la 1$, the available metal ions typically imply that our measurement of \xh\ is equivalent to a measure of $[\alpha/{\rm H}]$, since we most often find the metallicities constrained by ions of O, Si, and/or Mg. In some systems, the only real constraints on the ionization modeling and hence metallicities come from ions of C. Since we have found non-solar values of \ca\ in these absorbers (\citetalias{lehner13}, \citet{lehner16}, and \citetalias{lehner18}), assuming [C/H]$\, = [\alpha/{\rm H}]$ represents a bias that could shift the metallicities of those systems relative to the others. To account for this, we adopt a prior constraint on the distribution of \ca\ in systems where the metallicity is based predominantly on ions of C. We characterize the prior using results from those systems for which we could reliably measure the posterior PDF of \ca\ initially assuming a flat prior on \ca. We show in Fig.~\ref{f-carbalpha_pdf} the composite \ca\ PDF for all such absorbers. In \citetalias{lehner18b}, we will show that there is little dependence of this ratio with metallicity and \nhi, and therefore we can simply use this $\ca$ PDF as characteristic of all absorbers. The distribution is well fit by a Gaussian distribution with a mean and standard deviation of $\left\langle \ca\right\rangle = -0.046\pm0.349$. We applied this prior on the $\ca$\ ratio for absorbers with poorly-constrained $\ca$. 

\subsubsection{Summary of Prior Adoption}\label{s-mcmc-gauss-summary}

As described above, each absorber is modeled in a way that depends on the available constraints from the observation. The priors adopted on \logU\ and \ca\ may be relatively non-informative (flat) or constrained by the observed distributions of well-constrained absorbers. We summarize in Table~\ref{t-mcmc-met} for the SLFSs, pLLSs, and LLSs in our sample the number of systems that were modeled in each of the different combinations of $\logU$ and $\ca$ priors. 

\begin{deluxetable}{lrrr}
\tablewidth{0pt}
\tabcolsep=2pt
\tablecaption{MCMC Summary \label{t-mcmc-met}}
\tablehead{\colhead{Method} & \colhead{SLFS}  & \colhead{pLLS} & \colhead{LLS}\\}
\startdata
\hline
Flat priors on $\log U$ \& $\ca$	      & 55  & 32 & 2  \\
Flat prior on $\log U$ \& Gaussian prior on $\ca$  & 5   & 1  & 0  \\
Flat prior on $\log U$ \&  $\ca = 0$      & 14  & 7  & 10 \\
Gaussian prior on $\log U$ \& flat prior on $\ca$  & 2   & 3  & 1  \\
Gaussian priors on $\log U$ \& $\ca$		      & 72  & 14 & 1  \\
Gaussian prior on $\log U$ \& $\ca = 0$	      & 4   & 25 & 15 \\
\enddata
\tablecomments{In the SLFS, pLLS, LLS columns, each value indicates the number of absorbers.}
\end{deluxetable}

\subsection{A Detailed Example}
\label{s-strongLLS_MCMC_example}
\begin{figure}[tbp]
\epsscale{1.2}
\plotone{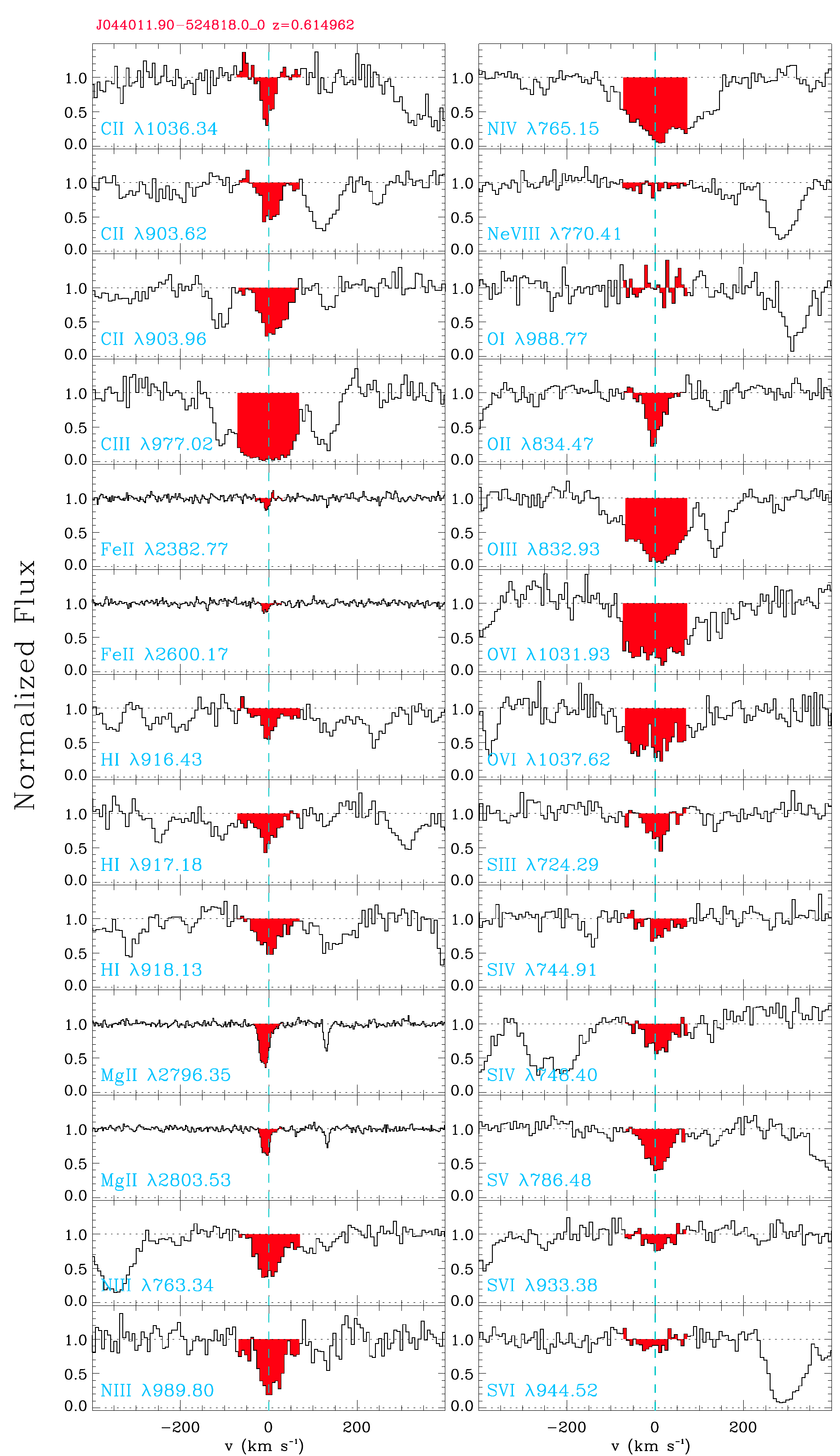}
\caption{Normalized absorption profiles as a function of the rest-frame velocity of the pLLS at $z=0.614962$ toward the sightline J044011.90-524818.0. These EUV and UV transitions are from \hst/COS G130M and G160M, while the NUV transitions (\feii\ and \mgii) are from VLT/UVES. The red regions show the integration range for each ion and atom (see \citetalias{lehner18} for more detail).
\label{f-strongLLS_plotstack}}
\end{figure}

To demonstrate our general procedure, we describe here the modeling of a sample absorber, the $z=0.614962$ system toward the sightline J044011.90-524818.0, to which we applied flat priors on the density ($n_{\rm H}$; and hence on $U$) and $\ca$ between the values listed in Table~\ref{t-mcmc_parameters}. Fig.~\ref{f-strongLLS_plotstack} shows the velocity profiles of the ions for this absorber. A detailed description of this system and summary of the column density measurements are given in \citetalias{lehner18}.\footnote{This absorber is also described in \citetalias{lehner13} (HE0439-5254, $z=0.6153$).} This absorber exhibits multiple velocity components in the metal ions and \hi\ (and the absorber at $v\simeq +110$ \km\ or $z = 0.615662$ is a SLFS included in the sample of \citetalias{lehner18b}). However, as discussed in \citetalias{lehner18}, we consider only the absorption associated with the strongest \hi\ absorption (here a pLLS with $\mlnhi = 16.20 \pm 0.01$), which is depicted in this figure by the red-shaded absorption. With absorption in both low and high ions, this absorber probes multiple gas phases (as many other absorbers in our sample), and no single-phase ionization model can reproduce simultaneously the column densities of all the ions in this case (see also \citetalias{lehner13}). The  low and some weak intermediate ions (e.g., \cii, \oii, \niii, \mgii, and \Siii) have a velocity structure similar to the \hi\ velocity transitions, with an absorption dominated by a single component (see red absorption in Fig.~\ref{f-strongLLS_plotstack}). The stronger intermediate ions (\oiii, \niv, and \ciii) show additional components, but their absorption is still dominated by the component at $z = 0.614962$ where the strong \hi\ absorption is observed,\footnote{Integrating the absorption profiles from $-35$ to $+50$ \km\ where the main absorption is observed would only decrease the column densities of \ciii\ and \oiii\ by 0.12 dex.} and we therefore elected in this case to include \oiii\ and \ciii\ in our model (a posteriori these ions do not constrain much the model because their absorption is strongly saturated, see below). However, in cases, where the main absorption of these strong intermediate ions were clearly shifted from the main component of \hi, these were not included in the ionization modeling (e.g., a striking example is the absorber at $z= 0.686086 $ toward J134100.78+412314.0, see Fig.~13.161 in \citetalias{lehner18} or Fig.~15 in \citetalias{lehner13}).  The ions we  use to model the cool ($T\sim10^4$ K), photoionized gas in this pLLS, and hence to determine its metallicity, are therefore the following low and intermediate ions: \cii, \ciii, \oi, \oii, \oiii, \niii, \mgii, and \Siii; the least constraining being \ciii\ and \oiii\ owing to being strongly saturated). For some of these species, the column densities were estimated using several transitions, providing extremely robust constraints.\footnote{Even though \feii\ has a secure measurement (two transitions being used to determine the column density), the ionization model fails to match the observed column, under-producing \feii\ by $\sim 0.7$ dex. As this is the only low ion not matched by the model, including it or not in the model did not change the results listed below. This is a rare example in the CCC sample where the \feii\ discrepancy could not be explained by dust depletion of $\alpha$-enhancement. Uncertainties in the atomic data for Fe could play a factor here, especially since  most of the accessible Fe ions are complex systems with heavy nuclei that require relativistic corrections to be modeled appropriately, and the multitude of energy levels for Fe provide an incredibly rich set of options for recombination (notably dielectronic recombination). } We refer the reader to \citetalias{lehner18} for the full discussion on how the column densities are determined for ions and atoms with multiple and single transitions (see also \citetalias{lehner13}). We note that a stronger weight was always put on the ions with multiple transitions than the ions with a single transition if there was some tension in the modeling. 

\begin{figure}[tbp]
\epsscale{1.2}
\plotone{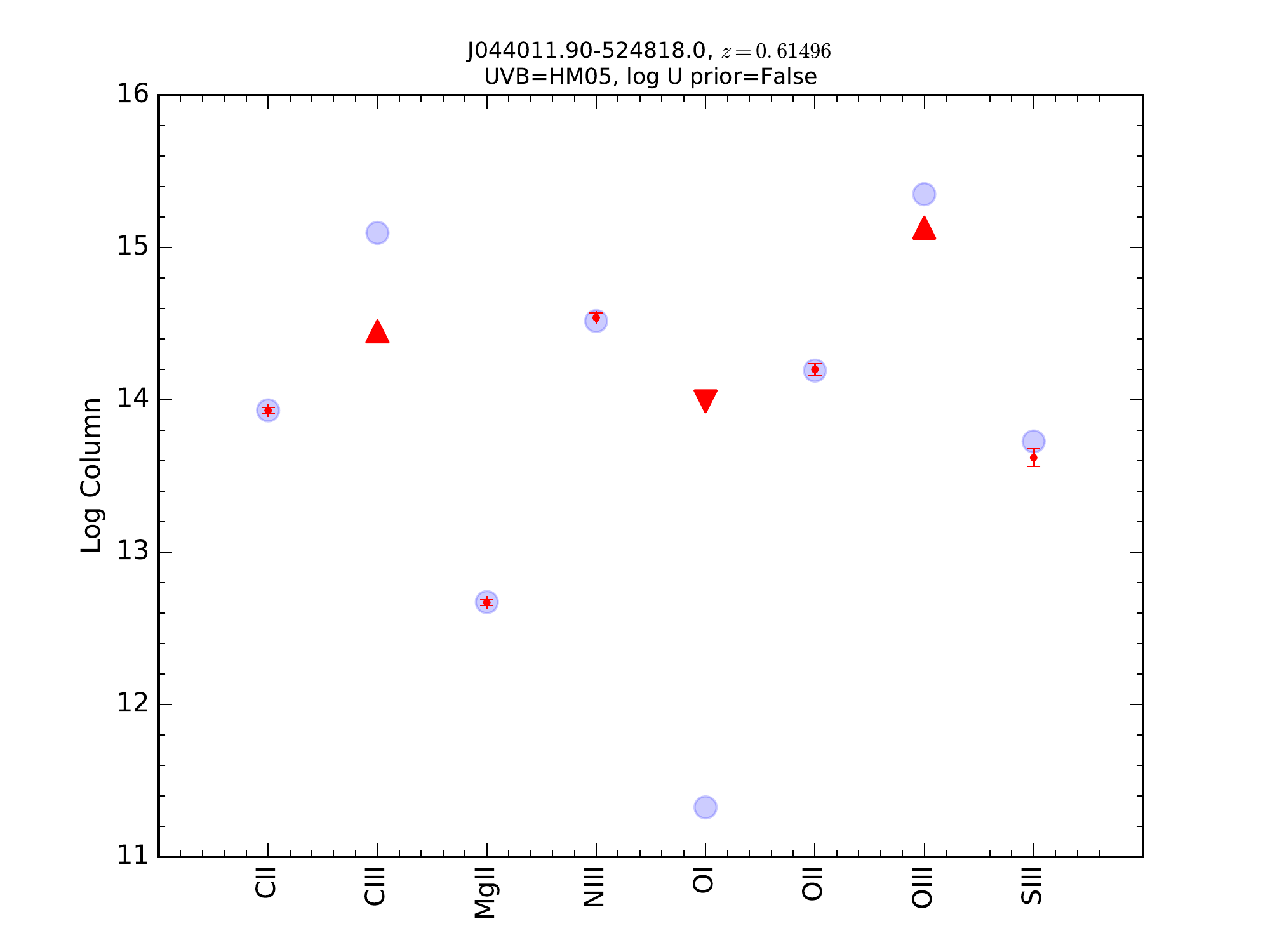}
\caption{Example of MCMC residual plot for the absorber at $z=0.614962$ toward the sightline J044011.90-524818.0. It shows the measured column densities for each ion (red) and the predicted column densities from the median MCMC model (blue). Triangles and downward triangles show lower (saturated transitions) and upper (non-detections) limits. 
\label{f-strongLLS_MCMC_output-residual}}
\end{figure}

\begin{figure}[tbp]
\epsscale{1.2}
\plotone{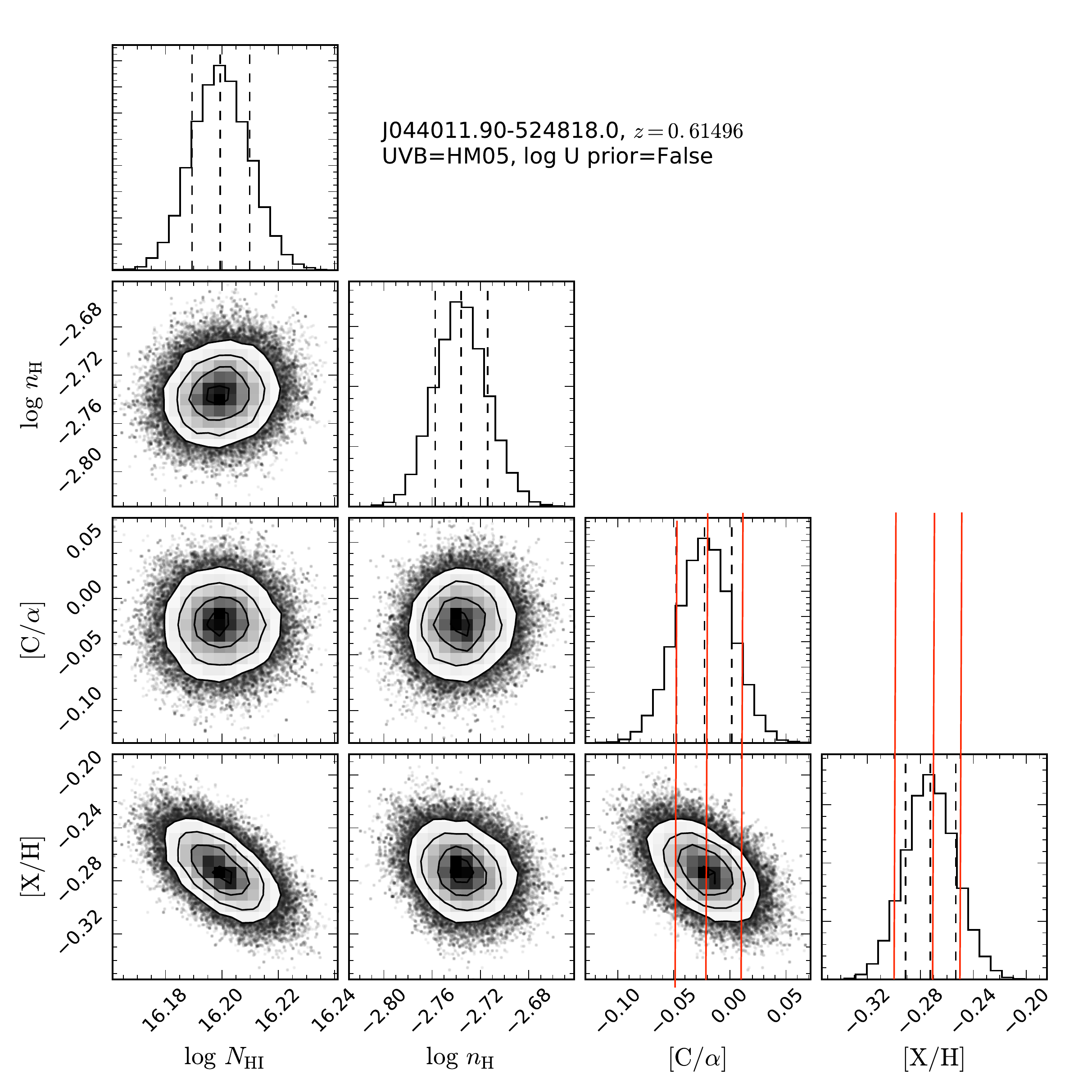}
\caption{Example of MCMC output corner plot for the sightline J044011.90-\allowbreak 524818.0. The histograms along the diagonal show the PDFs for \nhi, hydrogen number density ($n_{\rm H}$), [C/$\alpha$], and metallicity, respectively. The contour plots below the diagonal show the joint posterior PDFs of the given row and column. The dashed lines show in this case the 68\% CI. 
\label{f-strongLLS_MCMC_output-corner}}
\end{figure}

We perform the MCMC analysis as described above to model the atomic and ionic column densities using the following inputs: \nhi\ and redshift (with $1\sigma$ measurement errors) as Gaussian priors, $\xh$, $n_{\rm H}$, and $\ca$ as flat priors (see \S\ref{s-mcmc-method}). The output of this analysis includes three diagnostic plots: a representation of the behavior of the walkers as a function of step number, ``residual plots'' comparing the observed and best-fit (median) model ion column densities (Fig.~\ref{f-strongLLS_MCMC_output-residual}), and ``corner plots'' showing 2D distributions of the walkers in several planes sampling the full parameter space (Fig.~\ref{f-strongLLS_MCMC_output-corner}).

In Fig.~\ref{f-strongLLS_MCMC_output-residual} we compare the measured column densities for each ion (red) and the predicted column densities from the median MCMC model (blue). Upward triangles show lower limits (i.e., saturated transitions), while downward triangles show upper limits (i.e., non-detections). Red data points with error bars (sometimes smaller than the circles) denote well-constrained column densities. The predicted and observed column densities are in very good agreement in this case; and for many absorbers, we had a similarly good agreement. The agreement between the observations and model is quite remarkable in view of the extremely simplified geometry and model we employ. 

While this agreement was often as good as depicted in Fig.~\ref{f-strongLLS_MCMC_output-residual}, in some cases, we needed to iterate in our selection of the ions to be modeled. For example, nitrogen (\nii, \niii) was sometimes not well fitted, as expected since N and $\alpha$-elements do not have the same nucleosynthesis history. Non-solar $[{\rm N}/\alpha]$ can be common in low-metallicity gas (see \citetalias{lehner13}). Although we could have allowed this ratio to vary as we did for $\ca$, neither \nii\ nor \niii\ is detected as reliably or frequently as \ciii\ or \cii. Their measurements are not usually as constraining, and we simply elect to remove nitrogen if it does not fit the Cloudy photoionization model. Similarly, Fe does not always fit the observations owing to dust depletion or $\alpha$-enhancement. As shown in \citetalias{lehner18}, these effects are small since on average $\langle [$\feii/\mgii$]\rangle = -0.4 \pm 0.3$. However, as for N, if Fe was not well modeled, it was removed from the list of input ions. Finally, we occasionally had to remove some intermediate ions (e.g., \ciii, \oiii, \siiv) because clearly the bulk of the observed absorption was produced in a different phase (i.e., a hotter or lower-density phase) than the lower ionization states. In the appendix, we provide the input files for all of the absorbers, which include the specific list of ions used (and not used) in our models. 

In Fig.~\ref{f-strongLLS_MCMC_output-corner} we show the ``corner plot'' from the MCMC calculations for the same absorber. The histograms along the diagonal show the posterior PDFs for \nhi, $n_{\rm H}$, $\ca$, and $\xh$ marginalized over the other parameters. The vertical dashed lines represent the median and 68\% CI of each PDF. The contour plots below the diagonal show the joint PDFs for the parameters of the given row and column (e.g., the topmost contour plot shows the joint posterior PDF of $n_{\rm H}$ and \nhi, the latter of which is an input parameter).  For this absorber, we see that the metallicity posterior PDF is narrow and roughly Gaussian-distributed, with a median value $[{\rm X/H}]=-0.27$ and 68\% CI $[-0.29,-0.25]$. The $n_{\rm H}$ density PDF is narrow and well-behaved; a slight skewness in this PDF is commonly seen among our absorbers, with a slightly longer tail toward lower $n_{\rm H}$. The [C/$\alpha$] histogram is consistent with a slightly sub-solar $\ca$ value. Furthermore, the metallicity--density ($n_{\rm H}$) joint PDF (bottom-center contour) is roughly circular. For poorly-constrained absorbers, we usually see significant elongation in this parameter plane. In those cases, our adoption of a Gaussian prior on \logU\ (and therefore on density) serves to constrain the allowable metallicity range through the constraint on $n_{\rm H}$  (\S\ref{s-mcmc-gauss-u}). We emphasize that for all the absorbers, we checked that the output \nhi\ PDF did not deviate by $1\sigma$ from the original measurement; the solutions for poorly-constrained absorbers analyzed with flat priors can be driven to regions of parameter space that are inconsistent with the observed \hi\ column densities. 

We note that residual and corners plots as shown in Figs.~\ref{f-strongLLS_MCMC_output-residual} and \ref{f-strongLLS_MCMC_output-corner}, respectively, for all the absorbers in our sample are provided as supplement material (see Appendix for more details).

\subsection{Derived Metallicities from Photoionization Modeling}
\label{s-comp}

We summarize the results of our photoionization modeling in Table~\ref{t-strongLLS_metallicities} for all pLLSs and LLSs in the CCC database (the SLFSs are given in \citetalias{lehner18b}). For each pLLS/LLS, we list the sightline name, the absorber redshift ($z_{\rm abs}$), \lnhi, metallicity, density ($n_{\rm H}$), ionization parameter ($\log U$), and C/$\alpha$ ratio (when estimated). Each quantity is reported with the 68\% CI and median values, except in the cases where we derive an upper or lower limit, in which case we report the 80\% CI and median. We emphasize that these values are attempts to summarize the posterior PDFs provided by the MCMC sampling of the Cloudy models. While those PDFs can be well-behaved and nearly Gaussian, as for the absorber described in \S \ref{s-strongLLS_MCMC_example}, some are not as well behaved (see corner plots in the appendix). Prior to discussing these results in detail, we compare our newly-derived metallicities for common absorbers with the results of our previous surveys \citepalias{lehner13,wotta16} and other metallicity studies \citep[e.g.,][]{prochaska17} to assess any systematic difference between our approach and the results of those earlier works.

\startlongtable
\begin{deluxetable*}{lccccc}
\tabcolsep=3pt
\tablecolumns{6}
\tablewidth{0pc}
\tablecaption{Metallicities of the pLLSs and LLSs in the CCC Sample\label{t-strongLLS_metallicities}}
\tabletypesize{\footnotesize}
\tablehead{\colhead{Target}   & \colhead{$z_{\rm abs}$}   & \colhead{$\logNHI$}   & \colhead{$[{\rm X/H}]$}   & \colhead{$\log U$}   & \colhead{$[{\rm C/\alpha}]$}    \\
           \colhead{ }        & \colhead{ }               & \colhead{[cm$^{-2}$]} & \colhead{ }               & \colhead{ }          & \colhead{ }                    }
\startdata
J044011.90$-$524818.0     & 0.614962       & $16.20 \pm 0.01$       & $-$0.29, $-$0.27, $-$0.25          & $-$2.49, $-$2.47, $-$2.45  & $-$0.05, $-$0.02, +0.00             \\ 
J111908.59+211918.0       & 0.138500       & $16.22 \pm 0.02$       & $-$0.58, $-$0.56, $-$0.53          & $-$2.88, $-$2.81, $-$2.73  & $-$0.05, +0.07, +0.19               \\ 
J152424.58+095829.7       & 0.518500       & $16.22 \pm 0.02$       & $-$0.69, $-$0.66, $-$0.63          & $-$3.50, $-$3.45, $-$3.40  & $-$0.55, $-$0.50, $-$0.46           \\ 
J104117.16+061016.9       & 0.655390       & $16.24 \pm 0.02$       & $-$0.27, $-$0.22, $-$0.17          & $-$4.44, $-$4.37, $-$4.29  & $>$+0.74, $>$+0.90, $>$+0.98        \\ 
J215647.46+224249.8       & 0.621016       & $16.24 \pm 0.02$       & $-$2.07, $-$1.96, $-$1.84          & $-$2.95, $-$2.81:, $-$2.69 & \nodata                             \\ 
J101622.60+470643.3       & 0.727771       & $16.26 \pm 0.09$       & $-$1.61, $-$1.40, $-$1.18          & $-$3.91, $-$3.60:, $-$3.33 & \nodata                             \\ 
J102056.37+100332.7       & 0.465364       & $16.26 \pm 0.02$       & $<$$-$4.65, $<$$-$3.42, $<$$-$1.88 & $-$3.49, $-$2.98:, $-$2.45 & $-$0.42, $-$0.09:, +0.25            \\ 
J155232.54+570516.5       & 0.366020       & $16.28 \pm 0.02$       & $<$$-$4.56, $<$$-$2.89, $<$$-$1.24 & $-$3.47, $-$2.94:, $-$2.40 & $-$0.42, $-$0.08:, +0.26            \\
\enddata
\tablecomments{
The lower (``low'') and upper (``high'') bounds for each quantity represent the 68\% and 80\% CI\ for the detections and upper and lower limits, respectively. When a colon is present after the median value of $\log U$, this implies that a Gaussian prior was used to determine the properties of the absorber (see \S\ref{s-mcmc-gauss-u}). $^a$Proximate absorbers, i.e., absorbers with $\Delta v \equiv (z_{\rm em} - z_{\rm abs})/(1+z_{\rm abs})\, c < 3000$ \km\ (see \citetalias{lehner18} and \S\ref{s-close-prox}).
}
%
%
%
\end{deluxetable*}


\subsubsection{Comparison with \citetalias{lehner13}}
\label{s-comp-l13}

As discussed above, in \citetalias{lehner13}, we used a different methodology to derive the metallicity of the pLLSs and LLSs. \citetalias{lehner13} used similar techniques to derive the column densities of the absorbers and the same HM05 EUVB adopted here (for most of the absorbers in \citetalias{lehner13}; however, there were a few included from the literature in \citetalias{lehner13} that adopted a different EUVB). Similar to the approach we have adopted in CCC, \citetalias{lehner13} also used mostly the low and intermediate ions to estimate the metallicities. However, the metallicities derived in \citetalias{lehner13} used a consistency between models and the $\sim$1$\sigma$ bounds of the observed column densities to define the valid regions of parameter space (a ``$\chi^2$-by-eye'' type approach). The errors provided on $U$ and $\xh$ in that approach do not correspond to rigorous confidence intervals. Furthermore, metallicities for several of the absorbers compiled from the literature by \citetalias{lehner13} did not include error estimates. Obvious limitations of this method are that errors are not rigorously assessed and some $U$ and $\xh$ solutions could have been missed. 

\begin{figure}[tbp]
\epsscale{1.2}
\plotone{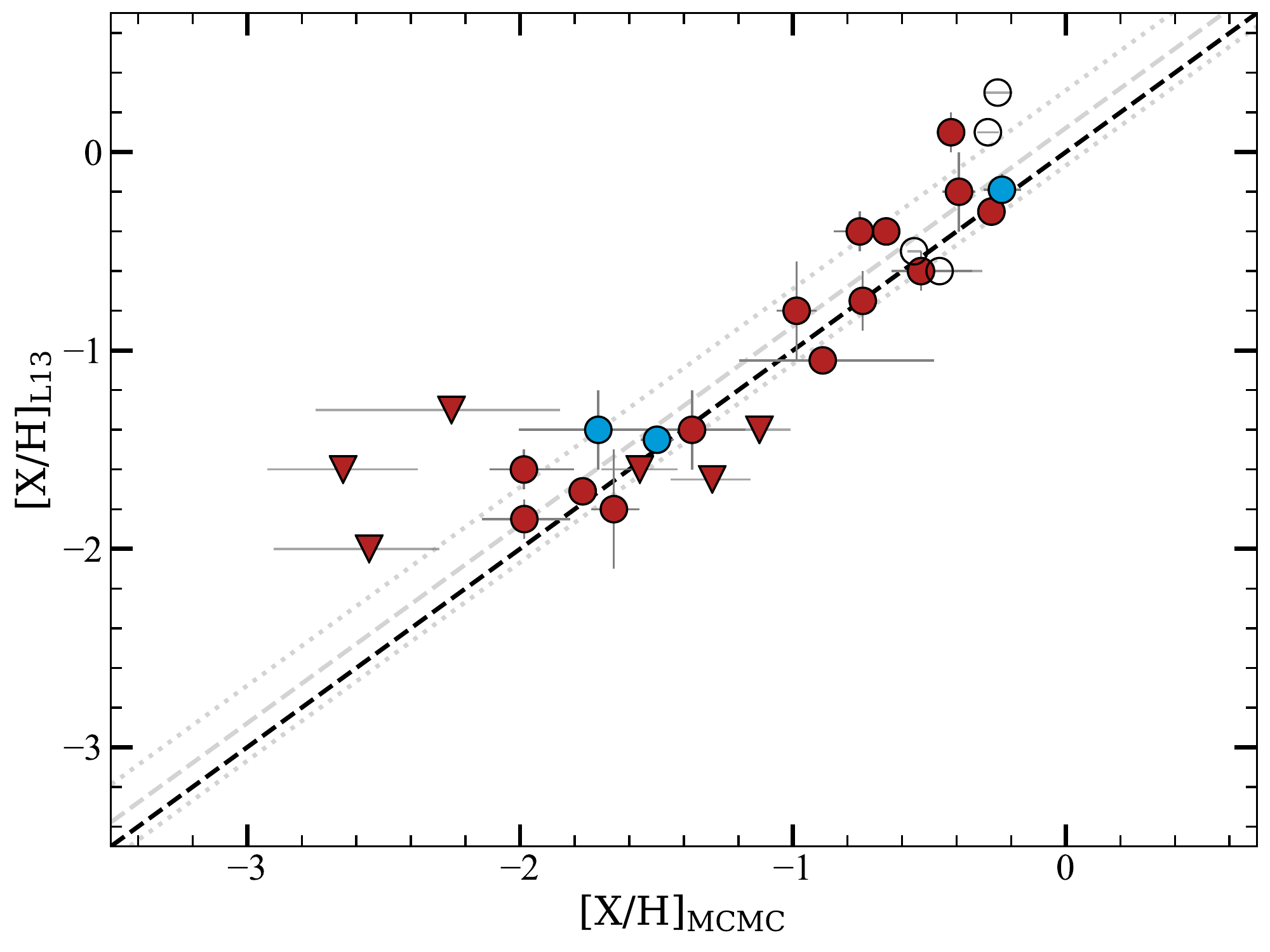}
\caption{The metallicities of the \citetalias{lehner13} sample as derived by \citetalias{lehner13} compared with those derived using the MCMC-driven analysis presented here. Blue data points represent absorbers for which the original column densities from \citetalias{lehner13} were used in the MCMC analysis. Red data points indicate absorbers for which new column density estimates were made (independent of the original measurements). Metallicities for which no error on the $y$-axis was given in the literature (see \citetalias{lehner13}) are denoted by open circles.  The black dashed line is the 1:1 relationship. The gray dashed and dotted lines are the mean and standard deviation of the difference between the two methods. \label{f-strongLLS_met_vs_met_L13}}
\end{figure}

We show in Fig.~\ref{f-strongLLS_met_vs_met_L13} the metallicity estimates derived using the present MCMC code ($x$-axis) with those estimated by \citetalias{lehner13} ($y$-axis). Limits on the \citetalias{lehner13} metallicities are denoted by triangles. Metallicities for which no error was given in the literature are denoted by open circles (see \citetalias{lehner13}). Blue and open circles represent absorbers for which the MCMC analysis was run on the original column density measurements presented in \citetalias{lehner13}, while red points represent absorbers for which column densities were re-estimated in \citetalias{lehner18} (i.e., that are independent of the original measurements in \citetalias{lehner13}; note that in a couple of cases, the UV transitions were not re-analyzed, but newly-available \mgii\ and \feii\ ground-based observations have become available). 

Overall there is a good agreement between the two sets of results, with the mean absolute difference between the two methods $\la$0.1 dex  (gray dashed lines).  We note that the lack of data around $\xh \approx -1$ corresponding to the dip in the bimodal metallicity distribution revealed in \citetalias{lehner13} is still present. However, all of the systems with upper limits in \citetalias{lehner13} are described by posterior PDFs on their metallicity, from which we derive a median and robust confidence interval. The new results for several of these systems make use of a Gaussian prior on $\log U$, an additional improvement in our approach that was not adopted in \citetalias{lehner13}.

\subsubsection{Comparison with \citetalias{wotta16}}
\label{s-comp-w16}

\citetalias{wotta16} presented a large sample of pLLS/LLS metallicity measurements based on the adoption of prior knowledge on the ionization parameter distribution in the population. They demonstrated this approach can yield accurate metallicities for pLLSs and LLSs at $z \la 1$ specifically when using only the observed ratio of \mgii /\hi\ without full constraints on the ionization model. Here we expand upon this approach to deriving ``low-resolution metallicities'' to assess metallicities in absorbers with a broader suite of ions; we make use of the \logU\ distribution as a prior in the Bayesian framework of our analysis. We adopt a Gaussian prior on \logU\ with absorbers for which the metal ion column density constraints are not enough to break the degeneracy between $\logU$ and $\xh$ (see \S\ref{s-mcmc-gauss-u}, \citealt{fumagalli16}). This includes the absorbers analyzed in \citetalias{wotta16} for which only \mgii\ and \hi\ are available to estimate the metallicity.

\begin{figure}[tbp]
\epsscale{1.2}
\plotone{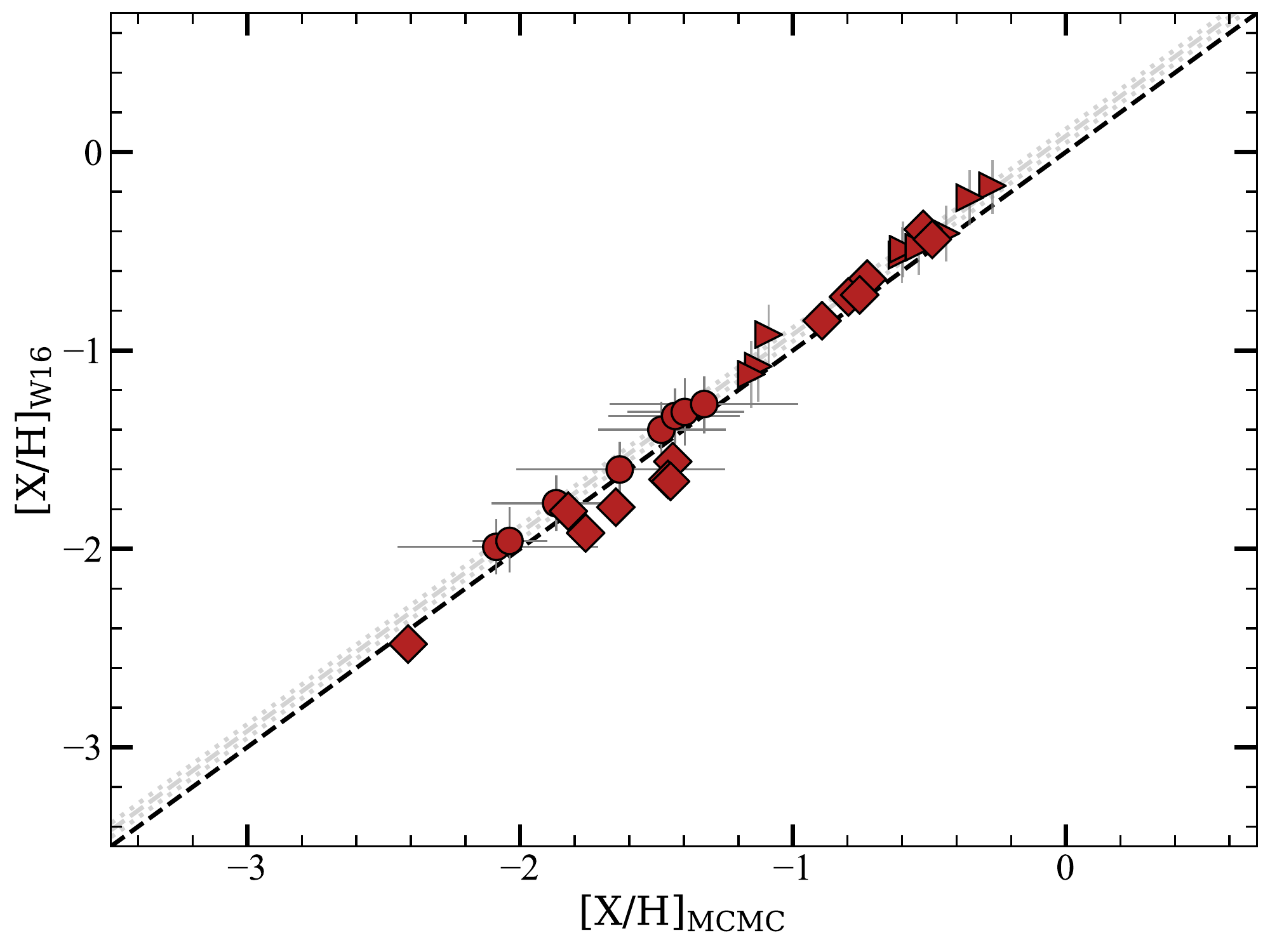}
\caption{The metallicities of the \citetalias{wotta16} sample as derived by \citetalias{wotta16} as a function of those derived using the Bayesian MCMC procedure. Both relied on the low-resolution method, where a Gaussian prior on \logU\ was applied. The diamond symbols represent simultaneous limits on the metallicities. The right-pointing triangles denote lower limits derived by the MCMC analysis; these represent the 95\% CI. The black dashed line is the 1:1 relationship.  The gray dashed and dotted lines are the mean and standard deviation of the difference between the two methods.
\label{f-strongLLS_met_vs_met_W16}}
\end{figure}

In Fig.~\ref{f-strongLLS_met_vs_met_W16}, we show the comparison between \citetalias{wotta16} ($y$-axis) and Bayesian MCMC ($x$-axis) analyses. In that figure, diamonds indicate simultaneous limits on the \citetalias{wotta16} and MCMC metallicity estimates. The exact placement of the MCMC limits in this figure depends on the CI chosen---here, we display the upper (lower) bound of the 95\% CI for the upper (lower) limits (we used 95\% CI in this figure to be consistent with \citetalias{wotta16}). Overall, there is a good agreement between the \citetalias{wotta16} and MCMC implementations. Several \mgii\ column densities that were detections in \citetalias{wotta16} were assigned lower limits after being remeasured with high-resolution data (see \S\ref{s-strongLLS_mgii_saturation_mods_vs_hires}). These points are seen in the upper-right of Fig.~\ref{f-strongLLS_met_vs_met_W16} as right-pointing triangles.

%

The small, systematic discrepancy between the detections (indicated by the circles) is caused by the different statistics used in our MCMC code compared with those adopted by \citetalias{wotta16}. In \citetalias{wotta16}, we used the expectation value (i.e., the mathematical average) of this distribution to compute the metallicities. In contrast, the MCMC approach adopts the median for $n_{\rm H}$ (which is related to the ionization correction factor through $\log U$). For a skew-normal distribution, the mean lies toward higher ionization corrections (and therefore higher metallicities) than the median by $\sim$0.1 dex; this is of similar order to the systematic discrepancy we see in Fig.~\ref{f-strongLLS_met_vs_met_W16}.

\subsection{Statistical and Systematic Errors}
\label{s-error}

As discussed in \citetalias{lehner13} and \citetalias{wotta16} \citep[see also][]{howk09,chen17}, the ionization corrections themselves and hence the metallicity PDFs are also subject to systematic errors. One of the large uncertainties is, of course, the shape of the photoionizing EUVB. Two widely used ionizing radiation fields are the HM05 and HM12 EUVBs \citep{haardt96,haardt01,haardt12}. A critical difference between HM05 and HM12 is the greatly reduced escape fraction of radiation from galaxies in the latter, leading to a harder EUVB spectrum. This generally gives higher metallicities (\citetalias{wotta16,lehner13}; \citealt{howk09}). Using a sample of 10 absorbers, we showed in \citetalias{wotta16} that changing the EUVB from HM05 to HM12 can increase the metallicities, on average, by $+0.3$ dex (the scatter ranging from 0.0 to $+0.6$ dex). \citetalias{wotta16} find no apparent change in that systematic with the metallicity of the absorber (i.e., low- and high-metallicity absorbers are affected the same way). 

Here we revisit the effects of changing the EUVB from HM05 to HM12, deriving metallicities for the 126 systems for which we could employ flat priors on $\logU$ (i.e., those with good constraints). We use the same methodology described in \S\ref{s-mcmc-method}, only adopting the HM12 EUVB rather than HM05. The MCMC models for 13 absorbers did not converge (mostly absorbers with $\xh \ga -1$) using the HM12 background; the metallicities for these absorbers could potentially be calculated by adopting a Gaussian prior on \logU , although it may be this EUVB is not appropriate for their conditions. These absorbers were not included further in our comparison. We note that when using the HM12 EUVB, we occasionally had to remove some ions that were used in the HM05 modeling to allow for the convergence of the models (a similar strategy had to be used in \citetalias{wotta16}). We also note that the harder HM12 spectrum often provides an agreement with the data that was not as good as with the HM05 EUVB; in particular, HM12 sometimes struggles to match the observed ratios of \siii/\siiii\ (or \mgii/\siiii), while this was not as much as a problem when using the HM05 EUVB. Similar difficulties were noted by \citetalias{wotta16} for 2 of the 13 absorbers analyzed with HM12. 

\begin{deluxetable}{lccc}
\tabcolsep=2pt
\tablecolumns{4}
\tablewidth{0pc}
\tablecaption{Comparison the metallicities derived using the HM12 and HM05 EUVB \label{t-hm12vshm05}}
\tabletypesize{\footnotesize}
\tablehead{\colhead{Sample} & \colhead{m} & \colhead{$\langle \xh_{\rm HM12} - \xh_{\rm HM05}\rangle $} & \colhead{Min.,Max.} }
\startdata
All absorbers		&  113  &$+0.37 \pm 0.19$&   $-0.15.+1.11$  \\
SLFSs			&   66  &$+0.40 \pm 0.19$&   $-0.15,+0.73$  \\
pLLSs			&   37  &$+0.38 \pm 0.17$&   $+0.03,+1.11$  \\
LLSs 			&   10  &$+0.16 \pm 0.12$&   $+0.00,+0.40$  \\
\enddata
\tablecomments{$m$ is the number of absorbers for a given sample. Min. and Max. are the minimumn and maximum difference values between HM12 and HM05.
}
\end{deluxetable}

\begin{figure}[tbp]
\epsscale{1.2}
\plotone{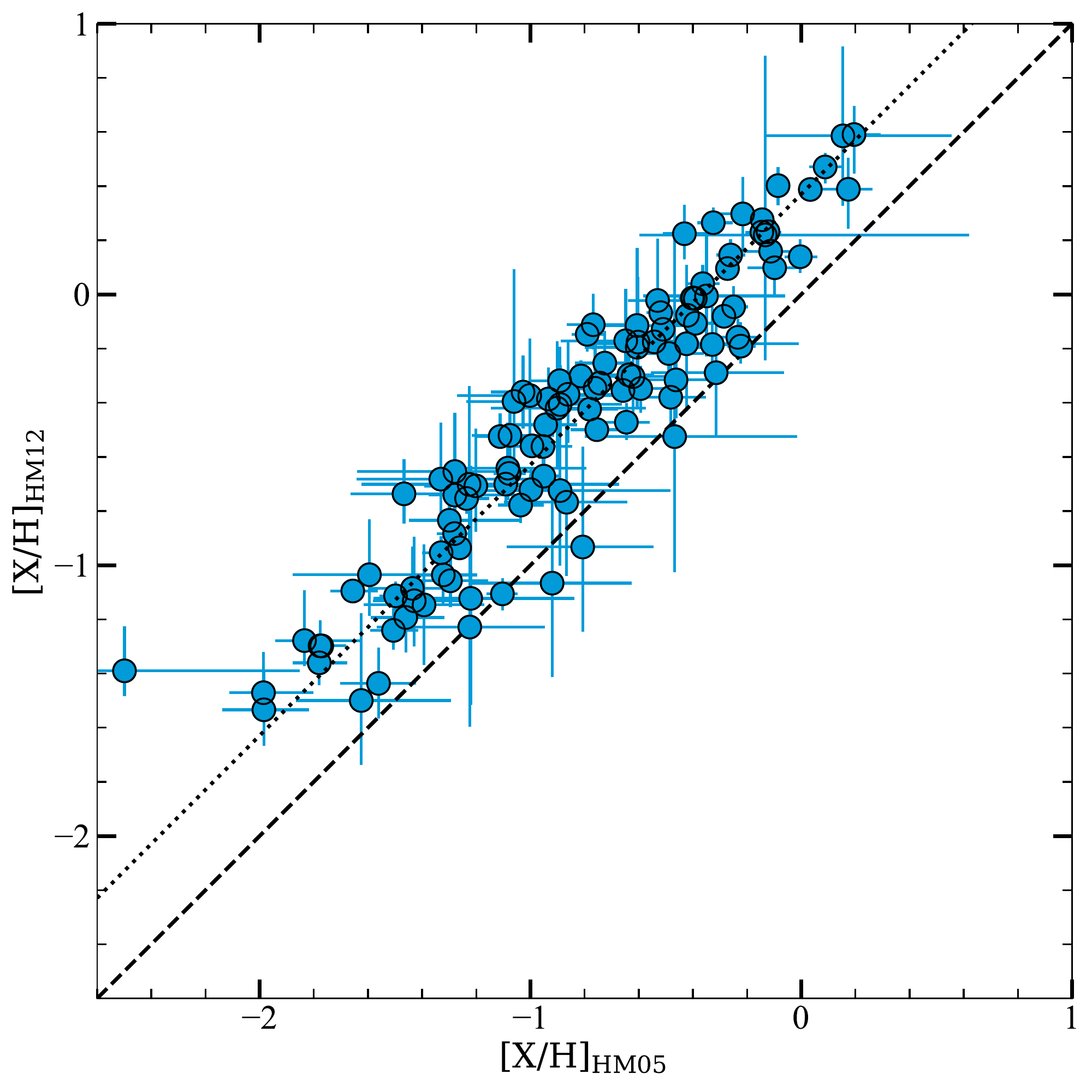}
\caption{Comparison of the median metallicities (with 68\% CI) of  CCC derived using the HM12 and HM05  EUVBs. The dashed line is the 1:1 relationship. The dotted line show the mean of the differences between the metallicities derived using HM12 and HM05. 
\label{f-hm12vshm05}}
\end{figure}
\begin{figure}[tbp]
\epsscale{1.2}
\plotone{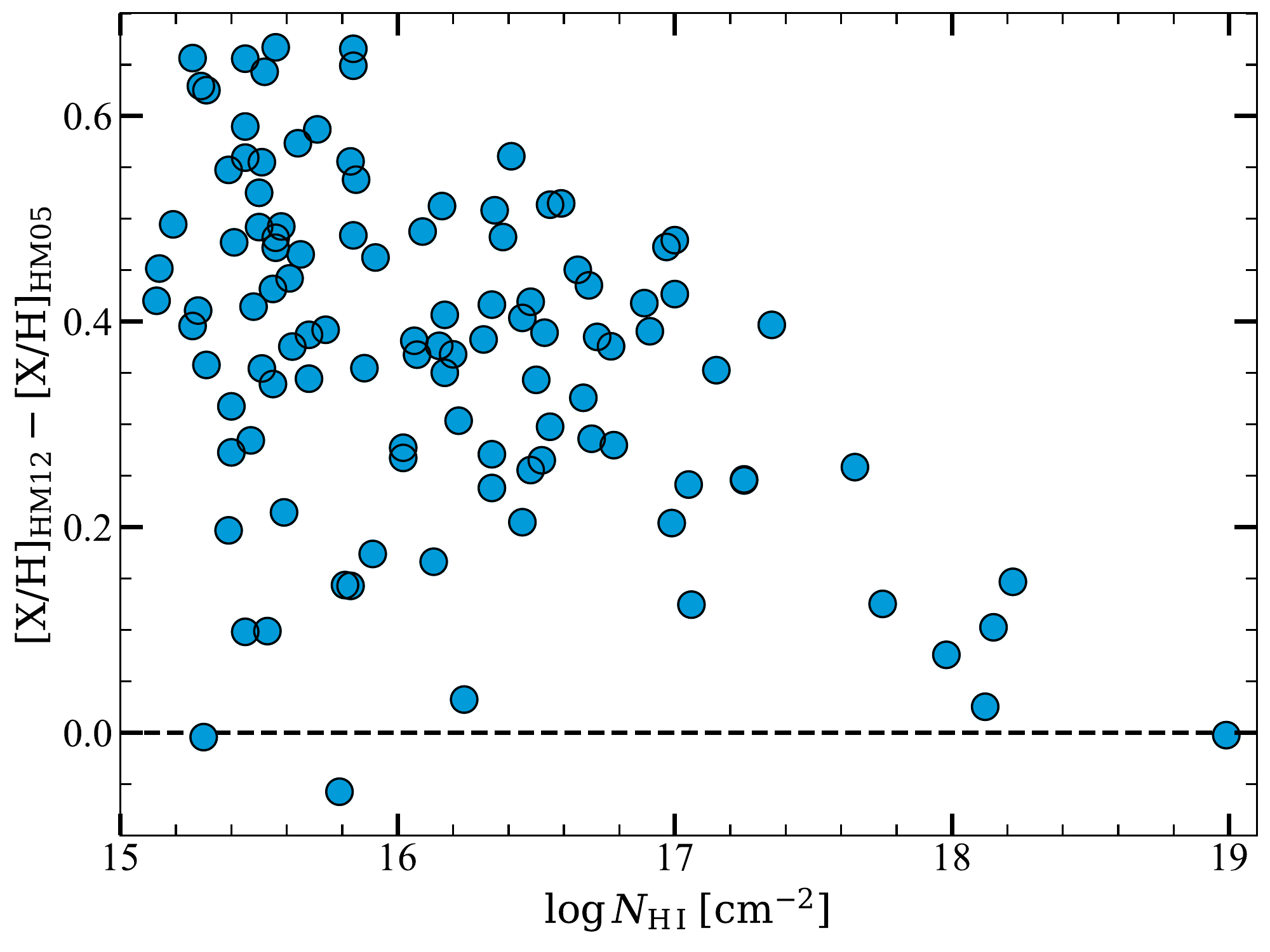}
\caption{Difference of the median metallicities of CCC derived using the HM12 and HM05 EUVBs against the \hi\ column densities of the absorbers. 
\label{f-hm12-hm05vsnh1}}
\end{figure}

In Fig.~\ref{f-hm12vshm05}, we show the comparison between the metallicities using HM12 and HM05. For most of the absorbers, we find that $\xh_{\rm HM12}\ge \xh_{\rm HM05}$, with a mean and standard deviation of  $\langle \xh_{\rm HM12} - \xh_{\rm HM05}\rangle =  +0.37 \pm 0.19$. This is quite consistent with the results from \citetalias{wotta16} described above \citep[as well as the results of ][]{howk09,werk14,chen17}. Similarly, if we analyze the absorbers presented in \citet{prochaska17} with their column densities, we find $\langle \xh_{\rm HM12}  - \xh_{\rm HM05}\rangle =  +0.26 \pm 0.19$ for those COS-Halos absorbers. The lower offset in the COS-Halos sample is likely due to the difference in the \hi\ distribution in the two samples, as we find that the mean calculated offsets depend on \nhi\ as we now demonstrate. 

We show in Fig.~\ref{f-hm12-hm05vsnh1} the difference of the median metallicities between HM12 and HM05 vs.\ \nhi; in Table~\ref{t-hm12vshm05} we list the mean differences of the metallicities derived from HM12 and HM05 for the SLFSs, pLLSs, and LLSs separately. There is a clear decrease in the scatter and the offsets between the two metallicity scales as \nhi\ increases, especially in the LLS regime compared to the SLFSs and pLLSs. This is likely due to (1) self-shielding becoming more important as the optical depth at the Lyman limit ($\tau_{\rm LL}$) increases and reaches $\tau_{\rm LL}>1$;  and (2) the required ionization corrections decrease with \nhi\ (this effect is seen in Fig. \ref{f-strongLLS_logU_vs_h1}, with \logU\ decreasing with \nhi).  Importantly, we finally note that there is no evidence that the offsets between HM05- and HM12-based metallicities are metallicity-dependent. 

In Table~\ref{t-hm12vshm05}, we also list the minimum and maximum metallicity differences between HM12 and HM05. Only one absorber has a median metallicity that is over 10 times different, but for that absorber the HM05 solution has large uncertainty (i.e., within 68\% CI, the HM12 and HM05 values are actually similar), and for that reason we do not show it in Fig.~\ref{f-hm12-hm05vsnh1}.

We find no other obvious systematics in our approach. Our use of $\ca$ as a free parameter, sometimes constrained with a Gaussian prior could in principle represent another source of systematic offsets between our results and other literature metallicities. To test this we compared the metallicities derived by \cite{prochaska17}, who adopt $\ca \equiv 0$, for pLLS/LLS absorbers with our own calculations for those absorbers using their reported column densities. We adopt the HM12 spectrum for this comparison, since it is the EUVB adopted by \cite{prochaska17}. We find excellent agreement between our results and theirs, with a mean difference $\langle \xh_{\rm COS\text{-}Halos}  - \xh_{\rm CCC}\rangle =  -0.05 \pm 0.23$. The slight offset is consistent with the mean offset from solar $\ca = -0.05$ (see \S\ref{s-mcmc-gauss-ca}). 

There are inherent systematic uncertainties associated with the ionization corrections employed to estimate the metallicities of absorbers with $15\la \mlnhi \la 19$. However, these do not affect strongly the general shape of the metallicity distribution; they shift the metallicity distribution by typically $\approx +0.4$ dex for the SLFSs and pLLSs and  $\approx +0.16$ dex for the LLSs (as one changes the EUVB from HM05 to HM12).  As argued in \citetalias{wotta16}, we consider the HM05 EUVB to be more suitable for our analysis since it often provides often solutions for singly- and doubly-ionized species more consistent with the observational constraints than HM12. 

\startlongtable
\begin{deluxetable*}{lccccc}
\tabcolsep=3pt
\tablecolumns{6}
\tablewidth{0pc}
\tablecaption{Closely Redshift Separated Absorbers \label{t-close}}
\tabletypesize{\footnotesize}
\tablehead{\colhead{Target}& \colhead{$[\mlnhi^1,\mlnhi^2]$}   & \colhead{$[z^1_{\rm abs},z^2_{\rm abs}]$} & \colhead{$\Delta v_c$}  & \colhead{$[(\xh^1),(\xh^2)]$}   \\
           \colhead{ }        & \colhead{[\cmm] } & \colhead{} & \colhead{(\km)} & \colhead{ }                 }
\startdata
J035128.56-142908.0    & $[16.38 , 16.55]$ & $[0.356924 , 0.357173]$ &     55.0     & $[(-1.33, -1.24, -1.13), (-0.34,-0.22,-0.09)]$ \\
J100535.25+013445.5    & $[16.52 , 16.36]$ & $[0.836989 , 0.837390]$ &     65.4     & $[(-1.62, -1.51, -1.39), (-2.27,-1.90,-1.61)]$ \\
J100902.06+071343.8    & $[16.45 , 18.68]$ & $[0.355352 , 0.355967]$ &    136.0     & $[(-0.44, -0.36, -0.28), (-0.95,-0.82,-0.68)]$ \\
J102056.37+100332.7    & $[16.48 , 16.26]$ & $[0.464851 , 0.465364]$ &    105.0     & $[<(-4.70, -3.54, -2.10), <(-4.65,-3.64,-1.88)]$ \\
\enddata
\tablecomments{Absorbers with $\Delta v_c \equiv (z^2_{\rm abs} - z^1_{\rm abs})/(1+z^1_{\rm abs})\, c < 500$ \km\ along the same sightline.
For the metallicities, the lower and upper bounds for each quantity represent the 68\% CI\ for detections and the 80\% CI\ for upper limits; the middle values are the median values. 
}
\end{deluxetable*}

\subsection{Proximate and Paired Absorbers}
\label{s-close-prox}

While not directly related to our individual metallicity measurements, the presence of some absorbers close to the target QSOs or close in redshift/velocity space to one another could potentially affect the shape of our derived metallicity distribution function for the population if these absorbers are unique. As noted in \citetalias{lehner18}, we do not reject {\it a priori}\ absorbers from our sample with redshifts near the redshifts of the QSOs against which they're observed, i.e., those with $\Delta v \equiv (z_{\rm em} - z_{\rm abs})/(1+z_{\rm abs})\, c < 3000$ \km, where $c$ is the speed of light. These are referred to as proximate absorbers. In Table~\ref{t-strongLLS_metallicities}, we mark the 5 proximate absorbers in our sample, all of which are pLLSs. For all but one, we employed a Gaussian prior on \logU\ to derive their metallicities. These absorbers do not radically differ from the intervening systems in their measurable properties. This statement applies, for example, to the pLLS at $z=0.4708$ toward J161916.54+334238.4 that we highlighted in \citetalias{lehner18} as having extremely strong absorption from intermediate to high ions. Since there are only 5 proximate pLLSs and since their metallicities are not conspicuous, we opted to include them in our sample. They cannot significantly skew our derived metallicity distribution for the whole population given they represent a small fraction of the total absorbers and their properties are not distinguishable from those of intervening systems.

In several cases we find absorbers closely separated in redshift space; we denote systems separated by $\Delta v_c \equiv |(z^2_{\rm abs} - z^1_{\rm abs})/(1+z^1_{\rm abs})\, c| < 500$ \km\ as ``paired absorbers.'' Where possible we consider these absorbers separately, but the ability to do so depends sensitively on the structure of the absorption. However, it is not {\em a priori} clear that they necessarily trace different halos; for that reason, we discuss these explicitly, as they could represent a double-counting of absorbers. We list in Table~\ref{t-close} the paired absorbers, giving their \nhi, \zabs, $\Delta v_c$, and $\xh$. Except for one LLS paired with a pLLS, all of these systems are paired pLLSs. This is, in part, an observational bias since at the COS resolution separating absorbers can be done more easily and robustly for absorbers with lower \nhi\ than with higher \nhi\ \citepalias{lehner18}. All the paired absorbers are intervening systems (i.e., not proximate). This table only includes pLLSs and LLSs; pLLS--SLFS pairs are discussed in \citetalias{lehner18b}. Table \ref{t-close} includes only 4 paired absorbers, all but one of which have distinct metallicities. We will discuss in more details the implications in \citetalias{lehner18b} with the entire sample, but it is clear from this small sample that metallicity variation exists on small redshift/velocity scales. For the absorbers where we could not separate the components, the metallicity is therefore a column density-weighted average on the metallicity. Since the metallicities of this sample are distinct, and because the sample size of paired absorbers is only 4, we elected to include each of these absorbers individually in our final sample.


\section{The Metallicity of the CGM probed by \lowercase{p}LLS\lowercase{s} Ss and LLS\lowercase{s} at $\lowercase{z}\la 1$}\label{s-met-pdf-cgm}

\subsection{The Metallicity PDF of the pLLSs}
\label{s-strongLLS_results-pLLS_MCMC}

\begin{figure}[tbp]
\epsscale{1.2}
\plotone{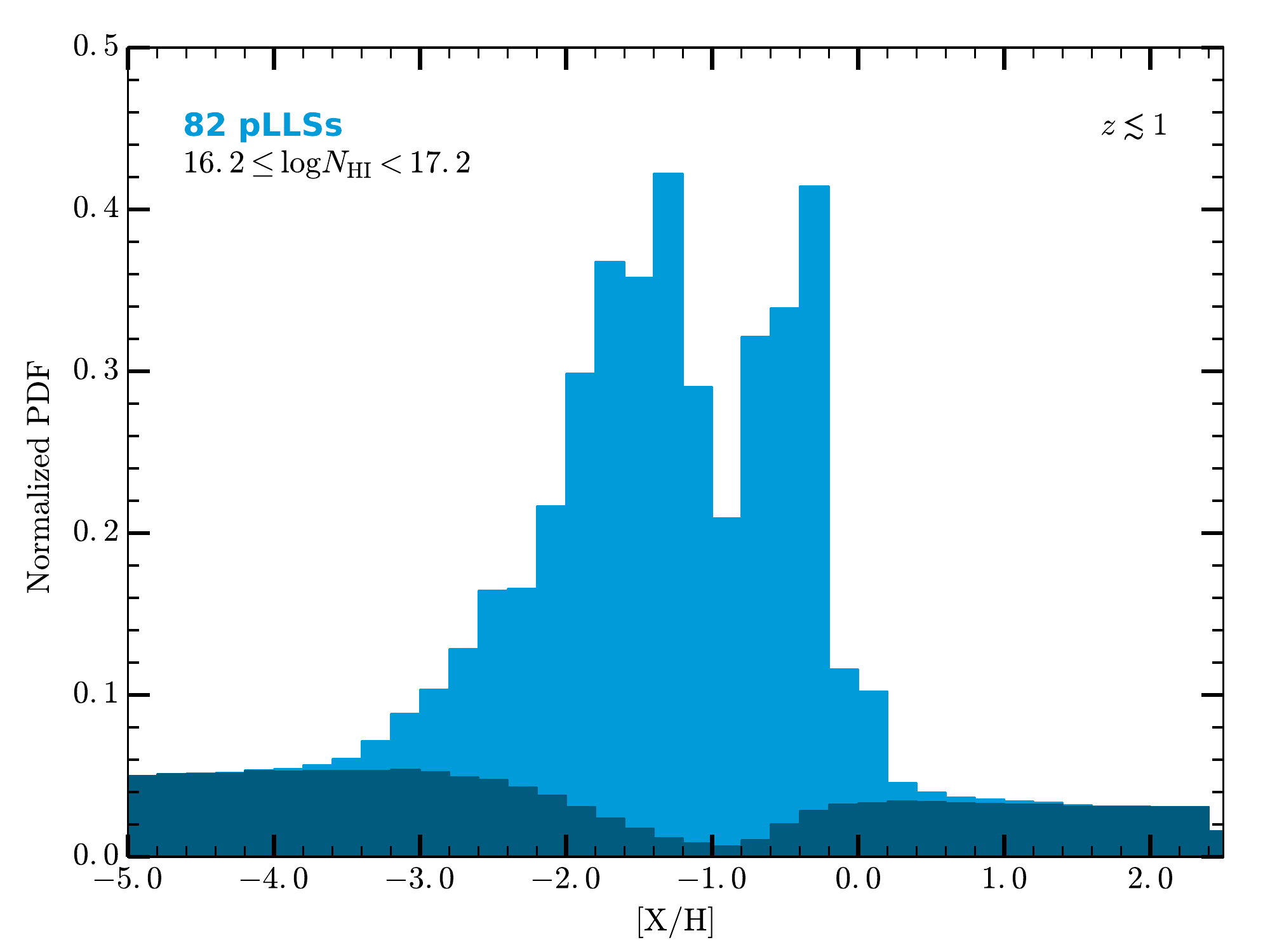}
\caption{Posterior metallicity PDF of the pLLSs at $z \la 1$. The darker regions indicate the contribution from the upper and lower limits. The metallicity PDF of the pLLSs appears bimodal, with peaks centered at  $[{\rm X/H}]\sim-1.7$ and $[{\rm X/H}]\sim-0.4$. 
\label{f-strongLLS_met_pdf-pLLS}}
\end{figure}

In Fig.~\ref{f-strongLLS_met_pdf-pLLS}, we show the metallicity PDF of the 82 $z \la 1$ pLLSs in our sample. This PDF is constructed by combining the normalized metallicity PDFs of all of the absorbers, in contrast to the distributions shown by \citetalias{lehner13} and \citetalias{wotta16} that employed single values for each absorber. However, similar to those earlier works, we find again  that the distribution has two peaks, centered at approximately $[{\rm X/H}]\sim-1.6$ and $[{\rm X/H}]\sim-0.4$. However, there are also some key differences. First, there is a long tail toward lower metallicities. This is a result of the upper limits being distributed to the lowest metallicities sampled by the MCMC method. While these upper limits existed in our previous surveys, they were placed in our displays at their highest allowed values. Second, the dip at $[{\rm X/H}]\sim-1.0$ is less prominent because (1) the tails of the PDFs for several individual absorbers extend into this region, and (2) a few new pLLSs have metallicities near $[{\rm X/H}]\sim-1.0$. We find that the median ($\pm$ the interquartile range) of the CCC pLLS metallicity PDF is $[{\rm X/H}]=\PLLSMCMCmedian\pm\PLLSMCMCmedianep$.

To test the presence of more than one peak in the metallicity distribution, we follow \citetalias{lehner13} who used the dip statistic \citep{hartigan85} and  the Gaussian Mixture Modeling (GMM) algorithm developed by \citet{muratov10}. For these procedures, we need to select a single, representative value for each absorber.  However, using the median values of the extended PDFs for these limits, which extend to what are unlikely extremes, would flatten the distribution, especially to low metallicities given the large number of many upper limits. The limiting metallicities for most absorbers are similar to well-constrained metallicities. We therefore treated the upper (lower) limits as values centered at the upper (lower) bounds of the 80\% CI for the statistical tests. Using the dip statistic, the chance of being bimodal is 95\%. The GMM also rejects a unimodal distribution at the $>$99.9\% confidence level. The GMM also estimates a significant separation of the peaks ($D= 3.29 \pm 0.50$), suggesting the pLLS metallicity PDF is consistent with having two peaks at $\xh = -1.72 \pm 0.09$ and $\xh = -0.38 \pm 0.06$. These results are fully consistent with our earlier results \citepalias{lehner13,wotta16}, but with a sample twice larger compared to the sample used in \citetalias{wotta16}. 

As discussed in \S\ref{s-error}, the overall effect of changing the EUVB ionizing radiation field is to increase the metallicities of the pLLSs by about 0.4 dex (see Table~\ref{t-hm12vshm05}). Hence the dip of the metallicity PDF would be $\xh \simeq -0.6$ and the two peaks would be at $\sim -1.4$ and $0.0$ dex if the HM12 EUVB was used instead. As shown in Fig.~\ref{f-strongLLS_mgii_vs_hi_2panels} and discussed in \citetalias{lehner18}, the column density of \mgii\ in the pLLSs in not uniformly distributed, instead having two well-separated clusters of data points, implying that the dip observed in the metallicity PDF is not an artifact of the ionization corrections we have applied. We also note that that reducing the sample to the 40 pLLSs that were modeled using only the flat prior on $\log U$ also shows the two peaks and the dip in the PDF at similar values.
 
\subsection{The Metallicity PDF of the LLSs}
\label{s-strongLLS_results-L13_W16_W17}

\begin{figure}[tbp]
\epsscale{1.2}
\plotone{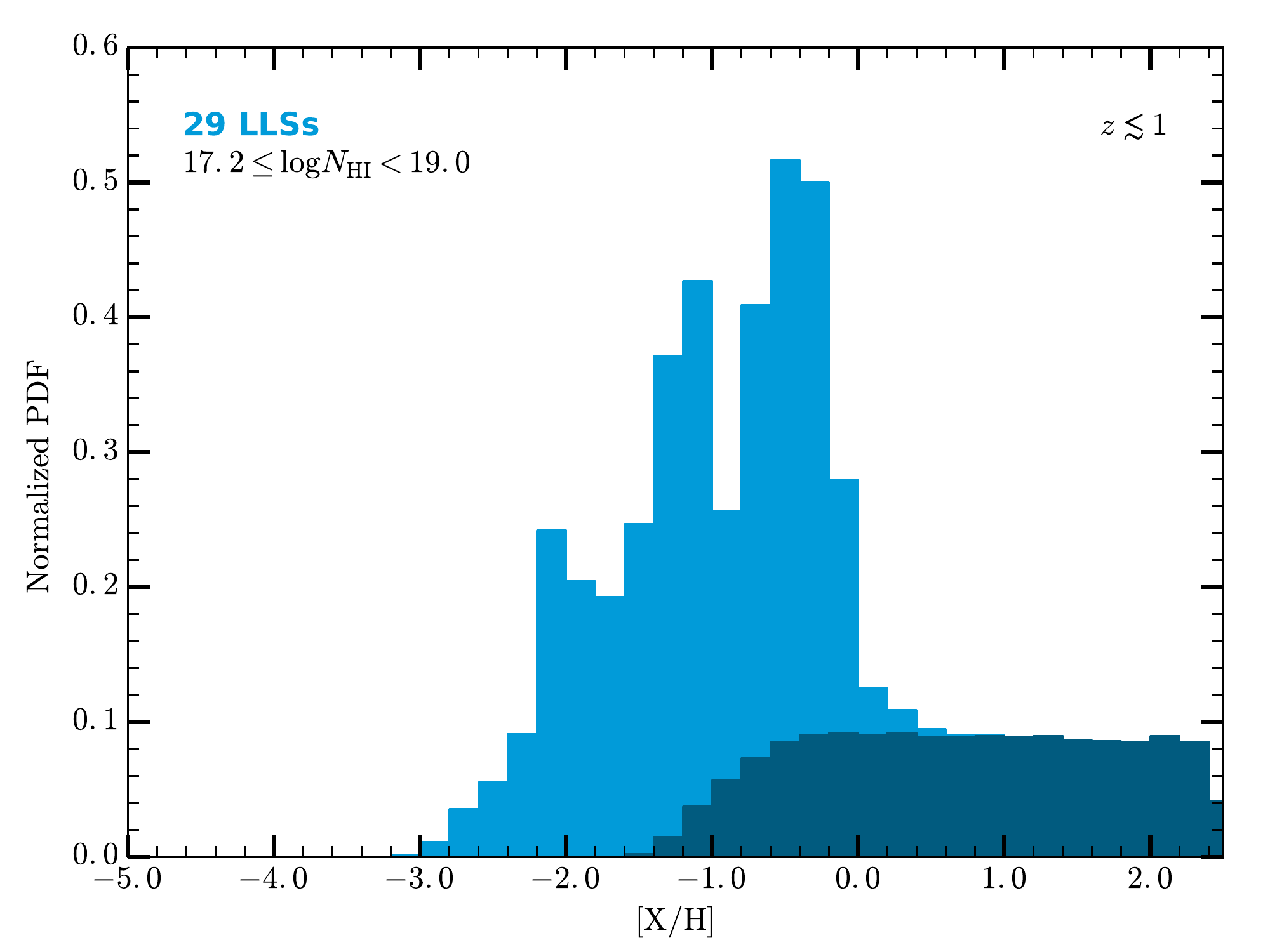}
\caption{Posterior metallicity PDF of the LLSs at $z\la 1$. The darker regions indicate the contribution from the lower limits. The PDF has at least two distinct peaks at $\xh \simeq-1.3$ and $-0.5$ with possibly another one at $\xh \simeq -2$.). 
\label{f-strongLLS_met_pdf_L13W16W17}}
\end{figure}

We found in \citetalias{wotta16} that the metallicity PDF of the LLSs may not follow that of the pLLSs, but the number of LLSs was quite low. We have now increased from 11 LLSs to 29 LLSs (a similar size sample to size of the combined sample of  pLLSs and LLSs used by \citetalias{lehner13}). In Fig.~\ref{f-strongLLS_met_pdf_L13W16W17}, we show the metallicity PDF of the 29 $z \la 1$ LLSs. It is a broad distribution with at least two distinct peaks at $\xh \simeq-1.3$ (slightly higher than the low metallicity peak of the pLLSs, although these two branches overlap) and $\xh \simeq-0.4$ (similar to the high metallicity of the pLLSs); there is possibly another peak at $\xh \simeq -2$, although it could be just an extension of the low metallicity branch as observed for the pLLSs. We note that the low resolution data used in the \citetalias{wotta16} survey are the primary driver for the probability at higher metallicities ($-0.1\lesssim[{\rm X/H}]\lesssim+0.6$). Specifically, lower limits on the \mgii\ column densities in several absorbers of \citetalias{wotta16} contribute lower limits to the metallicity PDF; this is one limitation of the ``low-resolution'' method presented in \citetalias{wotta16} (see \S\ref{s-strongLLS_mgii_saturation_mods_vs_hires}). Based on the metallicities derived from high-resolution data, it is unlikely, however, that the highest metallicities of the lower limits go beyond $\sim + 0.4$ dex solar, i.e., there is no strong evidence from our survey that the LLSs or pLLSs probe gas 2.5 times higher than the solar metallicity. 

\begin{figure}[tbp]
\epsscale{1.2}
\plotone{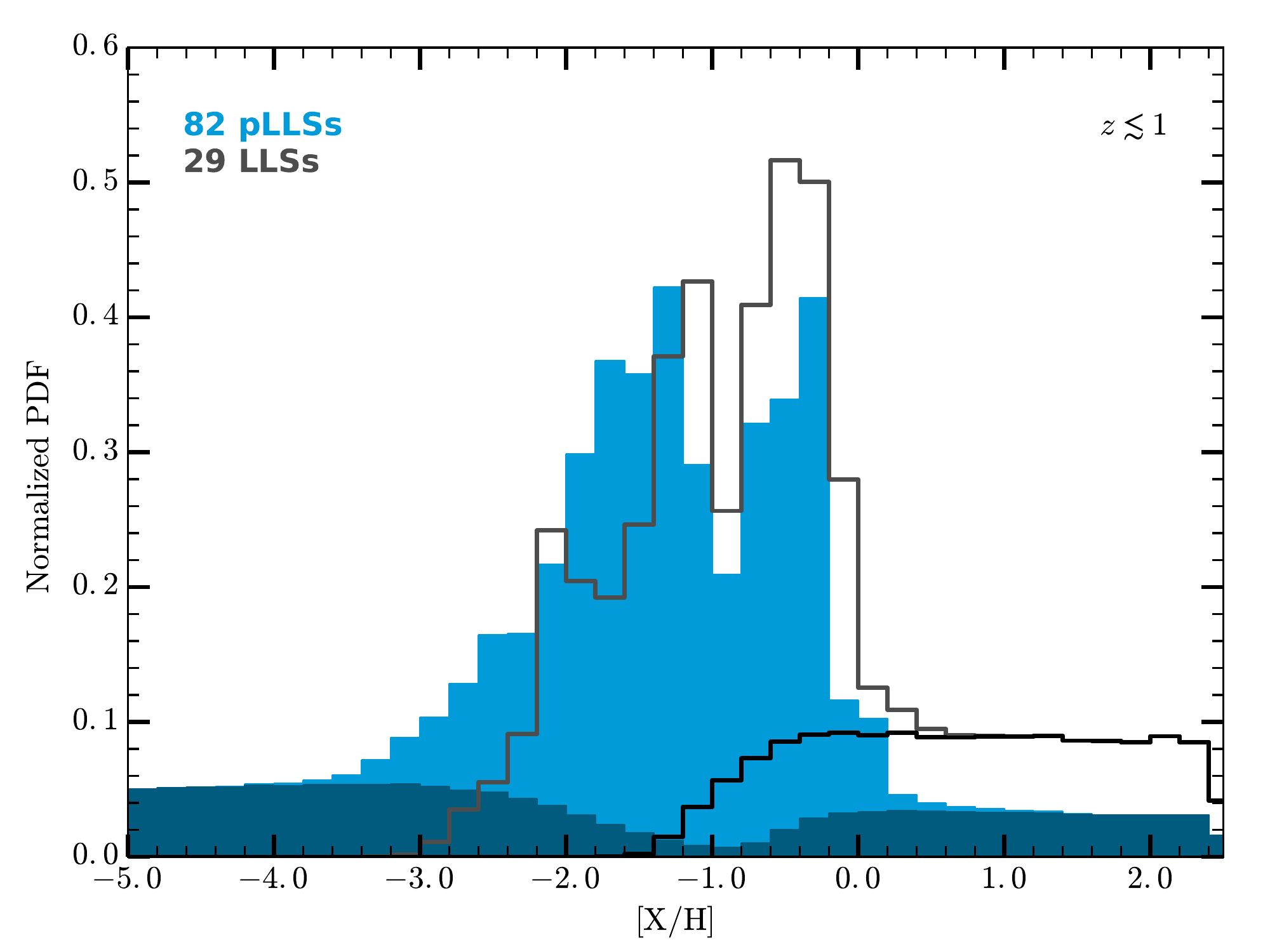}
    \caption{Posterior metallicity PDFs of the pLLSs and LLSs at $z\la 1$. The darker regions indicate upper and lower limits.
    \label{f-strongLLS_met_pdf_pLLS_LLS}}
\end{figure}

For a direct comparison with the pLLSs, we show in Fig.~\ref{f-strongLLS_met_pdf_pLLS_LLS} the metallicity PDFs for the pLLSs and LLSs. We find both pLLSs and LLSs probe a similar metallicity range ($-2.6\la \xh \la +0.4$) with a strong drop off above solar metallicity. There is a somewhat higher probability of LLSs with $\xh >-1$ than for the pLLSs. The probability at very low metallicities ($\xh \la -1.4$, see below) is, however, much higher for the pLLSs than for the LLSs.

To test for the presence of more than one peak in the metallicity distribution of the LLSs, we used the same strategy as for the pLLSs. However, in the case of the LLSs, there are only lower limits, and several of these lower limits are not particularly strong constraints owing to the use of the low-resolution spectra with only (saturated) \mgii\ to derive the metallicity \citepalias{wotta16}. Hence it is not surprising that the dip test does not reject the unimodality at the 89\% confidence level. On the other hand, if we adopted the median values for the lower limits, the unimodality would be rejected at the 99\% confidence level. The GMM still rejects a Gaussian-like unimodal distribution at the 92\% confidence level. However, the separation of the peaks $D = 2.18 \pm 0.98$ is less significant than for the pLLSs. The GMM finds two peaks at $\xh = -1.43 \pm 0.31 $ and $\xh = -0.54 \pm 0.12$. In summary, these statistical tests are less conclusive than for the pLLSs, even though by eye the metallicity PDF of the LLSs looks more complicated than a simple unimodal distribution.\footnote{As shown in \S\ref{s-error}, the effect of changing the EUVB ionizing radiation field to the HM12 would be to increase somewhat the metallicity by about 0.16 dex for the LLSs, but the overall shape of the metallicity PDF would be quite similar. However, in that case, the low metallicity peak of the metallicity PDF of the pLLSs would be similar to that of the LLSs. On the other hand, the high metallicity peak of the metallicity PDF of the pLLSs would be higher than that of the LLSs by about 0.5 dex.}

\subsection{Redshift Evolution of the Metallicity at $z\la 1$ }
\begin{figure}[tbp]
\epsscale{1.2}
\plotone{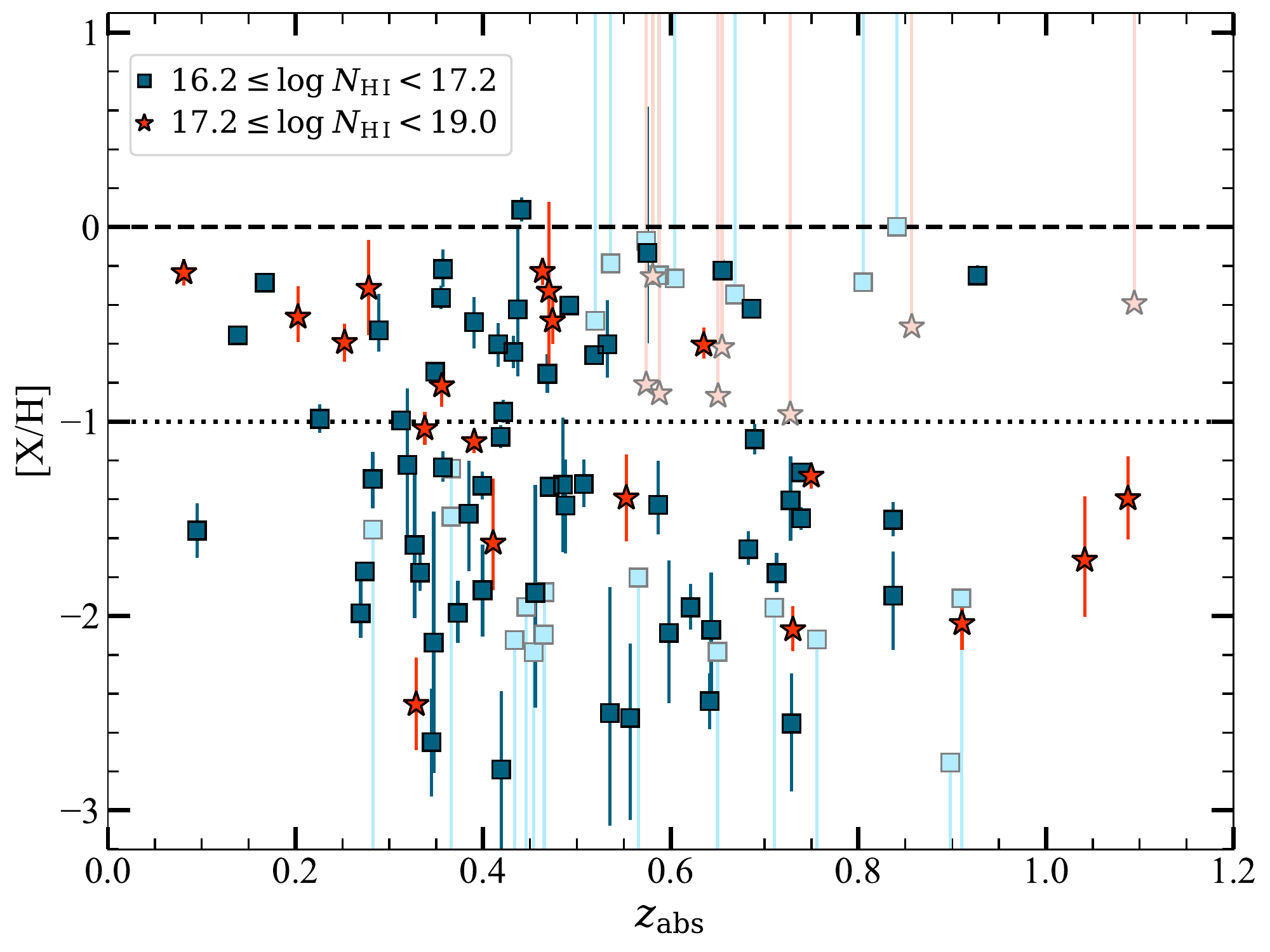}
\caption{Metallicities of the pLLSs and LLSs as function of redshift. The circles show the median values of the metallicity posterior PDFs are adopted with 68\% CI. The triangles give the upper and lower 80\% CI and the error bar gives the opposite 80\% limit (together encompassing the 80\% CI). The dashed line represents solar metallicity. The dotted line represents 10\% solar metallicity, where a dip in the metallicity of the pLLSs is observed. 
\label{f-met-vs-z}}
\end{figure}

In neither the samples of \citetalias{lehner13} nor \citetalias{wotta16} was there evidence for significant evolution of the metallicities of the pLLSs or LLSs with redshift at $z \la 1$. With a larger sample, we revisit this conclusion. In Fig.~\ref{f-met-vs-z}, we show the metallicities of the pLLSs and LLSs as a function of the redshift of the absorbers.  For the well-constrained metallicities, the central values represent the median of the posterior PDFs, and the error bars represent the 68\% CI. The triangles give the upper and lower 80\% CI and the error bar gives the opposite 80\% limit (together encompassing the 80\% CI). Overall, there is no clear evolution with redshift.   Both low and high metallicities are seen across the whole redshift range. As discussed above (and in \citetalias{lehner18}), most of the absorbers are at  $0.2 <z< 0.8$. Within this redshift range, about 44\% of the pLLSs+LLSs have $\xh \ge -1$. At $ z \ga 0.8$, the sample is much smaller, but a similar proportion of the systems are at low metallicities: 7 absorbers are at $\xh < -1$ and 5 are at $\xh >-1$. The sample for $ z \la 0.2$ is too small to draw conclusions on this (a total of 4 absorbers, 3 with  $\xh > -1$). Thus, there is not a good evidence for evolution in either the mean metallicity or the proportion of systems below $\xh < -1$. Applying the Spearman rank order test on absorbers with $0.2 \le z \le 0.8$ (i.e., over the redshift interval where most of absorbers are), the test confirms the visual impression that there is no statistical significant correlation or anti-correlation between the metallicity and redshift of the absorbers (the Spearman correlation coefficient is $r_{\rm S} \simeq -0.05$ and $-0.23$ with $p=0.70$ and 0.31 for the pLLSs and LLSs, respectively).

\label{s-met-red}

\begin{figure}[tbp]
\epsscale{1.2}
\plotone{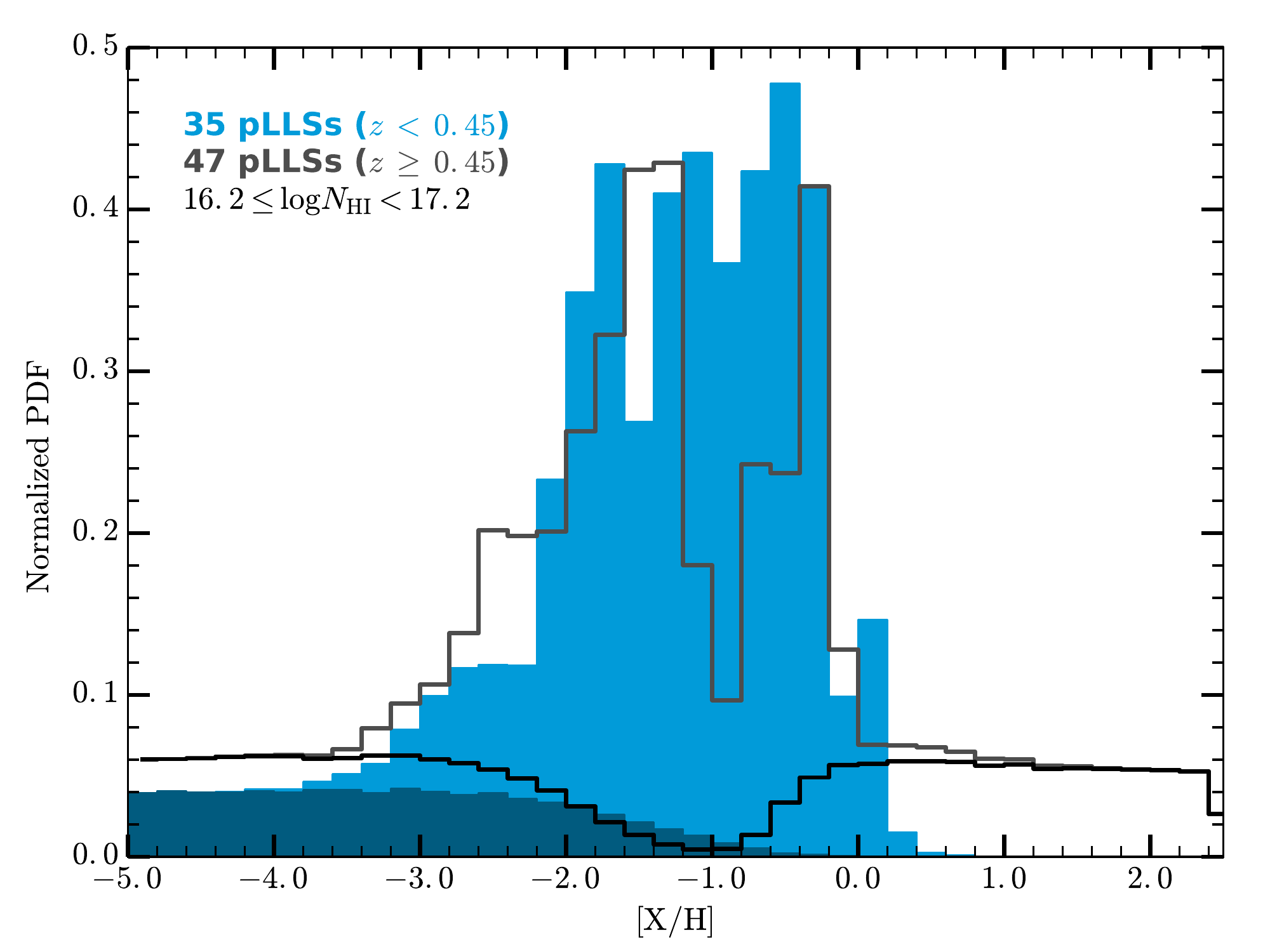}
\caption{Posterior metallicity PDFs of the pLLSs at $z\le 0.45$ ({\it blue histogram}) and $z>0.45$ ({\it gray solid line}). The darker regions and black line indicate the contribution from the upper and lower limits.  
\label{f-pLLS-pdf-vs-z}}
\end{figure}

 As shown in Figs.~\ref{f-strongLLS_met_pdf-pLLS} and \ref{f-strongLLS_met_pdf_L13W16W17}, our larger sample of both pLLSs and LLSs demonstrates, however, a filling of the distribution around a metallicity of $-1$ dex. Since the metallicity for each system is now included as a PDF, the effect is to smear out the distributions seen in \citetalias{lehner13} and \citetalias{wotta16}. In particular, the wings of the metallicity PDFs serve to fill in the gap seen $\xh \simeq -1$ in those earlier distributions. Therefore this filling in the dip of the PDF is expected. However, as shown in Fig.~\ref{f-met-vs-z}, some of the median values of the metallicities are also near the gap at $\xh \sim -1$. Considering a metallicity range defined by $\xh = -1.0 \pm 0.1$, there are 5 pLLSs and 3 LLSs in this interval (assuming all the 4 lower limits\footnote{Note that the large fraction of lower limits at $0.45 \la z \la 1$ is a direct result of several absorbers being part of the \citetalias{wotta16} sample where only \mgii\ was  used to derive the metallicity (see \S\ref{s-strongLLS_mgii_saturation_mods_vs_hires}).} for the LLSs extend beyond this range, which is likely based on the other measurements); all but one absorber (again excluding the lower limits) are at $z\le 0.45$. Considering both the pLLSs and LLSs, the sample sizes are 44 absorbers at $z\le 0.45$ and 67 at $z>0.45$. Yet, the smaller sample has the larger fraction of absorbers with $\xh = -1.0 \pm 0.1$, 9--27\%  compared to $<6\%$ at $z>0.45$ (all 90\% CI), implying a difference between these two samples and a statistically significant lack of absorbers with $\xh \simeq -1$ at $0.45 \le z \la 1$. In Fig.~\ref{f-pLLS-pdf-vs-z}, we compare the posterior metallicity PDFs of the pLLSs at $z< 0.45$ (age of the universe, $t_{\rm U}$ between 9 and 11 Gyr) and $z\ge 0.45$ ($6 \la t_{\rm U}\la 9$ Gyr). The dip near $\xh  \simeq -1$ at  $z\ge 0.45$ is quite apparent, but absent at $z<0.45$. So while the $\xh$--$z$ figure is largely a scatter plot at $z\la 1$, there is nevertheless some evidence of a mild trend in the metallicity of the pLLSs with redshift, where the metallicity PDFs change from a bimodal distribution at $0.45 \la z \la 1$ to a unimodal distribution at $z\la 0.45$. 

\subsection{Metallicity as a Function of \nhi}
\label{s-strongLLS_results-met_vs_h1}

\begin{figure}[tbp]
\epsscale{1.2}
\plotone{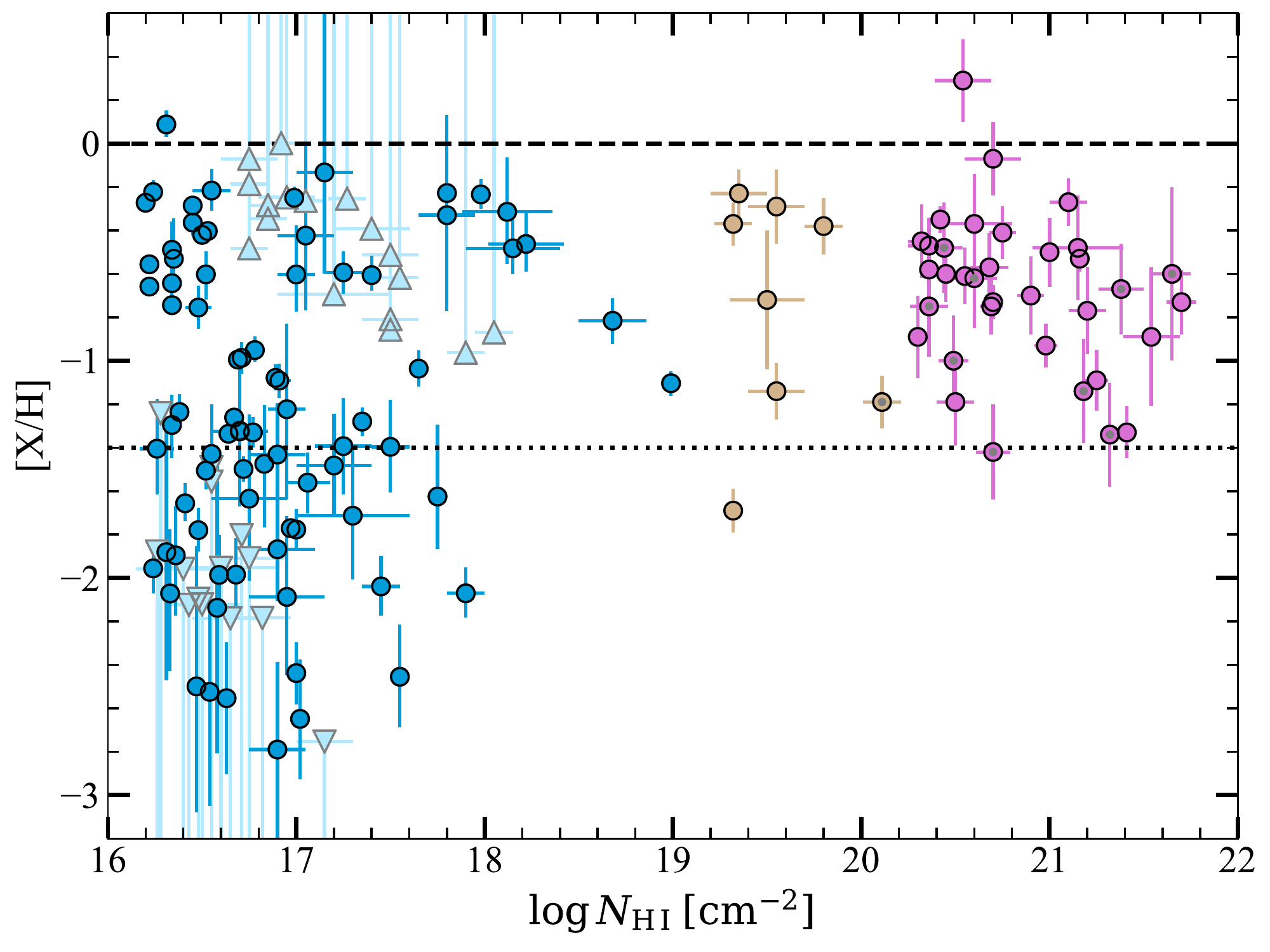}
\caption{Metallicities of the pLLSs, LLSs, SLLSs, and DLAs at $z\lesssim1$ as a function of \nhi. For the pLLSs and LLSs, the median values of the metallicity posterior PDFs are adopted with 68\% CI, except for lower and upper limits where the lower and upper values represent the minimum and maximum of the 80\% CI shown with the light colored lines. For SLLSs and DLAs, these represent the best estimated values with their 68\% errors (see \S\ref{s-sample-comp}).  The dashed line represents solar metallicity. The dotted line represents the very low metallicity gas, which is defined as the 2$\sigma$ lower bound of the DLA metallicities ($\xh <-1.4$). The SLLS and DLAs with a gray circle have their metallicities derived from \feii\ (and were corrected by $+0.5$ dex, see \S\ref{s-sample-dla}). 
\label{f-strongLLS_met_vs_h1}}
\end{figure}

We show in Fig.~\ref{f-strongLLS_met_vs_h1} the metallicities of the $z \la 1$ absorbers as a function of \nhi. The \hi\ column density is related (in an average sense) to the density of the gas as well as its typical location within a galaxy's halo. Thus, the observed relationship between metallicity and \hi\ column density can shed light on the origins of gas in the CGM. It is also a relationship that can be directly compared with simulation output (see, e.g., Fig.~12 in \citealt{fumagalli11a} or Fig.~7 in \citealt{hafen17}). Fig.~\ref{f-strongLLS_met_vs_h1} includes the pLLSs and LLSs from the CCC sample, the \hi-selected SLLSs (see \S\ref{s-sample-slls}), and the DLAs (\S\ref{s-sample-dla}) at redshifts $z \la 1$. For the pLLSs and LLSs with well-constrained metallicities (i.e., not including the lower and upper limits), the central values represent the median of the posterior PDFs, and the error bars represent the 68\% CI. For the upper and lower limits, the triangles give the upper and lower 80\% CI and the error bar gives the opposite 80\% limit (together encompassing the 80\% CI). For the SLLSs and DLAs, the best estimates with their $1\sigma$ error bars are shown. The horizontal dotted line at $\xh =-1.4$ represents the 2$\sigma$ lower bound of the DLA metallicities at $z\la 1$. We define absorbers below this metallicity as ``very metal-poor'' absorbers (see \citealt{lehner16}; \citetalias{wotta16}). 

As discussed in \S\ref{s-sample-slls}, with the \hi-selection, the SLLS metallicities have a similar distribution to that of the DLAs. This is distinct from the picture one finds when including  SLLSs that were selected based on the strength of the \mgii\ absorption (see the compilation in \citetalias{lehner13}), which show an overabundance of super-solar SLLSs. Building a larger sample of \hi-selected SLLSs at $z\la 1$ will be difficult, but would be extremely beneficial to understand the nature of these absorbers. We also note that despite the increase of LLSs in our sample, the information on the metallicities of LLSs in the range $18.2 \la \mlnhi < 19$ is still scant, owing to the difficulty in robustly determining \nhi\ in this range (and indeed there were 10 more candidates listed in Table~5 of \citetalias{lehner18} for which we were only able to place lower limits on \nhi). 

As first shown in \citetalias{lehner13} and then \citetalias{wotta16}, the range of metallicities covered by both the pLLSs and LLSs is much broader than that of the SLLSs or DLAs. There is an overlap in the metallicity interval $-1.4<\xh < 0$ for the pLLSs, LLSs, SLLSs, and DLAs, but  absorbers with \hi\ column densities $\mlnhi \ga 18$ very rarely have metallicities $\xh <-1.4$. We find the fractions of pLLSs with $\xh <-1$ and $<-1.4$ are 54--71\%  and 41--59\% (90\% CI), respectively, while for the LLSs these are 27--50\% and 13--32\%, respectively. While the frequency of low-metallicity LLSs is somewhat smaller than that of the pLLS (but see \S\ref{s-cdf-met-vs-nhi}), the frequency of the very metal-poor LLSs is significantly smaller, as also demonstrated in Fig.~\ref{f-strongLLS_met_pdf_pLLS_LLS}; additionally, all of the very metal-poor LLSs are found at $\mlnhi <18$ (but again there are only 6 LLSs with $18 \le \mlnhi <19$). While a smaller proportion of the LLSs than pLLSs are very metal poor, their fraction is still significantly higher than for the SLLSs and DLAs. Thus, the metallicity distributions of the pLLSs, LLSs, SLLSs, and DLAs are distinct, and very metal-poor gas is predominantly found in the pLLS regime and, to a lesser extent, in the LLS regime. 

Considering only the pLLSs and LLSs, the paucity of systems around $\xh =-1$ is quite striking with our sample of 82 pLLSs and 29 LLSs, i.e., a factor $\sim$4 times larger than the original sample of \citetalias{lehner13} where we first discovered the lack of systems near $\xh \simeq -1$ at $z\la 1$. With our larger sample, there is now some hint of a change in the metallicity PDFs of the pLLSs and LLSs as a function of $z$, since the few pLLSs and LLSs with  $\xh \simeq -1$ are mostly found at $z<0.45$ (see \S\ref{s-met-red}). 

\subsection{Cumulative Distribution of the Metallicity as a Function of \nhi}
\label{s-cdf-met-vs-nhi}

\begin{figure}[tbp]
\epsscale{1.2}
\plotone{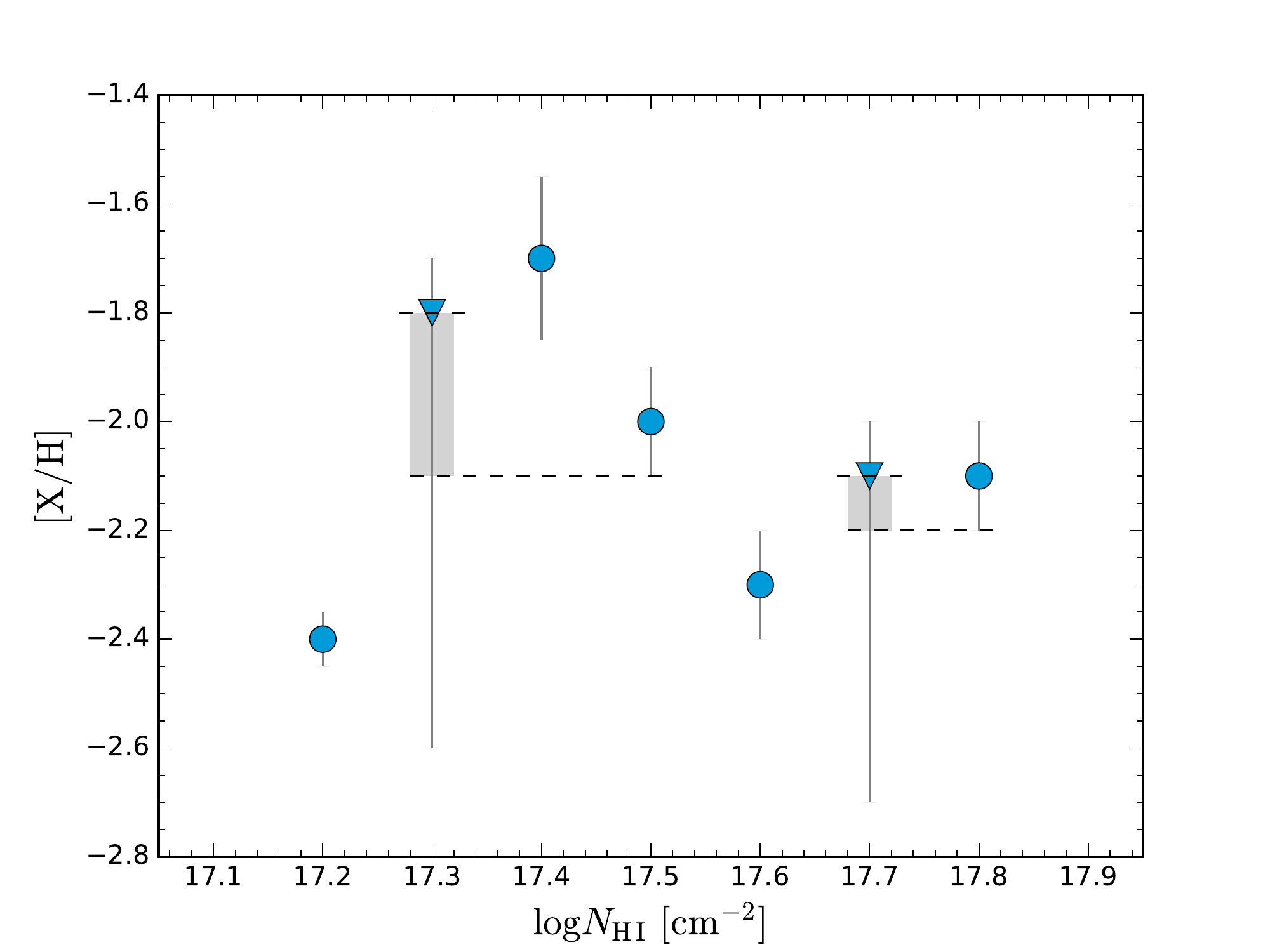}
\caption{Example of survival analysis-like technique for selecting walkers for the metallicity CDFs. The nearest detected metallicity equal to or lower than each upper limit is identified, and walkers between the (upper bound of the 80\% CI of the) upper limit and the lower bound of the nearest detection's 80\% CI are included in the CDF. The gray shaded regions depict the metallicity regime where walkers are kept when including the upper limits in this example, following this technique. The data in this figure are fabricated for demonstration purposes.
\label{f-survival_analysis_example}}
\end{figure}

Another way to characterize the metallicity distribution and its evolution with \nhi\ is through the use of cumulative distribution functions (CDFs) of the combined posterior metallicity PDFs. These can be used to assess, in particular, the probability of finding systems below various thresholds in metallicity. In cases where an absorber's metallicity PDF suggests it has an upper or lower limit on its metallicity, we add its contribution to the metallicity CDF by treating it in two different ways. First, we include the MCMC walkers as output by the models (i.e., they go to the extreme edges of our models as listed in Table~\ref{t-mcmc_parameters}). Second, we treat them in a way similar to censored points in a survival analysis. For an upper limit, we first compute the upper bound of its 80\% CI\ to use as the maximum of its adopted PDF. We then find the nearest detected absorber below this (i.e., a non-censored point in survival analysis) and use the lower bound of the detected absorber's interquartile range as the minimum of the upper limit's adopted PDF. We thus include in the final PDF (for the upper limit) the MCMC walkers with metallicities between these derived maximum and minimum metallicities. In Fig.~\ref{f-survival_analysis_example}, we demonstrate this process with fabricated data; the gray shaded regions depict the metallicity regimes where walkers are kept for each of the upper limits when including them in the metallicity CDFs using this second, survival analysis-like technique.\footnote{We do not use a flat-top distribution between the maximum and minimum in an effort to preserve the shape of the PDF at the upper bound of the limit. We also employ the lower bound of the nearest detection's interquartile range rather than the median to avoid excessively-narrow PDFs, should the derived upper limit and the nearest detection have similar metallicities, as it would be the case for the rightmost upper limit in Fig.~\ref{f-survival_analysis_example}.} We use an inverted process for including lower limits in the metallicity CDF. In the figures depicting our CDFs (see below), we shade the metallicity regime between the two approaches. 

\begin{figure}[tbp]
\epsscale{1.2}
\plotone{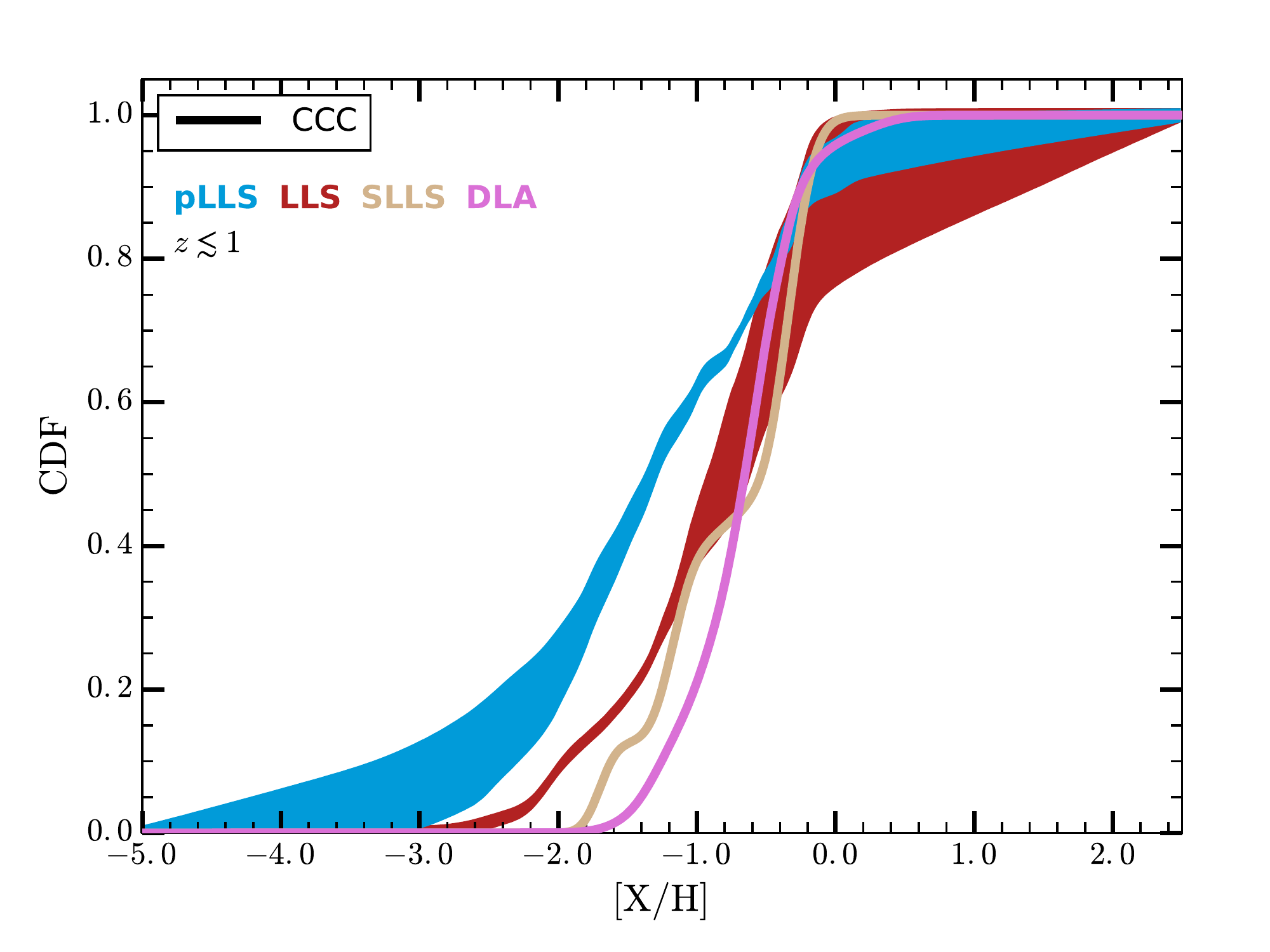}
\caption{The cumulative probabilities of the pLLS, LLS, SLLS, and DLA metallicity PDFs. 
\label{f-strongLLS_met_cdf}}
\end{figure}

In Fig.~\ref{f-strongLLS_met_cdf}, we show the CDFs of the pLLS, LLS, SLLS, and DLA metallicity PDFs. For the SLLSs and DLAs taken from the literature, we assumed each absorber's PDF is normally-distributed with a mean given by the reported central value and dispersion equivalent to the reported 1$\sigma$ measurement uncertainty. From Fig.~\ref{f-strongLLS_met_cdf}, the medians of the metallicity PDFs are $[{\rm X/H}]\simeq - 1.3$, $-0.8$, $-0.5$, and $-0.6$ for the pLLSs, LLSs, SLLSs, and DLAs, respectively. The CDFs  confirm that the distributions of the metallicity PDFs for the pLLSs and LLSs are both different from those of the SLLSs and DLAs. 

With the CDFs, the difference between the pLLS and LLS samples is even more obvious than shown in \S\ref{s-strongLLS_results-met_vs_h1}: the probabilities of pLLSs and LLSs with $[{\rm X/H}]\le -1.0$ are 62\% and 40\%, respectively; in the very metal-poor regime (below $[{\rm X/H}]=-1.4$), these are 48\% and 22\%, respectively, and the probability of the LLS metallicity PDF drops to $\sim$0\% for $[{\rm X/H}]<-2.4$, while it is still in the 10--20\% range for the pLLSs. 

While the result was tentative in \citetalias{wotta16}, it is now evident that the pLLSs have a significantly lower-metallicity population than the LLSs. This is unlikely to be an observational bias. As we argued in \citetalias{lehner18}, it is more difficult to separate velocity components in LLSs than pLLSs, and {\it a priori}\ it could be plausible that some metal-poor LLSs could be more often lost than pLLSs. However, there are several reasons why this should not create a bias against metal-poor LLSs. First, \citetalias{wotta16} show that overall the low-resolution data did not introduce a difference in the distribution of the pLLSs and LLSs compared to the high-resolution observations. Second,  \citetalias{wotta16} also show that 5/10 LLSs observed with the low resolution grating of COS have $\xh <-1$; i.e., the probability of finding low metallicity LLSs (with $\xh<-1$ is not small. However, the probability of very low metallicity LLSs with $\xh <-1.4$ is very small, with only 1/10 LLS in that sample. Third, most of the high-resolution \mgii\ observations have relatively simple velocity profiles with only one exception ($z=0.355967$ toward J100902.06+071343.8, see also \citealt{tumlinson11b}), where a complex velocity structure is observed and it could be possible that a metal-poor LLS could be hidden. Hence the lack of very metal-poor LLSs is unlikely to be a bias. There is therefore a clear trend from the metallicity PDFs, CDFs, and median metallicities that the metallicity distribution strongly evolves with \nhi, even between the pLLS and LLS regimes. It seems that the metallicity transition between the LLSs and SLLSs occurs in the $18 \la \mlnhi \la 19$ range because 6/6 LLSs in this range have $-1 \la \xh \la 0$, i.e., very similar to the SLLSs with 7/8 in a similar metallicity range.  

\section{Discussion}
\label{s-disc}
\subsection{Present State of Affairs}\label{s-disc-sum}

We have studied a sample of 82 \hi-selected pLLSs and 29 \hi-selected LLSs at $z\la 1$, quadrupling the original sample of \citetalias{lehner13}. Our sample of absorbers has been uniformly analyzed using the same EUVB (HM05) and the same Bayesian MCMC analysis technique for all the absorbers that needed an ionization correction to derive the metallicity (i.e., all the pLLSs and LLSs, and some SLLSs). This reduces significantly any possible systematic uncertainty from, e.g., the use of different EUVBs or different analysis techniques, when comparing the metallicity of absorbers. With this unprecedented, large sample of pLLSs and LLSs and uniform analysis, we strengthen several conclusions made in our previous surveys \citepalias{lehner13,wotta16} and uncover new trends. From the previous sections, we can summarize so far the main results as follows: 
\begin{enumerate}[wide, labelwidth=!, labelindent=0pt]
\item The pLLS metallicity PDF has two main peaks at $\xh \simeq -1.7 $ and $-0.4$  with  a strong dip at $\xh \simeq -1$, fully consistent with our first survey \citepalias{lehner13}.   
\item There is, however, some evidence that the metallicity PDF of the pLLSs may change from a bimodal distribution at $z\ga 0.45$ to a unimodal metallicity distribution at $z\la 0.45$.
\item The metallicity distribution of LLSs may be more complicated, with multiple clusters across different \nhi\ ranges. 
\item The PDFs of the pLLS and LLS metallicities probe a very wide range of metallicity from $\xh <-3$ to $\sim +0.4$ dex. This wide range of metallicities is not observed in the SLLSs or DLAs since since 60\% and 80\% of the SLLSs and DLAs are confined to $-1\la \xh \la 0$, respectively.
\item Very metal-poor gas with $\xh \la -1.4$ is only observed in the pLLSs and LLS regimes. This disparity between the pLLSs and LLSs on one hand and the SLLSs and DLAs on the other hand is not attributable to a redshift evolution effect.
\item While very metal-poor gas  is observed in the LLSs, the probability for such gas is much higher for the pLLSs than for the LLSs. 
\item Although the sample in the column density range $18\la \mlnhi\ \la 19$ remains small, the metallicity of the LLSs in this \nhi\ range is quite similar to the SLLSs, i.e., the metallicity distribution probably transitions in this \hi\ column density range to a DLA-like distribution with few very metal poor absorbers. 
\end{enumerate}

Below we discuss the possible origins of the  pLLSs and LLSs given the metallicities derived in this work, as well as compare with recent cosmological simulations that determined the metallicities of the gas as a function of \nhi. 

\subsection{Metallicity as an Indicator of Origins}\label{s-disc-origins}

The pLLSs and LLSs are known to probe the extended gas around galaxies at $z\la 1$. As discussed in \citetalias{wotta16}, the overdensities required to produce a pLLS or LLS at $z\la 1$ are consistent with those expected dense CGM gas rather than IGM gas. Furthermore, searches for galaxies in the fields of QSOs toward which a pLLS or LLS is observed reveal that almost every pLLS or LLS lies within 30--150 kpc of a galaxy (e.g., \citetalias{lehner13}; \citealt{lehner09,cooksey08}). So far, our ongoing multi-wavelength survey confirms these findings, always detecting at least one galaxy (and sometimes more than one) at the redshift of the pLLS or LLSs (\citealt{lehner17} and ongoing analysis the galaxy fields where pLLSs and LLSs have been observed). In all these cases, there is no evidence of pLLSs or LLSs at very small impact parameters ($\la 10$--20 kpc), a regime that is dominated by DLAs or SLLSs (see Fig.~9 in \citetalias{lehner13} and references therein). From a galaxy-selected absorber perspective, the COS-Halos survey has shown that pLLSs and LLSs are also prevalent in the inner CGM of galaxies, notably within $\rho \sim 160$ kpc of $L_*$ galaxies, with a tendency for SLLSs and DLAs to be at the smallest impact parameters \citep{werk14,prochaska17}. 

Thus the preponderance of the evidence favors the pLLSs and LLSs at $z \la 1$ all the probing extended gas around galaxies at impact parameters $\rho \sim30$--200 kpc. Hence Fig.~\ref{f-strongLLS_met_vs_h1} provides some insight on how the metallicity of dense gas varies around galaxies; as \nhi\ increases, and in particular changes from the pLLSs/LLSs to SLLSs/DLAs, the absorbing gas is typically found closer to galaxies.  According to cosmological simulations \citep[e.g.,][]{fumagalli11a,vandevoort12a,vandevoort12b,faucher-giguere15,hafen17,rahmati18}, pLLSs and LLSs also probe gas flows in the CGM of galaxies, both in and out of galaxies. With estimates of the metallicity of these absorbers, we can therefore start to decipher their mixture of plausible origins.

Enriched metal gas with $\xh \ga -1$ at $z\la 1$ is common at all column densities from the pLLS to the DLA regimes. This gas likely has many different origins, including outflows, tidally-disrupted gas, recycling gas. A better understanding of the galaxy--absorber properties may help differentiating these phenomena and determining if one of these processes dominate the observed populations of \hi-selected pLLSs and LLSs at $z\la 1$. 

As per the simulations referenced above, the gas in cold flow accretion should be very metal-poor (although still enriched at some levels by previous episodes of star formation). Based on Fig.~\ref{f-strongLLS_met_vs_h1}, very metal-poor gas with $\xh\la -1.4$ is not commonly found in the densest gas probed by SLLSs and DLAs, those probing the inner-most regions of the CGM. Instead our analysis shows that the incidence rate of very metal-poor gas is dominated by the pLLSs. Quite remarkably about 1/3 of the pLLSs (a 90\% CI of 20--39\% of pLLSs) have metallicities $\xh < -2$ (see also Fig.~\ref{f-strongLLS_met_pdf_pLLS_LLS}). Given the metallicity distribution of $z\simeq 2.5$--$3.5$ pLLSs and LLSs peaks at $\xh \simeq -2$ (see \citealt{lehner16,fumagalli16} and see also \citealt{cooper15,glidden16} for stronger \hi\ absorbers), this implies a population of absorbers in this column density range that has experienced little to no additional chemical enrichment over $\sim$6 Gyr. We emphasize that owing the evolution of the universe pLLSs and LLSs are not exact analogs at high and low $z$, instead pLLSs and LLSs at $z\sim 3$ should evolve into SFLSs and pLLSs at $z \sim 0.5$, respectively \citep{lehner16}. Nevertheless and as we will see \citetalias{lehner18b}, a large fraction of the SLFSs and pLLSs at $z\la 1$ have metallicities $\xh < -2$, implying little evolution over cosmic time for these structures. While at high $z$, some of the gas probed by the pLLSs and LLSs can have overdensities consistent with the IGM, as argued above, this is not the case in the $z\la 1$ universe: the very metal-poor gas at $z\la 1$ may have originated principally from the IGM, but it is now is within the gravitational reach of galaxies and likely bound. 

While the metal-enriched pLLSs and LLS ($\xh \ga -1$) confirm previous studies that metals can be found far from their formation sites \citep[e.g.,][]{prochaska11c,liang14,werk14,johnson15}, our surveys of \hi-selected absorbers have revealed a new population of very metal-poor gas ($\xh < -1.4$) probed by pLLSs and LLSs within $\sim$30--200 kpc galaxies. This very metal-poor gas has not been processed recently in galaxies or otherwise it would be more metal-enriched; it has also not been mixed with more metal-enriched gas, which would produce significantly higher metallicities than observed. Thus, this dense, cold, and very-metal poor gas resides within the halos of galaxies, but it has yet experienced little influence from their pollution despite the fact that galaxies have been enriching their halos for several billions of years via galaxy outflows \citep[e.g.,][]{shapley03,steidel10,weiner09,rudie12,liang14,martin13,werk13,rubin14,turner15}. Our survey shows that there is an important reservoir of such very metal-poor gas in the CGM of $z \la 1$ galaxies and a plausible source for this gas is the IGM.  Our results do not demonstrate whether the very metal-poor gas is new material accreting onto galaxies or has some other origin. Yet the simple fact that such gas resides within the gravitational potential well of galaxies makes it likely that galaxies could eventually accrete this near-pristine gas, no matter its origins.  

While it is now clear that there is very metal-poor gas in the CGM of galaxies, one might be surprised that there is not much evidence for it within 10--30 kpc, which would provide some indication that very metal-poor gas may actually reach the galaxies. The lack of evidence does not mean an absence of pLLSs and LLSs near galaxies. In fact, in our Milky Way, there is plenty of evidence of low \nhi\ absorbers ($\mlnhi \la 17.5$--18.5) within 10--20 kpc in form of ionized high-velocity clouds \citep[e.g.,][]{fox06,shull09,lehner11a,lehner12}, albeit typically at higher metallicities. The most plausible explanation for this conundrum is that the  covering factors of SLLSs and DLAs at impact parameters $\la 10$--50 kpc is substantial; any pLLSs or LLSs absorption can be easily lost in the presence of a high \nhi\ absorber. 

As we have previously argued (\citetalias{lehner13,wotta16}; \citealt{lehner17}), these very metal-poor pLLSs and LLSs have the expected physical properties (ionization state, densities) and chemical properties (metallicities) for cold flow accretion as observed in cosmological simulations (see, e.g., \citealt{fumagalli11a,vandevoort12b}). Below we confront the outputs of recent cosmological simulations with our results. However, it is quite plausible that other or multiple mechanisms are at play \citep[e.g., the precipitation models of][]{voit15}, which do not always make predictions for the metallicities and densities expected for flows through the CGM.

\subsection{Comparison with Cosmological Simulations}\label{s-disc-sim}

The results of many simulations suggest that cold flow accretion should be less frequent at $z \la 1$ than at $z>2$--3 \citep[][though see \citealt{nelson15} and \citealt{vandevoort11}]{hafen17,oppenheimer10,fumagalli11a,ubler14,stewart11a}. Above a transition galaxy mass $M_{\rm halo}\sim10^{12} M_{\odot}$, several simulations also have cold accretion becoming shock-heated \citep[e.g.,][]{birnboim03,keres05,oppenheimer10,stewart11a}, reducing the fraction of observed cold, low-metallicity gas. Given that the galaxies associated with the \citetalias{lehner13} sample range from $10^{11}\lesssim M_{\rm halo}/M_{\odot}\lesssim10^{12.5}$, it is surprising that we see such a high probability (\strongLLSfLowmetPLLSLLS) of cold gas in the low-metallicity regime of the $z \la 1$ pLLS+LLS metallicity PDF if these are probing cold flow accretion.\footnote{The range of the galaxies in the \citetalias{lehner13}  sample is $0.2<L/L^{*}<3.4$. Using the results of \citet{stocke13}, we calculate the halo mass range of these galaxies to be $10^{11}\lesssim M_{\rm halo}/M_{\odot}\lesssim10^{12.5}$.} 

To investigate this, we compare in detail our results to cosmological zoom results from the Feedback In Realistic Environments (FIRE) simulations \citep{hafen17} and the Evolution and Assembly of GaLaxies and their Environments (EAGLE) simulations \citep{rahmati18}. Both the FIRE zoom simulation and the EAGLE high-resolution cosmological hydrodynamic simulation of galaxy formation have reproduced successfully a variety of galaxy observables \citep[e.g.,][]{hopkins18,schaye15}; neither were, however, specifically calibrated to match the IGM/CGM observations. Therefore comparing the CGM properties to these types of simulations provide a new frontier to test these simulations. The FIRE and EAGLE simulations investigate the metallicities of absorbers of the pLLSs and LLSs (as well as the SLLSs ands DLAs for the FIRE simulations). These simulations have been the first to specifically quantify the metallicity PDF of the pLLSs and LLSs at $z\la 1$ and to confront their simulation results with the empirical results derived from our two previous surveys \citepalias{lehner13,wotta16}.  We emphasize that \citet{hafen17} use the EUVB model from \citet{faucher-giguere09} while \citet{rahmati18} use the \citet{haardt01} EUVB. Slight differences in the strength of the radiation field can change the \hi\ column density associated with similar structures, which may have a very minor impact on the comparison below. 

Both sets of simulations use the same method to calculate $\mlnhi$ and $[{\rm X/H}]$, where both \hi\ density and \hi-weighted metallicity are summed onto a two-dimensional grid.  We refer the reader to \citet{hafen17} and \citet{rahmati18} for the specific details on how FIRE and EAGLE compute these values and how resolution convergence of the gridding technique is tested.  However, it must be noted that this is not the same method used on the observed data where spectra are used.  Nevertheless, the gridding technique has been proven to reproduce the column densities derived from mock spectra through simulations \citep{altay11}.  

\subsubsection{Comparison with the FIRE Simulations}
\label{s-strongLLS_discussion-compare_FIRE_simulations}

The zoom FIRE simulations of \citet{hafen17} studied the CGM of using randomly-selected \hi\ absorbers around 14 simulated galaxies at $z \la 1$. The selection of their pLLSs and LLSs follows the  column density distribution function, which should be about similar to that of our survey (see \citetalias{lehner18}). This was the first study that simulated the metallicity PDFs of pLLSs and LLSs around $z \la 1$ galaxies, providing an extremely useful comparison with the empirical results derived here and in our previous surveys. In their Fig.~7, they show in particular the metallicities of the absorbers in the FIRE simulations as a function of \nhi. While this is similar to our  Fig.~\ref{f-strongLLS_met_vs_h1}, the clearest way to compare their simulation outputs with our results is to examine the CDF and PDF of metallicities over a range of column density.

\begin{figure}[tbp]
\epsscale{1.2}
\plotone{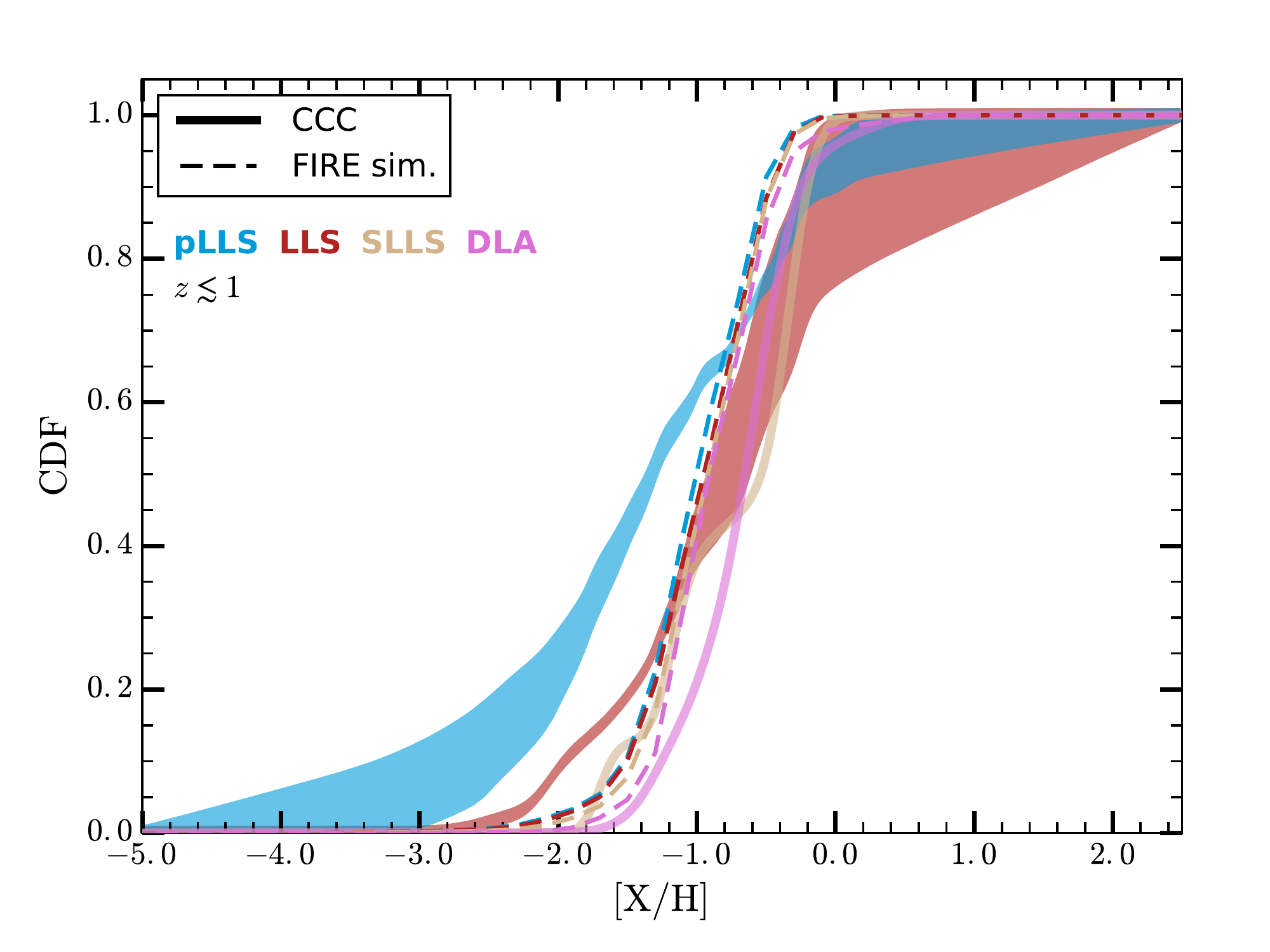}
\caption{The CDFs of the pLLS, LLS, SLLS, and DLA metallicity PDFs as a function of metallicity for the FIRE simulations of \citet{hafen17} and the CCC observations. 
\label{f-met_cdf-hafen}}
\end{figure}

\begin{figure}[tbp]
\epsscale{1.2}
\plotone{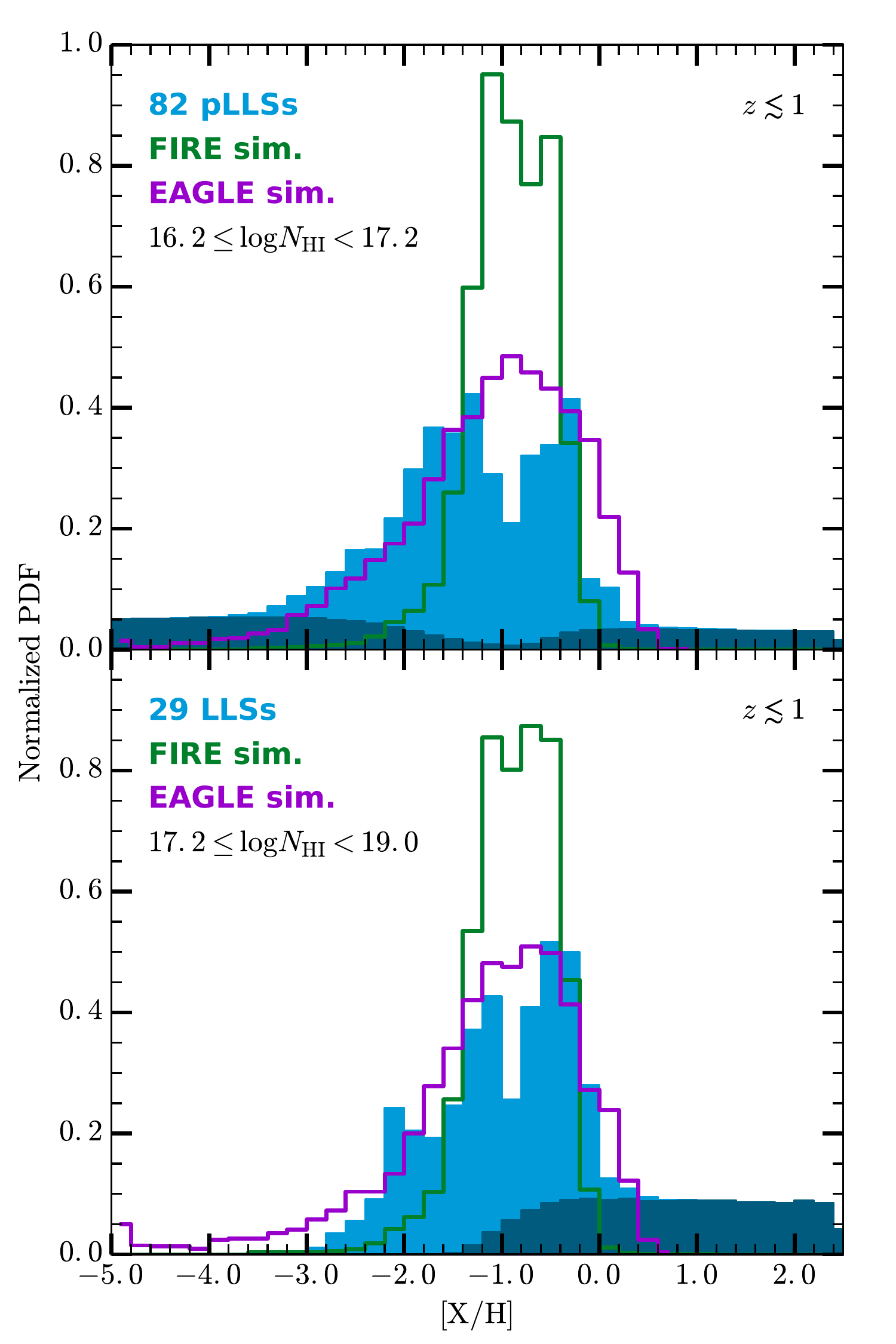}
\caption{The metallicity PDFs of the pLLSs (top) and LLSs (bottom) of from the FIRE  \citep{hafen17} and EAGLE (\citealt{rahmati18} but  matching the \hi\ column densities and redshifts of the CCC absorbers) simulations compared to the CCC results in blue.
\label{f-hafen_met_pdf}}
\end{figure}

In Fig.~\ref{f-met_cdf-hafen} we show the CDF for the FIRE simulations and the CCC observations for each class of \hi\ absorber. We also show in Fig.~\ref{f-hafen_met_pdf} the pLLS and LLS metallicity PDFs of the FIRE simulations and the CCC sample (Z.~Hafen 2018, priv.\ comm.). Based on these figures, there are marked differences between the observations and simulation outputs. First, the CDFs  in the FIRE simulations are strikingly similar across the range of \nhi, with little difference between the metallicity distributions of pLLSs, LLSs, SLLSs, and DLAs. This contrasts greatly with the observational results, where the metallicity PDFs differ significantly between the pLLSs and LLSs---both of which are different from the SLLSs and DLAs. Second, there appears to be very little low-metallicity gas in the FIRE simulations at any \nhi. The 5\% quantile of the simulated pLLS and LLS metallicity PDFs is at $[{\rm X/H}] \lesssim -1.5$; this is nearly the median of the observed pLLS PDF.  In contrast, the 5\% quantile of the CCC pLLS metallicity PDF  (at $[{\rm X/H}]\la -3$) is more than 1.5 dex (a factor $\sim30\times$) lower than in the simulations. The 5\% quantile of the CCC LLS metallicity PDF (at $[{\rm X/H}]\lesssim-2.2$) is $\sim$0.6 dex lower than in the simulations. Virtually none of the probability in the simulated metallicity PDFs resides below $[{\rm X/H}]\la -2$. These differences are not attributable to a lack of low-\nhi\ absorbers in the \citet{hafen17} sample since their \nhi\ distribution follows the standard column density distribution function (see above).  This suggests that the differences between their metallicity PDFs and ours are not caused by an \hi-selection effect.
%

Since the FIRE simulations zoom-in on a small number of galaxies (14), the galaxy selection could potentially bias the metallicity PDFs. However, the galaxies hosting the CCC absorbers likely have masses that are similar to those included in the FIRE simulations, at least based on a partially-analyzed and small subset of the CCC sample (mostly the pLLSs and LLSs in  \citetalias{lehner13}). The \hi\ absorbers in the FIRE simulations probe galaxies with $M_{\rm halo} \ge 10^{10} M_{\odot}$ (see their Fig.~3). While it is not yet clear what galaxy halo mass distribution the absorbers in our sample is, in almost every case we have been able to investigate in detail (14/15 of the \citetalias{lehner13} absorbers; see \citetalias{lehner13} and and ongoing analysis the galaxy fields where pLLSs and LLSs have been observed), we find an associated galaxy in the range $10^{11}\lesssim M_{\rm halo}/M_{\odot}\lesssim10^{12.5}$. Thus, while the sample size is yet small, it is evident that the CCC absorbers probe a wide range of galaxy masses that overlaps those of the FIRE simulations.

The lack of low-metallicity gas in the FIRE simulations could be caused in part by overly-strong feedback that enriches infalling gas; it may even be strong enough to carry enriched gas to near-IGM distances (which may fall back through the CGM in the future as intermediate- or high-metallicity inflow). Furthermore, strong feedback may cause the infall and outflows to over-mix, leading to an overabundance of metal-enriched gas (and leaving very little low-metallicity gas). This would be the scenario explored by \citet{liang16} (and see also \citealt{hummels13}), who study the effects of various feedback strengths on the galaxy and CGM properties in their simulations. They find that the feedback prescriptions producing realistic galaxies fail to match observed CGM properties, while those that reproduce the observed CGM properties overproduce the stellar masses of galaxies. The CGM resolution in these simulations is likely also an  issue. Recent high-resolution simulations  show that the CGM has physical structures on scales smaller than those typically resolved by typical cosmological hydrodynamic zoom simulations and especially for the cool gas probed by pLLSs and LLSs \citep{peeples18,vandevoort18}. Hence the physical structures like pLLSs and LLSs remain typically unresolved, causing  metals to be overmixed and the results to be resolution-dependent. We note, however, these recent high resolution simulations \citep{peeples18,vandevoort18,suresh18} explore only one single halo at $z=0$ or $z\sim 2$--3 and do not sample the cosmic variance of large scale structure. A larger sample of simulated galaxies would be needed to truly determine if resolution alone would solve most of these issues. 

At high metallicities, we find that the likelihood of absorbers in both \citet{hafen17} and this work decrease significantly at $[{\rm X/H}]\gtrsim0.0$. The outflows (and therefore the highest-metallicity LLSs) are expected to be comparable to the metallicities of the host galaxies' interstellar medium (see the discussion in \S5.2 of \citetalias{wotta16}). We might therefore expect the simulations and observations to agree at the highest metallicities if both samples trace some fraction of galactic outflow and are produced by similar galaxies. However, it may depend on the efficiency with which outflows dredge up (and carry into the CGM) surrounding interstellar medium. This ``mass loading'' factor in simulations could play a large role in determining the densities and metallicities of the outflows \citep{shen13,oppenheimer08}.

Between $[{\rm X/H}]\sim-1.1$ and $[{\rm X/H}]\sim-0.5$, there is a plateau in the metallicity PDF of the FIRE simulations; their pLLS+LLS metallicity PDF is consistent with a broad, unimodal distribution.\footnote{We note that the small peaks in probability at the edges of the \citet{hafen17} metallicity PDF derive from a statistical fluctuation due to the small number of simulated galaxies (14), and are accounted for by two $10^{12}M_{\odot}$ galaxy halos in their simulations (Z. Hafen 2018, priv. comm.), i.e., they are not statistically significant.} \citet{hafen17} conclude this is due to the variety of sources from which their absorbers arise: galaxies of different halo masses, tidally stripped gas from satellites, interstellar medium and stellar metallicities that span $>$2 dex. That is, any low-metallicity gas they find is dominated by a much larger population of intermediate- and high-metallicity gas.

While these simulations have some shortcomings, \cite{hafen17} identify origins of \hi-selected absorbers consistent with the interpretations we advanced above (and see also \citetalias{lehner13,wotta16}; \citealt{ribaudo11}). The pLLSs and LLSs in the FIRE simulations with $\xh \la -1$ to a significant degree probe gas that is inflowing onto galaxies (although there is also non-negligible fraction of metal-poor gas in the FIRE simulation that is associated with outflows). At higher metallicities, the absorbers are equally likely to probe outflowing and metal-enriched infalling (recycled) gas.

\subsubsection{Comparison with the EAGLE Simulations}
\label{s-strongLLS_discussion-compare_Eagle_simulations}

\citet{rahmati18} explored in detail the metallicities of strong \hi\ absorbers in the EAGLE simulations of \citet{schaye15}, specifically the EAGLE \emph{Recal-L025N0752} high-resolution (HiRes) volume. They specifically selected \hi\ absorbers from this 25-Mpc comoving volume for comparison against the sample of \citetalias{wotta16} \citep[see][]{rahmati18} and the CCC sample presented here, choosing sightlines matching the \hi\ column densities and redshifts for each of our absorbers.  

\begin{figure}[tbp]
\epsscale{1.2}
\plotone{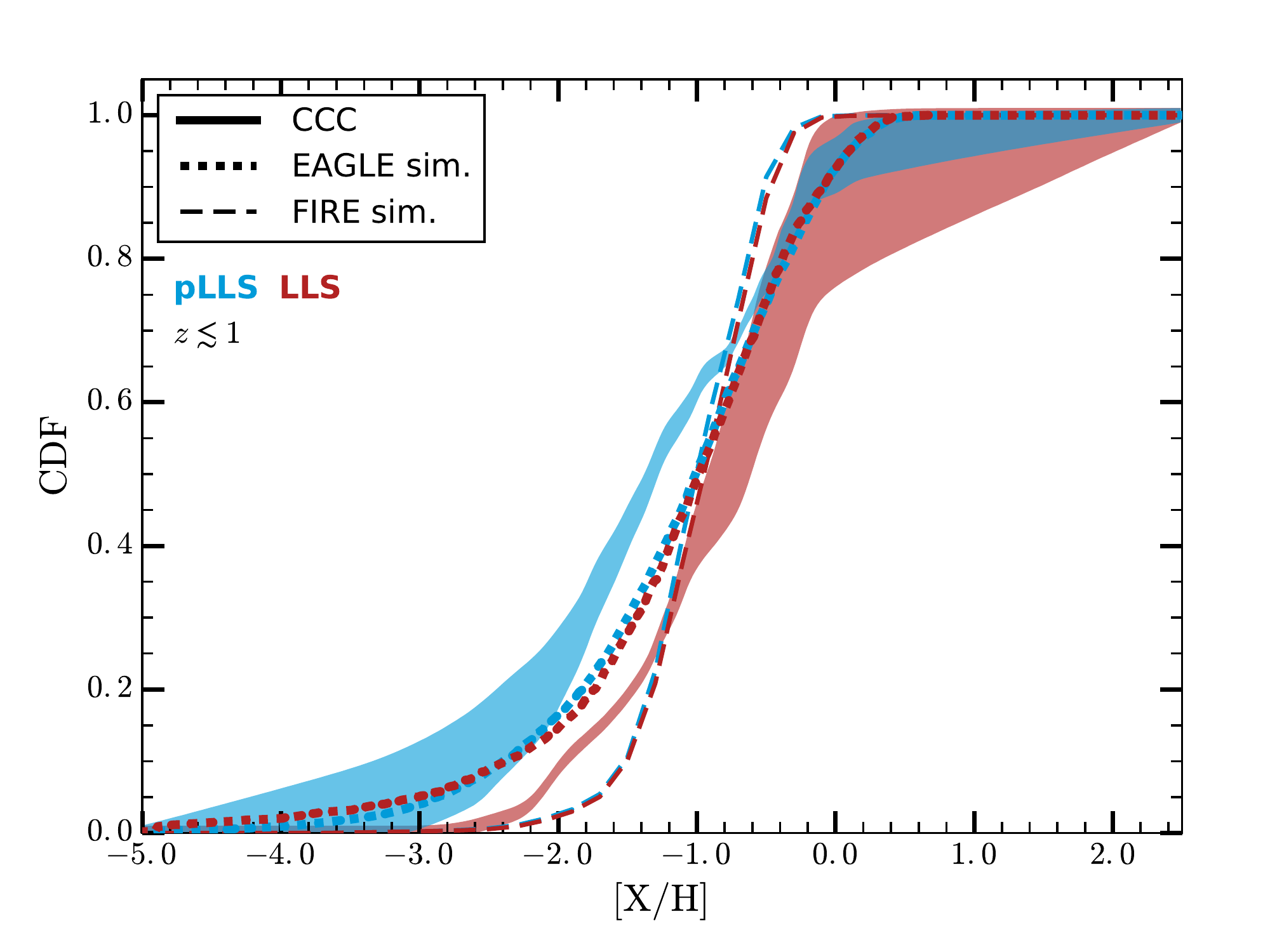}
    \caption{Comparison of the CDFs for the metallicity of the pLLS and LLSs from CCC, the EAGLE simulations (\citealt{rahmati18} but  matching the \hi\ column densities and redshifts of the CCC absorbers) and the FIRE simulations \citep{hafen17}.
    \label{f-met_cdf-rahmati-hafen}}
\end{figure}

In their Fig.~1, \citet{rahmati18} show the \hi-weighted metallicity CDFs of the absorbers in the EAGLE simulations as a function of redshift, and in their Fig.~6 they show CDFs in comparison to the \citetalias{wotta16} survey. We compare in Figs.~\ref{f-hafen_met_pdf} and \ref{f-met_cdf-rahmati-hafen} the metallicity PDFs and CDFs for the pLLSs and LLSs of this work and the EAGLE simulations (matching the observed \nhi\ and redshift distribution of the CCC absorbers). For the pLLSs, the EAGLE simulations match the CCC results much better at  $\xh<-1.2$, though the low metallicities are still underproduced. However, at  $\xh>-1.2$, the EAGLE metallicity PDF peaks in the dip of the observed metallicity PDF. These simulations may (depending on the nature of the lower limits) overproduce pLLSs at super-solar metallicities, showing a significantly higher number of metal-rich systems than the FIRE simulations. For the LLSs, the EAGLE simulations overproduce the number of low-metallicity absorbers.

From both Figs.~\ref{f-hafen_met_pdf} and \ref{f-met_cdf-rahmati-hafen}, it is evident that the EAGLE simulations produce absorbers with a much broader range in metallicities than in the FIRE zoom simulations, with many more lower- and higher-metallicity absorbers. In both simulations, however, the metallicity distributions of the pLLSs and LLSs are strikingly similar. This contrasts strongly with our empirical results, where the metallicity CDFs differ between the pLLSs and LLSs. However, while in the FIRE simulations there is virtually no difference in the CDFs of the pLLSs, LLSs, SLLSs, and DLAs (see Fig.~\ref{f-met_cdf-hafen}), in the EAGLE simulations, the metallicities of the SLLSs and DLAs increase relative to the pLLSs and LLSs as observed in the CCC. According to Fig.~1 in \citet{rahmati18}, the median metallicities of the pLLSs at $z = 0.5$ is about $\xh \simeq -1$, for the SLLSs it is $\xh \simeq -0.6$, and for DLAs is $\xh \simeq -0.4$.

\begin{figure}[tbp]
\epsscale{1.2}
\plotone{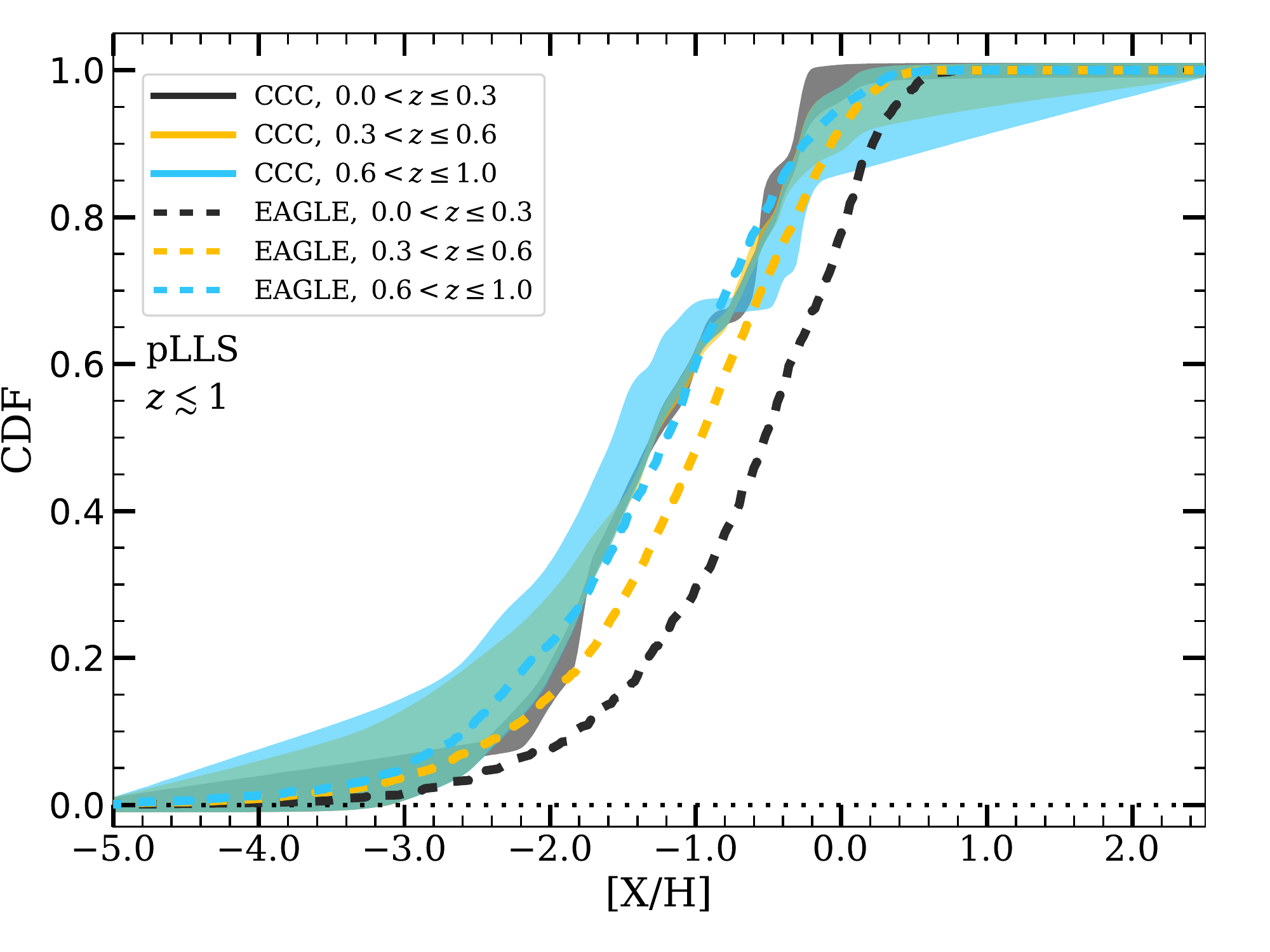}
    \caption{Comparison of the CDFs for the metallicity of the pLLS  from CCC and the EAGLE simulations in three different redshift intervals. A strong redshift evolution is observed in the EAGLE simulations that is not observed in the empirical CCC results. 
    \label{f-met_cdf-vs-z}}
\end{figure}

\citet{rahmati18} argued that the presence of a bimodality in the samples of \citetalias{lehner13} and \citetalias{wotta16} could be attributable to either small sample size or an evolution in the metallicity as a function of redshift. With the CCC, we have increased the sample of pLLSs and LLSs by a factor 4 compared with \citetalias{lehner13} and by a factor 2 compared with \citetalias{wotta16}, solidifying the observed dip in the metallicity distribution at $\xh\sim-1$ in the pLLS regime (which is also observed in the raw data, i.e., without any ionization modeling; see Fig.~\ref{f-strongLLS_mgii_vs_hi_2panels}).  \citet{rahmati18} note that there is strong metallicity evolution between $z=0$ and $z=1$ in their simulations: the  median pLLS metallicity  increases by 0.75 dex (nearly a factor 6) between $z=1$ and 0; for LLSs, it increases by 0.5 dex (factor 3) (see their Fig.~6). We revisit this finding in Fig.~\ref{f-met_cdf-vs-z} for the pLLSs using the CCC observations and the EAGLE simulations, where each simulated sightline matches the \hi\ column density and redshift of a CCC absorber. We select 3 redshift intervals so that there are enough absorbers in each bin to produce a metallicity CDF. Not surprisingly, the EAGLE simulations using the CCC sampling show a strong redshift evolution for the metallicities of the LLSs (a factor 5 between $z\le 0.3$ and $z> 0.6$ intervals at the median value). Such a strong evolution is not present in the CCC absorbers (Figs.~\ref{f-met-vs-z} and \ref{f-met_cdf-vs-z}). As we noted in \S\ref{s-met-red}, there is possibly a change in the functional form of the metallicity PDF of the pLLSs with redshift. This is observed in the CDFs near the inflection point around $\xh \simeq -0.9$, a flattening of the CDF that is somewhat more pronounced at $0.6<z\le 1$ than in the lower redshift intervals (see Fig.~\ref{f-met_cdf-vs-z}). The EAGLE simulations in the $0.6<z \le 1$ interval match best the CCC results, implying that the pLLSs and LLSs become far too metal-enriched in these simulations as $z$ decreases.

Hence while the EAGLE simulations produce a much larger fraction of low metallicity gas than is observed in the FIRE simulations, the EAGLE simulations also appear to have a similar issue: strong feedback that can cause the infall and outflows to over-mix, causing an overabundance of metal-enriched gas. This disagreement between simulations and CGM gas metallicities probed by pLLSs and LLSs may be due both to our incomplete knowledge of the physics (especially of feedback) and insufficient simulation resolution for the CGM \citep{vandevoort18,peeples18,suresh18}.

\section{Summary}
\label{s-strongLLS_summary}
With data from CCC, \citepalias{lehner18}, we have built the largest sample to date of \hi-selected pLLSs and LLSs at $z\la 1$ with 111 absorbers with $16.2 < \mlnhi < 19$. Here we estimated metallicities for 82 pLLSs and 29 LLSs in the CCC sample. An additional 152 SLFSs will be presented in \citetalias{lehner18b}. So far, searches of galaxies around sightlines with a pLLS or LLS observed in the spectrum of a $z\la 1$ QSO  have yielded nearly 100\% detection rate of at least one galaxy within projected distances of $\sim$30--200 kpc (see \citealt{lehner17}; \citetalias{lehner13}, and references therein), and hence these absorbers are good probes of the CGM of galaxies at low redshift. We have used a Bayesian formalism that exploits MCMC sampling of a grid of Cloudy photoionization models (where we assume photons from the HM05 EUVB provide the source of photoionization) to derive the posterior PDFs of the metallicities of the absorbing gas probed by the pLLSs and LLSs (as well as other quantities such the ionization parameter, densities, temperatures, etc., see  \citetalias{lehner18b}). The MCMC technique provides posterior PDFs, which can be combined and compared to output from cosmological simulations. This approach also robustly and uniformly estimates the confidence intervals on the metallicity of each absorber.  Our main findings can be summarized as follows.

\begin{enumerate}[wide, labelwidth=!, labelindent=0pt]

\item We show that a single-phase photoionization model is appropriate to match the column densities of  the low ions and often the intermediate ions seen in the pLLSs and LLSs at $z\la 1$. We show that $U$ decreases slightly from the lowest \nhi\ absorbers in the CCC through the LLSs. For a given class of absorbers, we show $\log U$ has a normal distribution, which can be used as a prior for absorbers with not enough constraints from the observed ionic ratios. 

\item We adopted the HM05 EUVB to model the CCC absorbers, but have also used the HM12 EUVB to study the systematics arising from the uncertainties in the EUVB. The HM12 EUVB yields higher metallicities,  with the correction typically larger for lower \nhi\ systems. The average HM12-derived metallicity is $+0.4$ dex higher than that derived using HM05. This is manifested as a global shift in the metallicities (i.e., it shifts $\xh$ in the same way for all absorbers). The ionization models produced with the HM12 EUVB are often in more tension with the observed column density ratios of adjacent ions than those produced with the HM05 EUVB. 

\item We show that the pLLS metallicity PDF at $z\la 1$ has two peaks at $\xh \simeq -1.7 $ and $-0.4$ and strong dip at $\xh \simeq -1$. There is some evidence that the metallicity PDF of the pLLSs may change with redshift, since the PDF of the pLLSs at $z\la 0.45$ appears to be consistent with a unimodal metallicity distribution. The metallicity PDF of the LLSs might be more complicated, with two peaks at $\xh \simeq -1.3$ and $-0.4$ and possibly a third one at $\xh \simeq -2$. 

\item The pLLS and LLS trace a very wide range of metallicities from $\xh <-2.8$ to super-solar. In contrast, DLAs and \hi-selected SLLSs have mostly metallicities in the range $-1.4 \la \xh \la 0$ over a similar redshift interval. Remarkably, very metal-poor gas with $\xh \la -1.4$ is observed frequently in the pLLS and LLS regimes and is largely absent in the gas probed by higher-\nhi\ absorbers. This disparity is not attributable to a redshift evolution effect. 

\item Very metal-poor gas  ($\xh <-1.4$) is observed in both pLLSs and LLSs, but it represents a higher portion of the population for the pLLSs (41--59\%) than for the LLSs (13--32\%). The metallicities seen in these absorbers are similar to that observed in pLLSs and LLSs at $z\sim 2$--3.5, which have a unimodal metallicity PDF peaking close to $\xh \simeq -2$. This dense, very metal-poor gas at $z\la1$ has remained poorly-enriched, even though it is currently found in the halos of galaxies, which have polluted the gas traced by the metal-rich pLLSs and LLSs to orders of magnitude higher abundances. The very metal-poor pLLSs and LLSs have properties consistent with those expected from the cold flow accretion in cosmological simulations, although it is possible that other processes may be at play. 

\item The comparison of our results with those from the FIRE simulations show that these simulations severely under-predict the amount of low-metallicity gas probed by the pLLSs and LLSs. The EAGLE simulations produce a much broader range of metallicities than FIRE at both low and high metallicities, and compare better with the CCC results. There is, however, a strong evolution of the metallicities of the  absorbers between $z\sim 0$ and 1 in the EAGLE simulations, which  is not observed in the CCC. The metallicity CDFs of the pLLSs and LLSs are also nearly identical between the FIRE and the EAGLE simulations. In both simulations, strong feedback seems to cause an overabundance of metal-enriched gas, although the lack of CGM resolution may also be part of the problem. 

\end{enumerate}

In  \citetalias{lehner18b}, we will show for the first time the metallicity PDF of the SLFSs and how the metallicities change over 7 orders of magnitude in \nhi\  ($15 < \mlnhi \la 22$). We will also explore the relative abundances in more detail (in particular $\ca$) and the physical conditions of these absorbers, including their densities and length-scales. 


\section*{Acknowledgements}
We greatly appreciate help from and thank Xavier Prochaska and Michele Fumagalli in assisting implementing the Bayesian MCMC software. We are extremely grateful to Zach Hafen for sharing and helping with the data from the FIRE simulations and for providing useful comments. We would like also to thank the anonymous referee for useful comments that helped improve the overall content of our manuscript.   Support for this research was provided by NASA through grant HST-AR-12854 and HST-GO-13846 from the Space Telescope Science Institute, which is operated by the Association of Universities for Research in Astronomy, Incorporated, under NASA contract NAS5-26555. This material is also based upon work supported by the  NASA  Astrophysical Data Analysis Program (ADAP) grants NNX16AF52G under Grant No.\ AST-1212012. KLC acknowledges support from NSF grant AST-1615296. This work was also supported by a NASA Keck PI Data Award, administered by the NASA Exoplanet Science Institute. Data presented herein were obtained at the W. M. Keck Observatory from telescope time allocated to the National Aeronautics and Space Administration through the agency's scientific partnership with the California Institute of Technology and the University of California. The Observatory was made possible by the generous financial support of the W. M. Keck Foundation. This research was supported in part by the Notre Dame Center for Research Computing through the Grid Engine software and, together with the Notre Dame Cooperative Computing Lab, through the HTCondor software. We specifically acknowledge the assistance of Dodi Heryadi and Scott Hampton.

\software{Astropy \citep{price-whelan18}, emcee \citep{foreman-mackey13}, Matplotlib \citep{hunter07}, PyIGM \citep{prochaska17a}}

\facilities{HST(COS), Keck(HIRES), LBT(MODS), VLT(UVES)}



\clearpage

\appendix

In this Appendix, we provide information regarding the supplemental files. First, we provide the MCMC input files in a machine-readable format in Tables~\ref{t-mcmc-plls} and \ref{t-mcmc-lls} for the pLLSs and LLSs in the CCC sample, respectively. In these tables (sorted by increasing right ascension), the first column provides the identification the absorber.  Columns 2 and 3 give the redshift of the absorber and its error; column 4 gives the ion or atom. Columns 5 and 6 report the column density of the ion and its $1\sigma$ error (for the purpose of the MCMC modeling, we have averaged the lower and upper error bars when they are not symmetric). Column 7 gives the flag indicating whether the measurement is a detection, an upper limit, or a lower limit (flag\,$ =0,-1,-2$, respectively). We only list in these tables the ions that were used in the MCMC photoionization modeling (for the full list of ions, one should refer to \citetalias{lehner18}).

Second, for each absorber studied in this paper, we show the residual and corner plots as shown in Figs.~\ref{f-strongLLS_MCMC_output-residual} and \ref{f-strongLLS_MCMC_output-corner}. The dashed lines in the corner plots  represent the 68\% CI for detections and the 80\% CI\ for upper and lower limits. From the corner or residual plots, one can determine readily which modeling was used: (1) if there is no entry for $\log U$ ($\log U\; {\rm prior} = {\rm False}$), then a flat prior on the ionization parameter was used; (2) if a value to $\log U$ is given, then a Gaussian prior on $\log U$ was used with the listed mean and dispersion values; (3) if $\ca$ is present, the absence of value indicates that a flat prior was used, otherwise a Gaussian prior was used on that ratio with the listed mean and dispersion values. The EUVB used in the modeling is also provided (in this case, HM05).

\startlongtable
\begin{deluxetable*}{lcclccr}
\tabletypesize{\scriptsize}
\tablecaption{Summary of the MCMC inputs for the pLLSs \label{t-mcmc-plls}}
\tablehead{
\colhead{ID}  & \colhead{$z_{\rm abs}$} & \colhead{$\sigma_{z_{\rm abs}}$}& \colhead{Ion} & \colhead{$\log N$} & \colhead{$\sigma_{\log N}$}  & \colhead{Flag} \\
\colhead{} & \colhead{} & \colhead{} & \colhead{} & \colhead{[cm$^{-2} $]}& \colhead{[cm$^{-2} $]}& \colhead{}\\
\colhead{(1)} & \colhead{(2)} & \colhead{(3)} & \colhead{(4)} & \colhead{(5)} & \colhead{(6)}& \colhead{(7)} 
}
\startdata
\hline
J011016.25-021851.0\_z0.399146 &   0.399146 &   0.000030 & \ion{H}{1}   &   16.77 &    0.02 & $  0$ \\
J011016.25-021851.0\_z0.399146 &   0.399146 &   0.000030 & \ion{C}{2}   &   13.26 &    0.24 & $ -1$ \\
J011016.25-021851.0\_z0.399146 &   0.399146 &   0.000030 & \ion{C}{3}   &   14.06 &    0.03 & $ -2$ \\
J011016.25-021851.0\_z0.399146 &   0.399146 &   0.000030 & \ion{Fe}{2}  &   12.44 &    0.24 & $ -1$ \\
J011016.25-021851.0\_z0.399146 &   0.399146 &   0.000030 & \ion{Fe}{3}  &   13.83 &    0.24 & $ -1$ \\
J011016.25-021851.0\_z0.399146 &   0.399146 &   0.000030 & \ion{Mg}{2}  &   12.16 &    0.12 & $  0$ \\
J011016.25-021851.0\_z0.399146 &   0.399146 &   0.000030 & \ion{N}{1}   &   13.64 &    0.24 & $ -1$ \\
J011016.25-021851.0\_z0.399146 &   0.399146 &   0.000030 & \ion{N}{2}   &   13.32 &    0.24 & $ -1$ \\
\enddata
\tablecomments{The flag column has the following definition: $0=$ detection; $-1 =$ upper limit (non-detection); $-2 =$ lower limit due to saturation of the line.}
\end{deluxetable*}

\startlongtable
\begin{deluxetable*}{lcclccr}
\tabletypesize{\scriptsize}
\tablecaption{Summary of the MCMC inputs for the LLSs \label{t-mcmc-lls}}
\tablehead{
\colhead{ID}  & \colhead{$z_{\rm abs}$} & \colhead{$\sigma_{z_{\rm abs}}$}& \colhead{Ion} & \colhead{$\log N$} & \colhead{$\sigma_{\log N}$}  & \colhead{Flag} \\
\colhead{} & \colhead{} & \colhead{} & \colhead{} & \colhead{[cm$^{-2} $]}& \colhead{[cm$^{-2} $]}& \colhead{}\\
\colhead{(1)} & \colhead{(2)} & \colhead{(3)} & \colhead{(4)} & \colhead{(5)} & \colhead{(6)}& \colhead{(7)} 
}
\startdata
\hline
J020930.74-043826.2\_z0.390472 &   0.390472 &   0.000100 & \ion{H}{1}   &   18.99 &    0.02 & $  0$ \\
J020930.74-043826.2\_z0.390472 &   0.390472 &   0.000100 & \ion{C}{2}   &   14.84 &    0.03 & $ -2$ \\
J020930.74-043826.2\_z0.390472 &   0.390472 &   0.000100 & \ion{C}{3}   &   14.55 &    0.20 & $ -2$ \\
J020930.74-043826.2\_z0.390472 &   0.390472 &   0.000100 & \ion{Mg}{2}  &   15.68 &    0.24 & $ -1$ \\
J020930.74-043826.2\_z0.390472 &   0.390472 &   0.000100 & \ion{O}{1}   &   14.76 &    0.24 & $ -1$ \\
J020930.74-043826.2\_z0.390472 &   0.390472 &   0.000100 & \ion{Si}{2}  &   14.09 &    0.03 & $  0$ \\
J020930.74-043826.2\_z0.390472 &   0.390472 &   0.000100 & \ion{S}{2}   &   14.28 &    0.24 & $ -1$ \\
J020930.74-043826.2\_z0.390472 &   0.390472 &   0.000100 & \ion{S}{3}   &   14.12 &    0.12 & $ -3$ \\
\enddata
\tablecomments{The flag column has the following definition: $0=$ detection; $-1 =$ upper limit (non-detection); $-2 =$ lower limit due to saturation of the line.}
\end{deluxetable*}

\phantom{}

\end{document}